\documentclass[a4paper,12pt]{JHEP3}

\usepackage{caption,graphicx,color}
\usepackage{amsmath,epsfig}
\usepackage{amssymb,amsfonts}
\usepackage{latexsym}
\usepackage[latin1]{inputenc}
\usepackage{float}
\usepackage{overpic}
\usepackage{slashed}
\usepackage{xcolor}
\let\normalcolor\relax
\usepackage{graphicx}
\usepackage{longtable}

\relax
\renewcommand{\theequation}{\arabic{section}.\arabic{equation}}

\def\be{\begin{equation}}
\def\ee{\end{equation}}

\newcommand{\de}{\partial}
\newcommand{\bear}{\begin{eqnarray}}
\newcommand{\bea}{\begin{eqnarray}}
\newcommand{\eear}{\end{eqnarray}}
\newcommand{\eea}{\end{eqnarray}}

\def\hri#1#2{\href{http://arxiv.org/abs/#1}{[ArXiv:#1]#2}}
\def\hre#1#2{\href{http://arxiv.org/abs/#1/#2}{[ArXiv:#1/#2]}}

\newbox\pippobox

\def\II{\relax{\rm I\kern-.18em I}}

\def\e{\epsilon}

\def\m{\mu}
\def\n{\nu}
\def\r{\rho}
\def\g{\gamma}
\def\s{\sigma}
\def\pa{\partial}

\def\sp{\;\;\;,\;\;\;}

\def\f{\varphi}

 \font\mybb=msbm10 at 12pt
 \def\bb#1{\hbox{\mybb#1}}

\title{Axion RG flows and the holographic dynamics of instanton densities}

\author{Yuta Hamada$^\flat$, Elias Kiritsis$^\natural$$^\flat$, Francesco Nitti$^\natural$, Lukas T. Witkowski$^\natural$
~\\
$^\natural$ \href{http://www.apc.univ-paris7.fr}{APC, AstroParticule et Cosmologie}, Universit\'e Paris Diderot, CNRS/IN2P3, CEA/IRFU,
Observatoire de Paris, Sorbonne Paris Cit\'e,\\
 10, rue Alice Domon et L\'eonie Duquet, 75205 Paris
Cedex 13, France\\
~\\
$^\flat$ \href{http://hep.physics.uoc.gr}{Crete Center for Theoretical Physics}, Institute for Theoretical and Computational Physics,
Department of Physics,  P.O. Box 2208,\\
University of Crete, 70013, Heraklion, Greece
}

\abstract{Axionic holographic RG flow solutions are studied in the context of general Einstein-Axion-Dilaton theories. A non-trivial axion profile is  dual to the (non-perturbative) running of the  $\theta$-term for the corresponding instanton density operator.  It is shown that a non-trivial  axion solution is incompatible with a non-trivial (holographic) IR conformal fixed point. Imposing a suitable axion regularity condition allows to select the IR geometry in a unique way. The solutions are found analytically in the asymptotic UV and IR regimes, and it is shown that in those regimes the axion backreaction is always negligible. The axion backreaction may become important in the intermediate region of the bulk.  To make contact with the axion probe limit solutions, a systematic expansion of the solution is developed. Several concrete examples are worked out numerically. It is shown that the  regularity condition always implies a finite allowed range for the axion source parameter in the UV. This translates into the existence of a finite (but large) number of saddle-points in the large $N_c$ limit. This ties in well with axion-swampland conjectures. }

\preprint{CCTP-2019-5\\ITCP-IPP 2019/5}

\begin{document}

\maketitle 

\section{Introduction, results and outlook}

Axions are pseudo-scalar particles which couple to topological densities. They were introduced for the first time to make the QCD $\theta$-term dynamical, in an attempt to solve the strong CP problem \cite{QCD-axion,Kim}. Since then, they have become ubiquitous in theories beyond the standard model where the term axion has come to refer to any pseudo-scalar particle which admits a shift symmetry, at least in perturbation theory. They have also been advocated as important components of the cosmological evolution of the universe, \cite{KN,Baer,Marsh} either as driving early time inflation, or being a component of dark matter or associated to dark energy.
More recently, they have been considered in the context  of the gauge hierarchy problem, \cite{relaxion}.
They have driven for some time now a program of experimental searches, \cite{GILLV,IR}, that has substantially expanded in the last few years.

They are also ubiquitous in string theory, \cite{ws}, where they originate from generalized gauge fields (forms) upon compactification.
String axions are correlated to Euclidean D-brane instantons \cite{OV,W,BCKW} which are also associated  to topological (instanton) densities.
D-brane instantons break the  shift symmetry non-perturbatively, and in cases with sufficiently low supersymmetry, \cite{BK,BCKW}, a  potential is generated.
 Its periodicity  is determined by the quantization of the associated topological charges.

What makes axions  special for  phenomenological applications  is the shift symmetry, which is exact in perturbation theory. This property is correlated with the fact that instanton densities are not renormalized and their correlators are free of UV divergences, \cite{panago}. Therefore, any potential for  axions has to be  generated non-perturbatively, which, depending on other parameters,  can keep axions light and their mass insensitive to UV physics. This property has been widely used in extensions of the standard model.

Axions play an important role in the  gauge/gravity  duality. In this context,  they are dynamical fields on the gravity side, dual  to the topological  density operator  on the field theory side. The prototypical example of this is the IIB ten-dimensional RR axion field dual to the $N=4$ super Yang-Mills (sYM) instanton density operator. By the holographic dictionary,  the leading (i.e. non-normalizable) term in the near-boundary  expansion of the axion  field is the field theory $\theta$-angle\footnote{Modulo an integer number of $2\pi$ shifts.}. Seen as a coupling constant,  the latter is usually not considered to run under the renormalization group (RG), as the instanton density operator is not perturbatively renormalized. However, in holography, a non-trivial vacuum expectation value (vev) for the instanton density operator drives a flow corresponding to a non-constant bulk axion field, and can be interpreted as non-trivial RG flow of the associated  $\theta$-angle, driven by non-perturbative effects. On the gravity side, this is captured by a holographic RG flow solution in which the  axion has a non-trivial bulk profile.

Similar finite (non-perturbative) renormalizations of the $\theta$ angle occur in non-holographic theories, with the original Seiberg-Witten theory as the simplest example, \cite{SW}.

Non-trivial bulk  axion RG flows can give rise to interesting phenomena in  holographic duals of QCD \cite{witten,iQCD,disect,VQCD}. One such phenomenon, that will be discussed further, later in this paper is the fact that holography provides a concrete framework and justification to the conjectured IR vanishing of an effective $\theta$-angle in QCD-like theories, \cite{witten,iQCD,VQCD}.
The special properties of instanton densities in QFT have also an impact in composite axion theories. As holography suggests, a bulk axion can be interpreted as the propagating effective field of states generated by an instanton density out of the QFT vacuum. Moreover, as argued in \cite{smgrav} such states have special properties, the most important being that their effective masses are insensitive to UV effects.
It is on the basis of this, that we expect that hidden theories may generate, beyond emergent gravity, also emergent axions, with properties that are distinct from conventional composite axion models, \cite{axion}.
In particular, in this framework, the emergent axion masses may be not be connected to QCD scales. The brane-world picture of such emergent axions is reminiscent of \cite{dudas}.
They may also be instrumental in connecting the self-tuning mechanism of the cosmological constant, \cite{self-tuning}, to the gauge hierarchy problem, \cite{hknw}.

Driven by these motivations, in this paper we provide  a systematic discussion of  bulk axion RG-flows in a  generic $(d+1)$-dimensional Einstein-axion-dilaton theory (which we assume to be dual to a $d$-dimensional QFT), whose Lagrangian enjoys an exact axion shift-symmetry (i.e.~the axion enters neither the potential nor the kinetic functions).  These theories may serve as bottom-up phenomenological models, or may be considered  as  proxies for  low-energy effective supergravities  emerging from top-down string theories. We shall restrict ourselves to effective bulk theories that contain a single standard scalar (dilaton) instead of many, and a single axion scalar (instead of many) without major loss of generality.\footnote{What is not included in our setup are RG flows/solutions where the fields depend on more than one internal coordinates. In the picture in which we only keep the holographic coordinate, such flows can be represented only upon the inclusion of an infinite number of KK-generated bulk fields.}  We shall consider holographic RG flow geometries, which display $d$-dimensional Poincar\'e invariance (and correspond therefore  to vacuum states of the dual QFT), of the general form
\be \label{intro5}
ds^2 = du^2 + e^{2A(u)} \eta_{\mu\nu}dx^\mu dx^\nu, \quad \f = \f(u), \quad a = a(u)
\ee
characterized by a scale factor $A(u)$, dilaton profile $\f(u)$ and axion profile $a(u)$, where $u$ is the holographic coordinate. We consider solutions  which have an asymptotic AdS-like boundary, and as such,  they are dual to field theories with a UV conformal fixed point, deformed by a relevant operator dual to the dilaton and,  generically, a $\theta$-angle.  In this context, we discuss  the full holographic RG flow geometry and compute the free energy as a function of the boundary sources, which include also the $\theta$-angle. In particular, we shall be interested in how the axion dynamics modifies the vacuum expectation value  of the relevant operator dual to  the dilaton, and how this  reflects onto the free energy of the solution.  We shall be interested  in the fully backreacted system, in which the effect of the axion on the metric and dilaton dynamics is fully taken into account.

The latter point deserves some comments. When axion running is  considered in holography, this is most often  discussed in the probe limit, i.e.~ignoring the backreaction of the axion on other bulk fields such as the metric, dilaton, etc. This is because, in known string theory examples, the axion backreaction is suppressed by $1/N_c^2$ and it is therefore subleading  in the large-$N_c$ limit. This corresponds to the gauge theory expectation since the $\theta$-term in the action, with $\theta\sim \mathcal{O}(1)$, is subleading in the large-$N_c$ limit, \cite{witten1}.  The axion running can still give $\mathcal{O}(1)$ contributions to quantities which are vanishing at leading order, such as the topological susceptibility, which indeed can be  matched to lattice results in phenomenological models \cite{iQCD,iQCD-data}. However the implicit assumption is that this does not lead to significant effects in the other sectors of the theory (e.g.~the dynamics of the running coupling and associated Yang-Mills Lagrangian operator) whose free energy is $\mathcal{O}(N_c^2)$.

Nevertheless, there are important exceptions where the axion backreaction becomes relevant: this is the case when the bulk axion becomes effectively  non-compact, and is allowed to take on arbitrarily large values, e.g.~$\mathcal{O}(N_c)$. In this case the axion contribution to  the bulk Einstein  equations  is unsuppressed. This is the case, for example, in axi-dilaton black holes with a linear axion profile \cite{Tada,Mateos,linear-axion}. In string theory, this can also occur in models with axion monodromy \cite{monodromy, monodromy2}, in which the axion decompactifies due to the coupling with extended objects. For large axion values, the axion backreaction is indeed important and hence cannot simply be ignored \cite{1011.4521, 1411.2032, 1602.06517, 1611.00394}.

The axion bulk profile is characterized by two coefficients $a_{UV}$ and $Q$,  which enter as the two integration constants of the second order axion equation of motion and  control the leading and subleading terms in the near-boundary expansion, which  schematically has the form
\be \label{intro1}
a(u) = a_{UV}  + Q \, e^{d u /\ell} + \ldots \qquad u \to -\infty,
\ee
where $u\to -\infty$ corresponds to the AdS boundary,  and $\ell$ is the (UV) AdS length. In the holographic dictionary,  $a_{UV}$ is related to the value of the $\theta$-term in the UV field theory (modulo 2$\pi$ shifts)  and $Q$ is proportional to the vacuum expectation value of the corresponding instanton density. Roughly speaking,  the probe limit corresponds to small $Q$.
 Due to the exact axion shift symmetry of the bulk Lagrangian, of the two parameters, only $Q$ enters non-trivially in the non-linear equations for the metric and dilaton. This does not mean however that the value of $a_{UV}$ does not affect the solution, as a relation between $Q$ and $a_{UV}$ may (and does) arise due to boundary conditions in the interior, as we discuss below.

 One important aspect of our analysis concerns  which boundary conditions one should impose  in the interior of the bulk geometry, where generically the metric scale factor vanishes at some coordinate value $u_{IR}$ (corresponding to the deep IR on the field theory side), which may be finite or infinity.  The requirement  that the geometry  is  regular is often too restrictive, as in Einstein-axion-dilaton models this only allows for solutions with an IR conformal fixed point and (as we shall see) constant axion. If we want to include solutions with a non-trivial axion profile, we must allow for  (apparently) singular geometries. This is not unusual in holography, as  singularities  may be  allowed  if they are mild enough (see e.g.~\cite{Gubser}) and in such cases they are expected to be resolvable using KK or stringy states. In the context of interest to us, these acceptable singularities correspond to a dilaton reaching infinity in the IR, provided its growth rate is below a certain bound which will be made explicit in Section 3. Known regular solutions as in \cite{witten}, when they are dimensionally reduced to match our setup, do indeed exhibit such a mild singularity in the lower dimensional picture, and which is automatically resolved if we restore the KK states (ie.~in the higher-dimensional description) .

The IR restrictions on the metric and dilaton however do not fix the behavior of the axion field in the interior. Indeed, even  if the axion backreaction is important in the IR, as we shall see,  a running axion never worsens  the  singularity in the interior. However,   in holography one should not expect additional freedom in the interior, as the solution should be completely fixed by the choice of  boundary sources plus some universal requirement in the IR.  This leads us to make an additional assumption  of {\em  axion regularity,} which consists in the requirement that the axion field vanishes at the IR endpoint,
\be \label{intro2}
\text{axion regularity:} \qquad a_{IR} \equiv a (u_{IR}) = 0.
\ee
This is motivated by  top-down string theory constructions, where the axion is a form field component along an  internal cycle, which shrinks to zero-size in the IR as in \cite{witten}. Single-valuedness then demands that the axion field vanishes at such points. Assuming this notion of axion regularity to hold in general leads to a consistent holographic interpretation of axion RG flows. In the probe limit, imposing equation (\ref{intro2}) results in a linear relation on the UV coefficients in (\ref{intro1}) of the type
\be \label{intro4}
Q = c\,  a_{UV}
\ee
where $c$ is a constant which depends only on the metric and dilaton profiles. However, as we shall see,  backreaction will turn (\ref{intro4}) into a  non-linear  relation. Interestingly,  the condition (\ref{intro2}) will also lead us to discard as unphysical a full class of solutions, in which $Q$ is fixed independently of $a_{UV}$. \\

Our main results are schematically summarized below.
\begin{enumerate}
\item {\bf  UV limit:} Regardless of the model and of the size of the axion leading term and subleading coefficients in the near-boundary expansion (\ref{intro1}), axion backreaction is always negligible in the UV. This is because the bulk axion has an exact shift-symmetry and does not enter the potential, thus making the leading term in (\ref{intro1}) drop out of the bulk field  equations. The remaining term, proportional to $Q$, vanishes relatively fast in the UV and does not contribute to leading order. In other words, near the UV, the geometry is always driven by the fields which contribute to the bulk potential, and the axion is not among them.  We provide an analytic expansion of the solution in the UV and  of the $\beta$-functions for the dilaton and axion couplings.

\item {\bf IR conformal fixed points:} In the IR, the backreaction of a running axion field can lead to important effects. In particular, a non-trivial axion is incompatible with a regular (holographic) IR fixed point (at which the dilaton reaches a constant value). The reason is  that the axion flow turns the asymptotic region close to the  Poincar\'e horizon into  a singularity of the bad kind. Therefore, the only solutions compatible with an axion  flow are those in which the dilaton runs to infinity (the boundary of field space) in the IR.

\item {\bf IR geometry:} Turning to solutions in which the dilaton diverges in the interior, we find that the solutions in the deep IR fall in two  classes:
\begin{itemize}\item {\em Leading} solutions, in which the axion field backreacts at leading order (without however worsening the singularity of the metric-dilaton solution);
\item {\em Subleading} solutions, in which axion backreaction enters only at subleading orders.
\end{itemize}
Both classes of solutions display ``acceptable'' singularities, whose details only depend on the large-field  behavior of the bulk potential. However, they have very different properties.

In leading solutions, the axion vev parameter $Q$ is completely fixed and independent on the axion source parameter $a_{UV}$. This makes it impossible, for a generic value of the source, to find solutions of this class  which satisfy our axion regularity condition (\ref{intro2}). Moreover these solutions are  inconsistent with any reasonable holographic interpretation (e.g. all $n$-point  correlation functions of the instanton density would vanish).

In subleading solutions, instead,  the axion can be set to zero in the IR for a generic value of the UV source, and this fixes $Q$ in terms of $a_{UV}$. In the fully backreacted solution,  the vev of the instanton density  is then (non-linearly) related to the source, which leads to a consistent holographic interpretation.

We are therefore led to discard the leading-type solutions and assume that only the subleading class is allowed in  holography. For solutions with this kind of IR asymptotics, the backreaction of a running axion is only important in the intermediate region of the bulk, far from the scaling UV and IR regions.

\item{\bf Free energy:} We perform a detailed calculation of the free energy (computed as renormalized bulk  on-shell action) of the RG flow solutions.\footnote{The general near-boundary renormalization of axion dilaton gravity theories was considered in \cite{Papa2}.}  In particular, we discuss its dependence on the parameters characterizing the axion profile. For small $Q$, when the axion can be considered as a  probe in the whole of the bulk geometry, we recover the  known large-$N_c$ result \cite{witten1,VW,witten},
\be \label{intro3}
{\cal F} (\theta) = N_c^2~{\cal F}_0 -\chi \, \text{Min}_{k} \left( \theta + 2\pi k \right)^2+{\cal O}(N_c^{-2}),
\ee
where $k$ runs over the integers, and $\chi$ is the topological susceptibility. The expression for $\chi$ is given as a function of the source and the vev of the operator deforming the theory away from the UV fixed point, and is of $O(1)$ in the large-$N_c$ expansion.

 The result (\ref{intro3})  coincides with what is also found in holography in the probe axion limit, \cite{witten,iQCD,VQCD}. Away from this limit, we generalize this result  to a more complicated, non-linear function of the $\theta$-angle when the axion source is large enough and backreaction becomes important.

\item {\bf Numerical examples:} In the fully backreacted case, analytic expressions of the solution can only be obtained in the asymptotic UV and IR regimes, but not in the intermediate region of the bulk. However, the connection between the UV parameters $a_{UV}$ and $Q$  in (\ref{intro1}), which is imposed by the IR condition  (\ref{intro2}), can only be established given the full solution, which can only be obtained numerically. In the last part of this paper,  we study numerically a large class of  examples, characterized by a UV fixed point and  parametrized in the IR by an exponential behavior of the bulk potential  and of the axion kinetic function at large dilaton, in which we study axionic RG flows. The numerics  match onto the analytical results in the asymptotic regions.

\item  {\bf Compactness of the source space}: Interestingly,  all the numerical  examples show the existence of a maximum allowed value of the axion source $a_{UV}$ in any given theory,
    \be
    |a_{UV}|\leq a_{UV}^{max}
    \label{i1} \ee
When  $|a_{UV}|> a_{UV}^{max}$   there are no regular solutions in the sense of the condition (\ref{intro2}). As $a_{UV}$ approaches its maximum allowed value, the vev parameter $Q$ diverges.
This property was observed before in \cite{VQCD}, and here we prove it  analytically and in full generality.

The connection of $a_{UV}$ to the UV $\theta$-angle is
\be
a_{UV}=c \, {\theta_{UV}+2\pi k\over N_c}
    \label{i2}\ee
where $\theta_{UV}\in[0,2\pi)$, $k\in \bb{Z}$ and $c$ a dimensionless number of ${\cal O}(N_c^0)$. To find $c$, one should know the detailed bulk-boundary correspondence. For example, in the conventional IIB normalization of the RR axion, $c=1$.

The existence of the bound (\ref{i1}) implies that for each value of $\theta_{UV}$, there is a finite number of holographic (saddle-point) bulk solutions, labelled by the integer $k$ in (\ref{i2}). The number of such solutions is ${\cal O}(N_c)$ and is reminiscent of a similar number of solutions found in the chiral Lagrangian.

The absence of regular solutions for a range of values for the source of a scalar field has been interpreted in the past (see \cite{CHS} for a recent example) as the necessity that other bulk fields must develop vevs (and therefore be non-trivial in the bulk) to allow for the existence of regular solutions.
Although this could be the case here, we believe that the axion case is special, because of the unusual identification of the axion source $a_{UV}$  with the QFT coupling $\theta_{UV}$ as in (\ref{i2}).
Moreover, the compactness of the allowed values of $a_{UV}$ is in agreement with what is obtained in theories like QCD from the chiral Lagrangian, \cite{witten1}, namely that the number of saddle points (metastable vacua) is $\sim {\cal O}(N_c)$. This is implying the existence of an upper bound on $a_{UV}$ that is $\sim {\cal O}(1)$.

\end{enumerate}

The analysis we have initiated in this paper leads to further questions and possible developments.

First, many of our results depend crucially on our assumption of axion regularity (\ref{intro2}). Although there is evidence that this is the right thing to impose, such evidence mostly comes from specific examples in string theory and from a reasonable interpretation of the holographic dictionary. Moreover, for axions that are internal components of generalized gauge fields this condition emerges from the standard condition of the vanishing of such components on Euclidean horizons.   It would be interesting to understand the axion regularity condition in a more general context, especially in the case where the non-perturbative effects of breaking of the PQ symmetry are taken into account.

One interesting result found in the numerical examples is that the axion source saturates to a maximum value as  $Q\to \infty$. This property was also  proven analytically in section \ref{Sec:Numerics}, although our bound in equation (\ref{Num20}) is not very ``efficient". This behavior was also seen in previous examples, \cite{VQCD}.
From the bulk theory point of view, this means that the total axion excursion in field space is always bounded, and at most of $\mathcal{O}(1)$ in bulk Planck units. This falls in line with observed restrictions on the axion field space in gravitational theories, \cite{ArkaniHamed:2006dz, Rudelius:2014wla, Montero:2015ofa, Brown:2015iha, 1602.06517, Long:2016jvd, Dolan:2017vmn, Hebecker:2017lxm}, and points in the same direction as the swampland ``distance'' conjectures \cite{Vafa:2005ui, Klaewer:2016kiy}. It would be interesting to understand this from a more fundamental perspective.
Already, as argued above, the compactness of the source space for the axion is in agreement with chiral Lagrangian arguments. Combining this with the IR condition (\ref{intro2}) and the monotonicity of the axion flow, it provides the bound on the axion variation. Such a line of thinking may lead to a solid proof of the axion swampland conjectures.

There are various possible phenomenological applications  of this work. One of them is in the context of holographic models of Yang-Mills/QCD. In particular, it would be interesting to understand how much the running of the $\theta$-term is compatible with observation and lattice results: we know that the $\theta$-term is small at low energy, but we do not know its UV value if there is a non-trivial running. This remark was already pointed out in \cite{iQCD}, but one may revisit that analysis in the fully backreacted case.

Finally, axionic RG flows  can be used as an important ingredient for holographic phenomenology in the context of model building. In particular, they should be added as an active  ingredient to the recently proposed self-tuning scenario of \cite{self-tuning}, in which a dilatonic brane-world is stabilized to a flat geometry  by an interplay of bulk and brane potentials. Adding a non-trivial axion may provide a way to obtain a large number of inequivalent vacua (parametrized by a different periodicity branch of the $\theta$-angle and therefore a different parameter $a_{UV}$)  over which to scan, with possibly different physical properties such as the Higgs mass and vev. This may help finding a stabilized vacuum with a naturally small value of the Higgs mass, \cite{hknw}. 

\vspace{1cm}
This paper is organized as follows. In Section \ref{general} we review how axions arise in several low-energy effective theories originating from string theory. In Section 3 we present the holographic setup, the field equations, and discuss the IR regularity conditions. In Section \ref{Sec:UV_and_IR} we solve the field equations in the asymptotic UV and IR regions, set up the probe limit as a systematic expansion around the constant axion solution, discuss the relation between the axion source and the field theory $\theta$-angle, and derive the holographic $\beta$-functions for the dilaton and axion couplings. Section 5 contains the calculation of the free energy as the bulk (renormalized) on-shell action, and shows how it reduces to known results in the probe limit. In Section \ref{Sec:Numerics} we present several examples in which the full solutions are worked out numerically. The  technical details about the asymptotic expansions and the probe approximation are left to the Appendix.
\vspace{1cm}

\section{Instanton densities and bulk axions
\label{general}}
Axions abound in string theory and their properties are well understood, \cite{ws}. They can be defined as pseudo scalars that have no potential in string perturbation theory.

The prototypical example is the RR axion of type IIB theory. It is involved in the non-trivial non-perturbative $\textrm{SL}(2,\mathbb{Z})$ symmetry of the theory along with the dilaton, and in the context of the holographic AdS$_5\times $S$^5$ solution, it is dual to the instanton density of four-dimensional $N=4$ sYM theory.
The relevant D$_3$ brane coupling is
\be
S_{CP-odd,D_3}\sim \int d^4x~a~ Tr[F\wedge F]
\label{2.0}\ee

The fact that the type IIB RR axion does not have a potential in perturbation theory is correlated with the fact that in $N=4$ sYM the instanton density does not have an anomalous dimension in perturbation theory, and does not have divergences in its correlation functions, facts that are known from the QFT side.
Of course, there are non-perturbative terms starting with the famous coefficient of the $R^4$ terms in the IIB string theory effective action, that are proportional to the $E_{3\over 2}(\tau,\bar \tau)$, Eisenstein series, \cite{gg,kp}, with $\tau=a+i e^{\f}$ where $\f$ is the string theory dilaton.
In the IIB theory, the axion has only a kinetic term, which is  of the form
\be
S_{IIB} \sim \int d^{10}x \,  e^{2\f}(\de a)^2 .
\ee

In the D$_4$ realization of a holographic YM theory, \cite{witten2}, the CP-odd coupling on D$_4$ branes,
\be
S_{CP-odd,D_4}\sim \int d^5 x ~A\wedge Tr[ F\wedge F]
\label{2.1}\ee
couples the RR bulk one form gauge field $A$ to the four-dimensional instanton density of the D$_4$-brane gauge-field strength, $F$, \cite{witten}.
Upon compactification of the 5th dimension on a circle, the 5th component of the RR one-form, $A_{\tau}$ behaves in four dimensions as a massless axion, whose shift symmetry is an avatar of the bulk gauge symmetry of $A$. Upon reduction to four dimensions, the  term in (\ref{2.1}) becomes
\be
S_{CP-odd,D_4}\sim \int d^5 x ~A_{\tau}~Tr[F\wedge F]
\label{2.2}\ee
and $A_\tau$ couples as the ten-dimensional axion to the four-dimensional YM instanton density.

Another occurrence is ABJM theory compactified on a circle, in which the CP-odd coupling of the D$_2$ brane to the IIA RR gauge field A is
\be
S_{CP-odd,D_2}\sim \int d^3 x ~A\wedge Tr[F]
\label{2.3}\ee
Like the previous example, upon compactification on a circle, the third  component of $A$, $A_{\tau}$,  becomes an axion from the two-dimensional
point a view, and the coupling (\ref{2.3}) a two-dimensional $\theta$-term,
\be
S_{CP-odd,D_3}\sim \int d^3 x ~A_{\tau}~Tr[ F]
\label{2.4}\ee

Finally, higher branes compactified on circles generate further axions and their couplings to the relevant  instanton densities.
For example the ten-dimensional RR axion $a$ couples to the six-dimensional instanton density of the D$_5$ brane,
\be
S_{CP-odd,D_5}\sim \int d^6x~a~Tr[F\wedge F\wedge F]
\label{2.5}\ee
and the RR-two form $C_2$ couples to the four-dimensional instanton density as
\be
S'_{CP-odd,D_5}\sim \int d^6x ~C_2\wedge Tr[F\wedge F]
\label{2.6}\ee
Upon compactification on a two-torus, this coupling becomes a standard four-dimensional instanton coupling with the internal components of the RR two-form $C_2$ along the torus playing the role of the axion.

In all these examples, the bulk axion coupling to instanton densities is either the string theory axion, or internal components of RR-forms. The underlying gauge invariance of these forms is the reason that the effective string axions have perturbative shift symmetries. In all cases, instanton corrections generate effectively a non-perturbative potential for the axions. Such potentials are however negligible in the $g_s\to 0$ limit (i.e.~the large $N_c$ limit) and can be ignored.

In all cases, the tree level action of the axions is of the form
\be \label{Y}
S_a = {1\over 2} \int Y(\f) g^{ab} \de_a a \de_b a
\ee
with the function $Y(\varphi)$ depending in general on the many scalar fields of the theory but not the axions. There may be topological couplings of the axions to other forms, but as we shall be seeking Lorentz invariant solutions, such couplings are usually not relevant.

To estimate the dependence of $Y$ in a simple case we consider an original gravitational theory that includes a $q$-form $C$ in $p+q+1$ dimensions along the lines discussed in \cite{GK2}, with action
\be
S_{p+q+1}=\int d^{p+q+1}x\sqrt{\hat G}\left[\hat R-{1\over 2(k+1)!}(dC)^2+\cdots\right]
\label{2.8}\ee
We compactify on $T^q$ and consider the ansatz
\be
ds^2=\hat G_{MN}dx^{M}dx^{N}=e^{-{2q\over p-1}\phi}g_{\m\n}dx^{\m}dx^{\n}+e^{2\phi}dx^Idx^I\sp C_{I_1I_2,\cdots,I_q}=a\epsilon_{I_1I_2,\cdots,I_q}
\label{2.9}\ee
where $x^I$, $I=1,2,\cdots,q$ are the internal torus coordinates and   $x^{\m}$, $\m=1,2,\cdots,p+1$ are the leftover coordinates.
Using the standard rules of dimensional reduction, \cite{GK2}, we obtain
\be
S_{p+1}=\int d^{p+1}x\sqrt{g}\left[R(g)-{q(p+q-1)\over d-1}(\pa \phi)^2-{e^{{-2q}\phi}\over 2}(\pa a)^2+\cdots\right]
\label{2.10}\ee
Normalizing the scalar field
\be
\phi=\sqrt{d-1 \over 2q(p+q-1)} \, \varphi
\label{2.11}\ee
we finally obtain
\be
Y(\f)\sim  e^{\sqrt{2q(p-1)\over p+q-1}\,\f}\;.
\label{2.12}\ee

The couplings (\ref{2.0})--(\ref{2.6}), being topological, are always purely imaginary, both in Euclidian and Minkowski signature.
This is correlated with the fact that the Euclidean two-point function of the  instanton density is (mostly) negative definite, due to reflection positivity for a pseudoscalar operator.
Therefore, if we consider the Euclidean AdS/CFT correspondence, and use a positive definite kinetic term in the bulk as in (\ref{A2}) then the correspondence is (up to normalization\footnote{These issues as well as normalizations are discussed in detail in \cite{CS}. Also, contact terms in the two point function of instanton densities are important.
In 4d YM, they are such that the topological susceptibility is positive, while the correlator is negative at finite distances, \cite{axion}.})
\be
a~~~\leftrightarrow~~~i~Tr[F\wedge F]
\label{2.7}\ee
Another property stems from the fact that RR-forms have unusual dilaton dependence in their tree level action. This translates into the fact that the correct scaling of the $\theta$-angle in 4d YM is that $\theta/N_c$ should be of ${\cal O}(N_c^0)$ in the large $N_c$ limit, \cite{disect}.

\section{Axion effective actions and RG flows}

Given the general considerations in the previous section, from this point on we take  a bottom-up perspective. We  study the axion dynamics in a  general setting which is in line with the expectations  from string theory and top-down holographic models. To this end,   we consider an Einstein-axion-dilaton theory in a $d+1$-dimensional bulk space-time parametrized by coordinates $x^a\equiv (u, x^\mu)$. The most general two-derivative action compatible with axion shift symmetry  reads,
\be\label{A2}
S = M_p^{d-1} \int d^{d+1}x \sqrt{-g} \left[R - {1\over 2}g^{ab}\de_a\f\de_b \f -{1\over 2}Y(\f)g^{ab}\pa_a a\pa_b a-V(\f)\right] + S_{GHY},
\ee
where $M_p$ is the bulk Planck scale,  $g_{ab}$  the bulk metric,  $R$  its associated Ricci scalar,  $V(\f)$ is some bulk scalar potential  and $Y(\f)$ a function controlling the axionic kinetic term, which we take to be everywhere positive\footnote{In the case of IIB dual to $N=4$ sYM, the function  $Y\sim {\cal O}(N_c^{-2})$.}. $S_{GHY}$ is the Gibbons-Hawking-York term at the space-time boundary (e.g.~the UV boundary if the bulk is asymptotically $AdS$). The fields $\f$ and $a$ are taken to be dimensionless.

As usual, the metric is dual to the QFT energy-momentum tensor and $\f$ is dual to a leading relevant scalar operator in the QFT.
If the QFT is YM theory, $\f$ would correspond to the $Tr[F^2]$ operator that controls the main coupling of YM theory. The axion is dual to a pseudoscalar operator in the QFT that has very special properties. Such operators are instanton densities in QFT.

The  bulk field equations are given by:
\bea
&& R_{ab} -{1\over 2} g_{ab} R = {1\over 2}\de_a\f\de_b \f +{Y\over 2}\de_a a\de_b a-  {1\over 2}g_{ab}\left( {1\over 2}g^{cd}\de_c\f\de_d \f+{Y\over 2}(\pa a)^2 + V \right),  \label{FE1}\\
&& \de_a \left(\sqrt{-g} \, g^{ab}\de_b \f \right)- {\de V \over \de \f}-{Y\over 2}(\pa a)^2 =0 \sp  \de_a \left(\sqrt{-g}\, Y\, g^{ab}\de_b a \right)=0 \label{FE2}
\eea

We  consider the case where the bulk space-time has $d$-dimensional Poincar\'e invariance, so that the solution is  dual to the ground state of a Lorentz-invariant QFT.
 In the domain-wall (or Fefferman-Graham) gauge,  the metric and scalar field are:
\be\label{FE7}
ds^2 = du^2  + e^{2A(u)} \eta_{\mu\nu}dx^\mu dx^\nu \, , \qquad \f = \f(u) \, , \qquad  a=a(u) \, ,
\ee
with
\be
2(d-1)\ddot A+\dot\f^2+Y \dot a^2=0 \, ,
\label{a1}\ee
\be
d(d-1)\dot A^2-{\dot\f^2\over 2}-{Y \dot a^2\over 2}+V =0 \, ,
\label{a2}\ee
\be
\ddot\f+d\dot A\dot \f-V'-{Y'\over 2}\dot a^2=0 \, ,
\label{a3}\ee
\be
\pa_u(Y~e^{dA}~\dot a)=0 \, ,
\label{a4}\ee
where a dot stands for a $u$-derivative while a prime stands for a $\f$-derivative.

\subsection{The first-order formalism\label{first}}

The axion equation (\ref{a4}) can be  integrated once to give
\be
\dot a ={Q\over Y e^{dA}} \, ,
\label{a5}
\ee
where $Q$ is an integration constant with dimension of mass.
We shall now  introduce three scalar functions of the dilaton field,  $W(\f)$, $S(\f)$ and $T(\f)$ in terms of which the full set of equations of motion reduce to a system of first-order differential equations, which we find more suitable to analyze the system.

The functions $W(\f)$, $S(\f)$ are defined in to give rise to first order flow equations for the metric and dilaton,
\be
\dot A(u)=-{W(\f(u))\over 2(d-1)},
\label{a6}\ee
\be
 \dot \f(u)=S(\f(u)),
 \label{a7}
 \ee
We  also define a function $T(\f)$ such that
 \be
 T(\f(u))\equiv {Q^2\over e^{2dA(u)}} \, .
 \label{a27} \ee
The quantities $W(\f)$, $S(\f)$  and $T(\f)$, as  functions of the dilaton as the independent variable,  are defined piecewise in any region in which the solution $\f(u)$ is monotonic. In the absence of the axion, the function $W(\f)$ reduces to the familiar superpotential of Einstein-Dilaton gravity \cite{DeWolfe:1999cp}, and $S = W'$.

Written in  terms of $T$, equation (\ref{a5}) becomes
\be
\dot a = {\rm sign}(Q) {\sqrt{T}\over Y} \, .
\label{a5-ii}
\ee
Note that,  as $Y\geq 0$ and $e^{dA}\geq 0$, the sign of $Q$ determines the monotonicity properties of the axion evolution, which do not change along the flow.

Using the definitions (\ref{a6}, \ref{a7}, \ref{a27}, \ref{a5-ii}),  the equations of motion  \eqref{a1} and \eqref{a2} become equations for $W,S$ and $T$,
 \be
 S^2-W'S+{T\over Y}=0 \, ,
 \label{a8}
 \ee
 \be
 {T'\over T}={d\over d-1}{W\over S} \, ,
 \label{a8-2}
 \ee
 \be
 {d\over 4(d-1)}W^2-{S^2\over 2}-{T\over 2Y}+V=0 \, .
 \label{a9}
 \ee
This is the main set of equations we shall investigate in this paper. The two equations in \eqref{a8} are  first order differential equations while the equation in \eqref{a9} is algebraic. Therefore, the solutions for $W$, $S$ and $T$ will depend on two integration constants. One of them can be taken to be  $Q$, which then enters in  the axion flow equation (\ref{a5}).  The second one will be denoted by $C$, and later we shall relate it to the vev of the operator associated to $\f$.  Then, solving for $a$, $A$ and $\f$ by integrating \eqref{a5-ii}, \eqref{a6} and \eqref{a7} will introduce three further integration constants. However, the integration constant associated with $A$ just redefines the constant $Q$, and is hence not a physical parameter  (equivalently, it can be chosen so that the boundary metric has unit normalization, thus fixing the scaling ambiguity in  other parameters such as $Q$)

To summarize, we have four physical integration constants, two of which come from solving for $W$, $S$, $T$ and another two which arise from integrating \eqref{a5} and \eqref{a7}. In the holographic context, the former two ($Q$ and $C$) correspond to vevs of operators in the dual QFT, while the latter two are identified with the corresponding sources.

 We can rewrite the equations \eqref{a8} and \eqref{a9} in various ways.
  Applying a $\f$-derivative to (\ref{a9}), and using both equations (\ref{a8}) and (\ref{a8-2}) we obtain
  \be
  {d\over 2(d-1)}WW'-SS'-{d\over 2(d-1)}{WT\over SY}+{TY'\over 2Y^2}+V'=0.
   \label{a28}\ee
  We may now solve either (\ref{a8}) or (\ref{a9}) for $T$, resulting in
  \be
  T=2Y\left( {d\over 4(d-1)}W^2-{S^2\over 2}+V\right)=Y(W'S-S^2)
  \label{a29} \ee
  and use these expressions in (\ref{a9}) and (\ref{a28})   to obtain equations where $T$ has been eliminated:
  \be
  -W'S+{d\over 2(d-1)}W^2+2V=0,
   \label{a24}\ee
  \be
  -SS'+{d\over 2(d-1)}WS+{Y'\over 2Y}S(W'-S)+V'=0.
  \label{a30}\ee
 Equation (\ref{a30}) is also equivalent to the Klein-Gordon equation (\ref{a3}), which was not included in the system (\ref{a8}-\ref{a9}).

 We can solve (\ref{a24}) algebraically for $S$ as
 \be
 S=2{V\over W'}+{d W^2\over 2(d-1)~ W'},
  \label{a31}\ee
 and substitute into (\ref{a30}) to obtain

\bea
&& 2(dW^2+4(d-1)V)^2~W''+ \nonumber \\
+&& 2(d-1)\left[(dW^2+4(d-1)V){Y'\over Y}+4(d-1)V'\right]W'^3- \label{aa4}   \\
-&& (dW^2+4(d-1)V)W'\left[(dW^2+4(d-1)V){Y'\over Y}+2(4(d-1)V'+dWW')\right]=0, \nonumber
\eea
which can be taken to be the master equation of the system. Solving for $W$ introduces two integration constants, which correspond to $Q$ and $C$ mentioned above.

Notice that the analogue of the holographic $c$-theorem, i.e.~$\dot{W} >0$, also holds here. Indeed,
\be
{dW\over du}={dW\over d\f}{d\f \over du}=W'S\geq 0,
\label{a32a}\ee
as the last inequality is implied by $T\geq 0$ in (\ref{a8}).

\subsection{Regularity}\label{Sec:Regularity}

Here we collect expressions for the curvature invariants of our geometry in terms of the functions $W$, $S$ and $T$. This will permit us to translate conditions arising from requiring regularity of the geometry into constraints on $W$, $S$ and $T$.

The Ricci scalar and the square of the Ricci tensor can be computed from the equations of motion to be
\be
R=-{d\over d-2}V+{1\over 2}\left(S^2+{Q^2\over Ye^{2dA}}\right)=-{2\over d-2}V+{d\over 4(d-1)}W^2 \, ,
\label{a18}\ee
\be
R_{\m\n}R^{\m\n}={d~V^2\over (d-2)^2}+{V\over d-2}\left(S^2+{Q^2\over Ye^{2dA}}\right)+{1\over 4}\left(S^2+{Q^2\over Ye^{2dA}}\right)^2
\label{a19}\ee
$$
={d~V^2\over (d-2)^2}+{V\over d-2}\left({d~W^2\over 4(d-1)}+V\right)
+{1\over 4}\left({d~W^2\over 4(d-1)}+V\right)^2
$$
$$
={d^2+4d-4\over 4(d-2)^2}V^2+{d^2~V~W^2\over 8(d-1)(d-2)}+{d^2\over 64(d-1)^2}W^4 \, ,
$$
\be
R_{\m\n\r\s}R^{\m\n\r\s}=4d(\ddot A+\dot A^2)^2+2d(d-1)\dot A^4=2d(d-1)(W^2+4V)^2+{d\over 8(d-1)^3}W^4. \,
\label{a20}\ee
It is clear from the expressions above that regularity implies that $W$ has to be regular along the flow. Another observation from equations (\ref{a18}), (\ref{a19}) is that a singularity can appear when $e^A\to 0$ which usually occurs at the IR end of the flow. In that case,  if $Y$ is regular a singularity is unavoidable.

One immediate consequence is that a regular asymptotically $AdS$ interior in the IR,\footnote{When the axion is  trivial, such an IR asymptotics   may be realized as the flow reaches a minimum  of the dilaton potential.} corresponding to an IR conformal fixed point at finite $\f=\f_0$, is incompatible with a non-trivial axion flow. This
can be seen comparing the first and second lines in equation (\ref{a19}): for  $e^{A} \to 0$ as $\f \rightarrow \f_0$, and assuming  $V(\f_0)$ and  $Y(\f_0)$  remain finite,  $W(\f)$ must necessarily diverge at $\f_0$. However this is inconsistent with a  regular IR fixed point, which requires  $W(\f_0)$ to be finite.

The conclusion above generalizes what is  known to be the case in the type IIB theory. From the axion evolution equation (\ref{a5}), the  solution is
\be
a(u)=a_{UV}+Q\int_{u_{UV}}^u {du\over Y e^{dA}},
\label{a21}\ee
where $a_{UV}$ is an integration constant and $u_{UV}=-\infty$. This integration constant is the ``source", and will be related to the $\theta$-angle of the dual  QFT in section \ref{FT}. When $Q=0$, the  constant solution is always a solution and the dual coupling ($\theta$-angle) does not flow.
The solution reduces to the solution in the absence of the axion, and the flow is regular.
If this is the unique regular solution, it corresponds to a marginal coupling.
In the case of ${\cal N}=4$ SYM and IIB supergravity, we have $Y=e^{2\f}$ and the dilaton has no potential. Therefore, the dilaton flow that could be generated by a non-zero $\dot a$ is singular and consequently the dilaton is constant along the RG flow. Then (\ref{a21}) would be singular if there is a non-trivial flow, and for this reason $\theta$ is an exactly marginal coupling (modulus).

{ To have non-trivial flows of the axion, we must have that $Y e^{dA}$ vanishes sufficiently fast at the end of the flow, and the stronger condition that $Y e^{2dA}$ is finite at the end of the flow. As we assume $Y$ is finite for finite $\f$, it follows that for this condition to hold $\f$ has to reach infinity at the end of the flow. This may appear problematic, since $\f\to \infty$ generically implies divergent $W$ and $V$, which in turns implies a singular IR. However singularities of this kind,  reached at infinity in field space, can sometimes be acceptable in holography \cite{Gubser}. This will be discussed in detail in section \ref{IR} where  we shall find that turning on a non-trivial flow for the   axion does not make the singularity worse than what it is  for a  constant axion. }

In the IR, the value of the axion satisfies
\be
a_{IR}=a_{UV}+Q\int_{u_{UV}}^{u_{IR}} {du\over Y e^{dA}},
\label{a21a}\ee
As we shall argue in section \ref{sub}, the correct IR regularity condition turns out to be
\be
a_{IR}=0
\label{a21b}\ee
in accordance with previous studies, \cite{iQCD,VQCD}.
Then (\ref{a21}) becomes
\be
a(u)=Q\int^{u}_{u_{IR}} {du\over Y e^{dA}}\sp  a_{UV}=Q\int_{u_{IR}}^{u_{UV}} {du\over Y e^{dA}}.
\label{a21c}\ee

\section{The UV and IR structure of the flows}\label{Sec:UV_and_IR}

In this section, we shall investigate the UV and IR asymptotics of the axion-driven solutions. This is important in order to understand the near boundary structure where the QFT correlators are defined as well as the structure of the on-shell action and its IR divergences.

\subsection{The UV (near-boundary) structure}

A UV fixed point is as usual, associated to a maximum of the scalar potential $V(\varphi)$. By a shift of $\varphi$ we can translate it to zero, which we shall identify with the UV fixed point of the flow.
 In the vicinity of that maximum,  the bulk functions $V(\varphi)$, $Y(\varphi)$
can be expanded in a regular power series\footnote{This is the case in all known supergravity examples that are low energy limits of string theories.} in $\varphi$.
The functions $W(\varphi)$, $S(\varphi)$ and $T(\varphi)$ can also be expanded in a series for small $\f$, but this type of series turns out to be a trans-series, that contains also non-analytic powers.
Near an extremum of the potential we have
\be
V=-{d(d-1)\over\ell^2}-{1\over2}{m^2\over\ell^2}\f^2 + V_3\f^3 +\mathcal{O}(\f^4) \, ,
\quad
Y= Y_0 + Y_1\f + \mathcal{O}(\f^2) \, .
\label{UV1}\ee
with as usual
\be
\Delta_{\pm}={d\over 2}\pm \sqrt{{d^2\over 4}-m^2\ell^2}
\label{UV29}\ee
For a maximum,  $m^2>0$,  ${d\over 2}<\Delta_{+}<{d}$ and $0<\Delta_{-}<{d\over 2}$.

 The length scale $\ell$ in (\ref{UV1}) will be identified with the AdS radius at the UV boundary of the geometry.

The expansions for $W,S,T$ for small $\f$, i.e.~the near-UV expansions,  can be found using similar techniques as in \cite{exotic}, \cite{curved}. The leading terms in this expansion are universal. As in the standard case of purely dilatonic flows, there are two branches for the solutions for $W,S,T$ depending on the coefficient of the leading $\f^2$ term in $W$ being ${\Delta_+\over 2}$ (plus-branch) or ${\Delta_-\over 2}$ (minus-branch) \cite{exotic}. More specifically, the universal terms in the expansion of the two branches are
\be \label{leadingexp}
W_\pm = {2(d-1) \over \ell} + {\Delta_\pm \over 2\ell} \f^2 + \ldots, \quad S_{\pm} = {\Delta_{\pm} \over \ell} \f +\ldots, \quad T = 0 + \ldots.
\ee
These terms are the same as one finds by  solving, perturbatively around $\f=0$,  equations (\ref{a8}-\ref{a9}) with  $T=0$ and  $S = W'$, which reduce to the superpotential equation
\be\label{supeq}
{d\over4(d-1)}W^2-{1\over2}W'^2+V=0
\ee
The order of the subleading terms in (\ref{leadingexp}) depends on the details of the potential and on the values of $\Delta_{\pm}$ and contain the integration constants which determine the vev of the operators dual to $\f$ and $a$, as we discuss below.

The full analysis is performed in appendix \ref{UV}, where the reader can find all the details.  Here we summarize some of the important results.

\begin{itemize}
\item {\bf Minus-branch} \\
The near-boundary expansions for the minus-branch of the solutions for $W, S$ and $T$ have two integration constants $C$ and $q$.
The first is related to the vev of the QFT operator dual to the scalar $\f$, while
the constant $q$ determines the vev of the QFT operator dual to the axion. These constants enter  in the expansion around $\f=0$ at subleading order compared to the universal terms written in equation (\ref{leadingexp}). Schematically, the leading terms which depend on $C$ and $q$ appear as follows:
\bea
&& W_{-(C,q)}(\f) = W_0(\f) + \ldots+ {C}|\f|^{d\over\Delta_-} + \ldots +{q\over 2d Y_0}|\f|^{2d\over\Delta_-} + \ldots , \label{w-}\\
&& S_{-(C,q)}(\f) = W_0'(\f) + \ldots+ {Cd\over\Delta_- }|\f|^{{\Delta_+\over\Delta_-}} + \ldots +  {q \, Y_1\over 2Y_0^2(d+\Delta_-)}|\f|^{2d\over\Delta_-}
+ \ldots \label{s-}\\
&& T_{-(C,q)}(\f) = {q\, |\f|^{2d\over\Delta_-}}+ \ldots   -{2Cd^2 \over (\Delta_+-\Delta_-)\Delta_-^2} |\f|^{\Delta_+ - \Delta_-\over\Delta_-} + \ldots \label{t-}
\eea
where $W_0(\f)$ has a universal, analytic expansion independent of $C,q$, starting as in equation (\ref{leadingexp}).
 For more complete expressions the reader is referred to equations (\ref{aUV2})-(\ref{aUV4}).

\item {\bf Plus-branch} \\
The near-boundary expansions for the plus-branch have a single integration constant $q$, with the same interpretation as above, but no analogue of the constant $C$.  These solutions correspond to an  RG flow driven by the vev (rather than by a source) of the operator dual to $\f$. Schematically, we have
\bea
&& W_{+(q)}(\f) = W_{0}(\f) + \ldots  +{q\over 2d Y_0}|\f|^{2d\over\Delta_+} + \ldots , \label{w+}\\
&& S_{+(q)}(\f) = W_0'(\f) + \ldots + {q  \,Y_1 \over 2Y_0^2(d+\Delta_+)}|\f|^{2d\over\Delta_+}
+ \ldots \label{s+}\\
&& T_{+(q)}(\f) = {q\, |\f|^{2d\over\Delta_+}}+  \ldots \label{t+}
\eea
The complete expansions can be found in equations (\ref{aUV5})-(\ref{aUV7}).
\end{itemize}

In both branches the near boundary behavior of the bulk fields  $A, \f$ and $a$ can be computed from the first order flows, (\ref{a5-ii})-(\ref{a7}) and the solutions found for $W,S,T$. They are given in (\ref{aUV10}), (\ref{aUV11}) and (\ref{aUV23}) for the minus branch and in (\ref{aUV12}), (\ref{aUV13}) and (\ref{aUV24}) for the plus branch.

In the minus-branch the integration constant of the first order flow for $\f$  is denoted by $\f_{-}$, and  in the standard dictionary  it is interpreted as the source for the operator dual to $\f$; in the plus-branch the integration constant is $\f_+$.
The vev of the operator  dual to $\f$ is given, in the two branches, by
\be \label{vev-C}
\langle O\rangle_-= C \, (M_p \ell)^{d-1}{d\over \Delta_-}|\f_-|^{\Delta_+\over\Delta_-}, \qquad  \langle O\rangle_+ = (M_p \ell)^{d-1} (2\Delta - d) \f_+
\ee
In both branches, the integration constant $q$ is  related to the constant $Q$ controlling the axion flow, which was introduced in \eqref{a27}, by
\be \label{aUV28-main}
Q^2=q \, {1\over\ell^2} \left(\ell |\f_\pm|^{1/\Delta_\pm}\right)^{2d}.
\ee

The UV value of the axion is determined through equation (\ref{a21a}), which can be written using (\ref{a7}-\ref{a27}) as
\be
a_{UV}={\rm sign}(Q)\int^{\f_{UV}}_{\f_{IR}} d\f~{\sqrt{T}\over YS}+a(\f_{IR}),
\label{UV8}\ee
and will be related to the field theory $\theta$-angle in subsection \ref{FT}.

From this we can read off the expectation value of the operator  $O_a$ dual to the axion in the plus- and minus-branches (labeled by $\pm$) as
\be
\langle O_a\rangle_{\pm}= {1\over N_c} {\rm sign}(Q) (M_p \ell)^{d-1} \, \sqrt{q}\,  |\f_{\pm}|^{d\over\Delta_{\pm}}.
\label{UV26}\ee

\subsection{The IR structure\label{IR}}

We shall now investigate the IR asymptotics of the bulk solutions.
As we have seen in the section~\ref{Sec:Regularity}, when the flow ends at a finite value $\f=\f_0$, $Y$ must diverge there for the flow to be regular. Therefore, $Y$ should have a pole with a more severe divergence at this finite value of $\f$. This is however not the behavior one usually obtains from effective theories of string theory.
We conclude that a flow with a non-trivial axion must occur only in cases where $\f$ runs to the boundary of its field space,\footnote{This was indeed the behavior in Improved Holographic QCD(IHQCD), \cite{iQCD} a bottom-up holographic theory constructed to emulate the dynamics of YM in 4 dimensions. This is also the behavior in V-QCD, \cite{VQCD}, which emulates the dynamics of QCD In the Veneziano limit.} i.e.~$\f\to \pm \infty$.

Then, for $\f\to \infty$ we parametrize the asymptotic behavior of the various functions that enter the gravitational action in (\ref{A2}) as
\be
V\simeq -{V_\infty\over\ell^2}e^{b\f}\sp W\simeq {W_\infty\over\ell} e^{w\f}\sp S\simeq {S_\infty\over\ell} e^{s\f}\sp Y\simeq Y_\infty e^{\g \f}
\label{a32}\ee
with $V_\infty, W_\infty,Y_\infty$ positive and dimensionless constants.\footnote{The scale $\ell$ is inserted for convenience, even though one should keep in mind that this is a length scale which defines the UV asymptotics.} {  We use $b\geq 0$ to parametrize the leading large-$\f$ behavior of the bulk potential. We also take $\gamma>0$ as required for $Y(\f)$ to diverge in the IR.} This is indeed the qualitative behavior one obtains from string theory as explained in section \ref{general}.

 We then solve (\ref{aa4}) asymptotically.
  As in the non-axionic case there are the generic singular solutions with
 $W\sim S\sim e^{\pm\sqrt{d\over 2(d-1)}\f}$ which solve (to leading) order the equations with $V=0$, and they are then corrected to full solutions.
These generic solutions violate the Gubser bound and are therefore not acceptable.
The potentially regular solutions have $w={b\over 2}$ to leading order and the coefficients satisfy the following algebraic equation:
\be
W_\infty\left[8 (d-1) V_\infty + (b^2 (d-1) - 2 d) W_\infty^2\right] \left[(4 b (d-1) V_\infty +
   \g (4 (d-1) V_\infty - d W_\infty^2)\right]=0 .
\label{a33}   \ee
This has two solutions which we denote by $W_\infty^{(L,S)}$ and which are given by
\be
W_\infty^{(L)}=\pm2\sqrt{(d-1)(b+\g)\over d\g}\sqrt{V_\infty}\, ,
\label{a35}\ee
\be
W_\infty^{(S)}=\pm
\sqrt{8V_{\infty} \over {2d \over d-1}-b^2}\, .
\label{a34}\ee
The two signs are equivalent under a reversal of the direction of the flow.

From (\ref{a31}) we subsequently obtain, to leading order as $\f\to \infty$,
\be
S\simeq {1\over\ell} S_\infty e^{{b\over 2}\f}\sp S_\infty={dW_\infty^2-4(d-1)V_\infty\over (d-1)b W_\infty}
\label{a36}\ee
and
\be
T\simeq {1\over\ell^2} T_\infty e^{(b+\g)\f}\, ,
\label{a37a}\ee
with
\be
T_\infty=Y_\infty{(dW_\infty^2-4(d-1)V_\infty)\left[((d-1)b^2-2d)W_\infty^2+8(d-1)V_\infty\right]\over 2b^2(d-1)^2W_\infty^2}\, .
\label{a37}\ee
where in expressions (\ref{a36}) and (\ref{a37}) $W_{\infty}$ is either one of  (\ref{a34}) or (\ref{a35}) depending on the chosen solution.

\subsubsection{The leading solution}

We shall start our analysis with the solution in (\ref{a35}).
By substituting it into (\ref{a36}) and (\ref{a37}) we obtain
\be
S_\infty=\pm 2\sqrt{d V_\infty\over (d-1)\g (b+\g)}={4V_\infty\over\gamma}{1\over W_\infty^{(L)}} \sp T_\infty=2 {(b^2 (d-1) - 2 d +(d-1) b\g)\over (d-1) \g (b + \g)}V_\infty Y_\infty
\label{a39}\ee
Note that this solution has a non-trivial axion profile to leading order as $T_{\infty}^{(L)}\not=0$.

The combination $S^2+{T/Y}$ that enters the curvature invariants in \eqref{a18}, \eqref{a19} and \eqref{a20} is given by
\be
S^2+{T\over Y}\simeq  {1\over\ell^2} {2 b\over \gamma}~e^{b\f} \, ,
\label{a40}\ee
i.e.~it grows with $\f$ like $V \sim e^{b \f}$. Hence, for large $\f$,  the contribution to the curvature invariants due to this term is proportional to the contribution from the remaining terms (containing $V$). This  implies that the singularity-behavior at large $\f$ of flow solutions with non-trivial axion profile is qualitatively similar to that of flows without an axion.
Therefore if the asymptotics of the potential satisfy the Gubser bound, \cite{Gubser,thermo}, which here corresponds to
\be
b\leq \sqrt{2d\over d-1} \, ,
\label{a41}\ee
then the solutions, although singular, are expected to have a resolvable singularity.

Solving (\ref{a6}) and (\ref{a7}) to leading order we obtain
\be
e^{{b\over 2}\f}\simeq {2\over bS_\infty(u_{IR}-u)/\ell}\,\,,\qquad  e^A\simeq
e^{A_{IR}}\left({u_{IR}-u\over \ell}\right)^{b+\g\over d b},
\label{a42}\ee
where $u_{IR}$ (the position of the IR singularity) and $A_{IR}$ are  integration constants. {The former is arbitrary and it reflects the freedom left by translation invariance in $u$ of the equations; the latter is a constant shift of $A(u)$ which can be fixed once and for all in the UV.  }

From (\ref{a27}) we obtain
\be
Q^2=e^{-2dA_{IR}}{T_\infty\over\ell^2}\left({2\over b S_\infty}\right)^{2(b+\g)\over b}.
\label{a43}\ee
Therefore, the solution fixes the value of $Q$. This is the analogue of the regularity condition in the IR.

Notice that the value of $Q$ is fixed independently of  the value of the source $a_{UV}$. Hence,  for a generic value of $a_{UV}$,  from (\ref{a21a}) we observe that there is no way we can make $a_{IR}=0$, which we will identify as the axion regularity condition in the next subsection \ref{sub}.

Even if we were to relax this condition, these solutions are still problematic, as we shall now argue. Indeed, for any fixed dilaton source $\f_-$,
we would obtain a  solution (with $a_{IR}\neq 0$)  whose only free parameter is $a_{UV}$ while the vev is uniquely fixed by (\ref{a43}) together with (\ref{UV26}).  This is problematic in the context of holography:
It is normal that the vev is fixed by regularity, but the vev must depend on the source\footnote{Otherwise all correlators vanish.}.
The reason this is happening here, is that the source is obtained from an exact constant solution to the equations of motion for the axion linked to the fact that this bulk field has an exact  shift symmetry\footnote{We know from string theory that this shift symmetry is broken by Euclidean D-brane instantons.  If a  potential is generated by such instantons, then it is non-perturbative in the large-N expansion. In highly supersymmetric cases like N=4 sYM no such potential is generated as discussed already in section \ref{general}.}.
This is different from the behavior seen in iHQCD, \cite{iQCD} and VQCD, \cite{VQCD}.
For these reasons, we shall discard this solution in the following, as it is not properly holographic.

\subsubsection{The subleading solution\label{sub}}

If we choose the solution (\ref{a34}) instead, we obtain
\be
S_\infty=
{b\over2} W_\infty^{(S)} \sp T_\infty=0.
\label{a38}\ee
To leading order, this corresponds to a solution with a trivial (constant) axion and $S = W'$.  Thus, to leading order the axion backreaction is negligible. This however is a starting point for obtaining a solution with a non-trivial axion.  The regular axion solutions discussed and tested in \cite{iQCD} and \cite{VQCD} are indeed of that type.
Such type of behavior in the IR was linked to dangerous irrelevant operators in systems at finite density, \cite{GK,DG}.

To set up the perturbative analysis,  we write
\be \label{sub1}
\ell W \simeq W_\infty^{(S)} e^{{b\over2}\varphi} + \delta W,
\quad
\ell S \simeq {b\over2}W_\infty^{(S)}\, e^{{b\over2}\varphi} + \delta S,
\ee
where $W_\infty$ is given in (\ref{a34}), and $\delta W$ and $\delta S$ are small perturbations in the IR. $V$ and $Y$ are the same as in \eqref{a32}.
For details on how the solutions are obtained we refer the reader to appendix \ref{aB}, and instead proceed in the following to give the results for the subleading terms:
\be
\label{sub8}
\delta W=
	-{D\over2}{ 1  \over {b\over2}+\g-{d\over (d-1)b}}e^{-\left({b\over2}+\g-{2d\over (d-1)b}\right)\f}
+\cdots \, ,
\ee
\be
\delta S=
	-{D\over2}{{b\over2}+\g \over {b\over2}+\g-{d\over (d-1)b}}
	e^{-\left({b\over2}+\g-{2d\over (d-1)b}\right)\f}
	+\cdots \, ,
\label{sub9}
\ee
 where $D$ is an arbitrary constant, which is the IR manifestation  of the integration constant $Q$ which controls the axion flow  in equation (\ref{a21}).

The leading non-vanishing contribution to $T$ is
\be
\ell^2 T=
{b\over2}D W_\infty^{(S)} Y_\infty \,e^{{2d\over(d-1)b}\f}+\cdots.
\label{sub10}\ee

As we have assumed  $b\geq 0$, demanding that the expressions (\ref{sub8}-\ref{sub9}) become indeed small as $\f \to +\infty$ results in the requirement
\be
\gamma\geq {2d\over(d-1)b}-b={2d-(d-1)b^2\over (d-1)b},
\label{sub11}\ee
In the case of equality, the subleading solution above  coincides with the non-trivial solution found in (\ref{a39}).
Using the results (\ref{sub1}) and (\ref{sub8}-\ref{sub10}) we can integrate the first order flows for the bulk fields $a,\f,A$ in the IR regime. The results  can be found in appendix \ref{aB}, equations (\ref{asub12}), (\ref{asub22}) and (\ref{asub25}).

This solution however seems to have another flaw: the parameter $D$, associated to the vev term in the near-boundary axion expansion, is not constrained by any IR regularity condition. Therefore, the axion solution seems to have two arbitrary integration constants, contrary to the holographic dictum that one of the them (the ``vev'') should be related to the other (the ``source'').

This situation has been seen before in solutions that involve (generalized) gauge fields in string theory. For example, at finite density we have non-trivial solutions for the $A_t$ component of gauge fields that seem to be regular and have two arbitrary integration constants. Moreover the vev-related constant (that in this example is the charge density) is unrelated to the source (the chemical potential). The reason is again the same. The gauge invariance of the bulk theory implies that there is always an exact constant solution for $A_t$ and it is the source.

However, in this case we know that there is a subtle IR regularity condition for $A_t$ that links the vev to the source. For example in the presence of an IR horizon or end of space (in Euclidean space they are both the same thing) $A_t$ must vanish there.
Similarly,  the many scalar axions that appear in string theory upon compactification (as discussed in section \ref{general}) correspond to generalized gauge field components and we therefore expect that we should have a similar vanishing regularity condition in the IR.\footnote{It was shown in \cite{iQCD} that such an IR condition gives results compatible with what we expect from YM.} The black $D_4$ example of Witten is realizing this boundary condition if interpreted from the 5-dimensional point of view, \cite{witten}.

We are therefore led to postulating that the proper IR regularity condition for such solutions is
\be
 a_{IR}=0
\label{sub21a} \ee

Equation (\ref{sub21a}) implies a correlation between the sign of the integration constant $Q$ in the axion equation (\ref{a5-ii}) and the source $a_{UV}=a(u\to-\infty)$.
If $a_{UV}>0$, then $Q<0$ so that the solution decreases until it vanishes in the IR. If $a_{UV}<0$, then $Q>0$ so the axion increases until it vanishes in the IR.

Imposing this IR regularity condition , the source $a_{UV}$ is obtained as
\be
a_{UV}=-{\rm sign}(Q)\int^\infty_0 d\f~{\sqrt{T}\over YS}.
\label{sub21}\ee
where we have used our conventions that $\f_{UV}=0$ and $\f_{IR}=\infty$.

As there is a single integration constant $Q$ controlling the axion flow, the two parameters  $D$ and $q$ refining, respectively,  the IR and UV expansions, must be related to each other. Indeed, such a relation is derived in appendix \ref{aB}, and it reads:
\be
{q\over D}
=\lim_{\f_{UV}\to0\atop \f_{IR}\to\infty}
{{b\over2}W^{(S)}_\infty Y_\infty e^{{2d\over(d-1)b}\f_{IR}} \over |\f_{UV}|^{2d/\Delta_-} \exp\left({d\over d-1}\int^{\f_{IR}}_{\f_{UV}}d\f\,{W\over S}\right)}.
\label{sub14}\ee
One can check that the limit is finite and non-vanishing. The  relation (\ref{sub14}) is non-linear, since  the first order functions $W$ and $S$ depend non-linearly on $q$.
The integration constant $A_{IR}$, which sets the scale of the warp factor,   is also determined by $q$ and is given in (\ref{asub26}).

\subsection{Small $q$ perturbation theory\label{q}}

We have found that the regular axion solution is subleading in the IR regime.
The same is also true in the UV regime. Because the axion is dual to a dimension $d$ (marginal) operator, its (constant) source does not affect the flow, while its vev does. The vev however starts at order  $e^{4u}$ as $u\to -\infty$  in the UV, and is therefore subleading compared to the other bulk fields.
As we shall also see explicitly in section \ref{Sec:Numerics} which collects numerical results from axion driven flows, the axion backreacts on the rest of the fields only in the intermediate regime of the flow.

Before turning to exact (numerical) solutions including the full backreaction, in this section we study the system analytically in the probe approximation, i.e. under the assumption that the axion backreaction on the metric and dilaton is small not only in the asymptotic regions, but over the entire geometry.  To set up a systematic expansion valid for a small backreacting axion\footnote{Notice that small backreaction does not require small $a(u)$ but rather small $\dot{a}(u)$, since a constant axion does not backreact on the geometry}, it is  convenient to develop a perturbation theory for small $q$,  the UV parameter  appearing in the near-boundary expansions (\ref{w-}-\ref{t-}) and related via equation (\ref{aUV28-main}) to the $Q$ parameter controlling the axion flow.
This will allow us to set  up the boundary conditions both in the UV and in the IR, thanks to equation (\ref{sub14}).
For small $q$, we shall be able to  derive an analytical formula for the leading order  correction to the non-backreacted solution with $q=0$.

Assuming small $q$, the functions $W, S$ and $T$ can be expanded as
\be
W=W^{(q0)}+ q W^{(q1)} + q^2 W^{(q2)}+\mathcal{O}(q^3),
\label{q1}\ee
\be
S=S^{(q0)} + q S^{(q1)}+ q^2 S^{(q2)}+\mathcal{O}(q^3),
\label{q2}\ee
\be
T=q T^{(q1)} + q^2 T^{(q2)}+\mathcal{O}(q^3),
\label{q3}\ee
where $S^{(q0)}=W'^{(q0)}$.
In appendix \ref{aq} we show that $W^{(q1)}$, $S^{(q1)}$ and $T^{(q1)}$ can be expressed in terms of $W^{(q0)}$ and $S^{(q0)}$. Here we quote the final result,
\be
\ell^2 T^{(q1)}=\lim_{\f_{UV}\to0} |\f_{UV}|^{2d/\Delta_-}\exp\left({d\over d-1}\int^{\f}_{\f_{UV}}d\f'\,{W^{(q0)}\over S^{(q0)}}\right).
\label{q4}\ee

\be
\ell W^{(q1)}=
-\lim_{\f_{UV}\to0}{1\over2} |\f_{UV}|^{2d/\Delta_-}e^{{d\over2(d-1)}\int^\f_{\f_{UV}}d\f'{W^{(q0)}\over S^{(q0)}}}
\int^{\infty}_{\f}d\f' {e^{{d\over2(d-1)}\int^{\f'}_{\f_{UV}}d{\f''}{W^{(q0)}\over S^{(q0)}}}\over Y \ell S^{(q0)}},
\label{q9}\ee
\be
S^{(q1)}=
{d\over2(d-1)}{W^{(q0)}\over S^{(q0)}}W^{(q1)}-{T^{(q1)}\over2YS^{(q0)}},
\label{q10}\ee
Therefore the corrections at order $\mathcal{O}(q^1)$ can be written in terms of the $\mathcal{O}(q^0)$ solutions $W^{(q0)}$ and $S^{(q0)}$.

In this way, from equation (\ref{sub14})   we obtain the linearized relation between the integration constants $q$ (from the UV) and $D$ (from the IR),
\be
D=
q \lim_{\f_{UV}\to0\atop \f_{IR}\to\infty}
{|\f_{UV}|^{2d/\Delta_-}e^{{d\over d-1}\int^{\f_{IR}}_{\f_{UV}}d\f'\,{W^{(q0)}}}\over{b\over2}W_\infty Y_\infty e^{{2d\over(d-1)b}\f_{IR}}}
+\mathcal{O}(q^2),
\label{q20}\ee

{ As the axion backreacts,  the integration constant $C$ which appears in $W$ at order $\f^{d/\Delta}$, and  that controls the vev of $O$ as in (\ref{UV26}), also gets modified. Indeed, there will be terms in $W^{(q1)}, W^{(q2)}$, etc, which will contribute to that  order.  Therefore, the integration constant $C$ becomes effectively a function of $q$, and we can develop it in a small-$q$ expansion, }
\be
C=C^{(q0)}+ q\, C^{(q1)} + q^2 C^{(q2)}+\mathcal{O}(q^3)\, ,
\label{q18}\ee
and calculate the ${\cal O}(q)$ correction  to the UV constant $C$:
\be
C^{(q1)}=
-\lim_{\f_{UV}\to0}{1\over2} |\f_{UV}|^{d/\Delta_-}\int^\infty_0 d\f' {e^{{d\over2(d-1)}\int^{\f'}_{\f_{UV}}d{\f''}{W^{(q0)}\over S^{(q0)}}}\over Y \ell S^{(q0)}},
\label{q17}\ee

Finally, requiring axion regularity as explained in section \ref{sub},  we can obtain from \eqref{sub21} the small $q$ perturbation of the source of the axion $a_{UV}$:
\begin{align}
\nonumber a_{UV} &=
-{\rm sign}(Q)\sqrt{q} \lim_{\f_{UV}\to0} |\f_{UV}|^{d/\Delta_-} \int^\infty_0 d\f {e^{{d\over2(d-1)}\int^{\f}_{\f_{UV}}d\f' {W^{(q0)}\over S^{(q0)}}}\over Y \ell S^{(q0)}}+\mathcal{O}(q) \\
&=2 \, {\rm sign}(Q) \sqrt{q} \, C^{(q1)}+\mathcal{O}(q).
\label{q19}\end{align}

\subsection{The axion source and the UV value of the $\theta$-angle.\label{FT}}

The source of the axion solution $a_{UV}$ in (\ref{a21}) and (\ref{UV8}), is related to the $\theta$-parameter $\theta_{UV}$ of the UV CFT as
\be
\theta_{UV}+2\pi k=N_c a_{UV},
\quad k\in \mathbb{Z}, \quad 0\leq\theta_{UV}<2\pi.
\label{UV9}\ee
It should be remarked that the relation of $\theta_{UV}$,  that is a field theory coupling with values in $[0,2\pi]$, and $a_{UV}$ that is a bulk source is many-to-one. There is an infinite number a priori of real values of $a_{UV}$ that give rise to the same value of $\theta_{UV}$, corresponding to the different values of the integer $k$ in (\ref{UV9}).
Different values of $k$ correspond to the same $\theta_{UV}$ but different values of $a_{UV}$,  and therefore to different saddle points of the bulk theory. They do not have the same energy, and the dominant saddle point is the one with the lowest energy, \cite{Witten-th,witten,iQCD}.
The saddle-points with different values of $k$ are separated by axionic D-branes, which are branes that couple magnetically to the axion, \cite{witten,disect}.

One of the important questions that we address in this paper is the range of values that $a_{UV}$ can take. We  find, that although, $a_{UV}$ as a boundary condition can take any real value, there are ``regular" solutions\footnote{The notion of regularity here is generalized \`a la Gubser, as explained  earlier. Such solutions are actually mildly singular (with a ``good" singularity in the definition of \cite{Gubser}). They are expected to be resolvable in string theory. A textbook example of this is is the Witten solution,  \cite{witten}, of the black D$_4$ theory, which is regular in 6 bulk dimensions but is singular, with a good singularity once we compactify to five dimensions.} only for a compact set of $a_{UV}$ values
\be
a_{UV}\in \left.\Big[0,a_{UV}^{max}\right)
\label{za}\ee
For $k=0$ in (\ref{UV9}), $a_{UV}$ is infinitesimal at large $N_c$.
In that case there is always a regular solution as shown in \cite{iQCD}.
However, when backreaction of the axion is taken into account this ceases to be true anymore, as was already seen in \cite{VQCD}.
In section \ref{Sec:Numerics} and equation (\ref{Num20}) we show that there is always a finite upper bound $a_{UV}^{max}$ for a regular solution to exist.

From (\ref{UV9}) we obtain
\be
a_{UV}={\theta_{UV}+2\pi k\over N_c}\sp k\in \bb{Z}
\label{z}\ee
and from this and (\ref{za}) we
deduce that the number of distinct (saddle-point) solutions with the same $\theta_{UV}$ is equal to the number of possible values the integer $k$ can take in (\ref{z}).
This number is
\be
\bb{Z}\ni n=\left\lfloor{N_c a_{UV}^{max}\over 2\pi}\right\rfloor
\label{z2}\ee
where $\lfloor z \rfloor$ is the maximum integer smaller than or equal to the real number $z$.
For large $N_c$ this is a large number, that should be compared to a similar number emerging from the chiral Lagrangian, see section 5 of \cite{VQCD}.

\subsection{The holographic $\beta$-functions\label{beta}}

The first order flow equations derived in section  \ref{first} can be used to define  the holographic $\beta$-functions along the lines discussed in \cite{exotic}:
\be
\beta_{\f}(\f) \equiv {d\f\over dA}=-2(d-1){S(\f)\over W(\f)} \, ,
\label{b1} \ee
 \be
 \beta_{\theta}(\f)\equiv {da\over dA}=-2(d-1){\sqrt{T}\over W~Y}=-2(d-1){\sqrt{W'S-S^2}\over W\sqrt{Y}} \, .
\label{b2}\ee
It should be remarked that, as expected, the $\beta$-functions do not depend on the axion field, as it is protected by the perturbative Peccei-Quinn symmetry.

Using (\ref{a24}), the $\beta$-functions can be expressed in terms of the functions $V(\phi)$ and $Y(\f)$ and the superpotential $W(\f)$.
\be
\beta_{\f}(\f)=-d{W\over W'}-4(d-1){V\over WW'} \, ,
\label{b3}\ee
\be
\beta_{\theta}=-2(d-1){\sqrt{{dW^2\over 2(d-1)}+2V-\left({d\over 2(d-1)}{W^2\over W'}+2{V\over W'}\right)^2}\over W\sqrt{Y}} \, .
\label{b4}\ee

Using \eqref{aUV2}, \eqref{aUV3} and \eqref{aUV4}, the UV expansions of the $\beta$-functions are given by
\be
\beta_{\f}= \bigg(- \Delta_- \f + {\cal O}(\f^2)\bigg)- {C d\over \Delta_-} |\f|^{\Delta_+\over\Delta_-}\bigg\{1 + \left( {d\over8} - {\Delta_-^2\over2d}\right) {\f^2\over d-1}+ {\cal O} \left(\f^3, C |\f|^{\Delta_+-\Delta_-\over\Delta_-}\right)\bigg\}
\label{b5}\ee
$$
- q |\f|^{2d\over\Delta_-} {Y_1 \over2Y_0^2(d+\Delta_-)}  \bigg\{ 1 + \mathcal{O}\left(\f, C |\f|^{\Delta_+-\Delta_-\over\Delta_-}, q |\f|^{{2\Delta_+\over\Delta_-}+1}\right) \bigg\},
$$
\be
\beta_\theta= - {\sqrt{q} \over Y_0} |\f|^{d\over\Delta_-}
\bigg[
\bigg(1+{\cal O}(\f) \bigg) - {C d^2 \over (\Delta_+ - \Delta_-) \Delta_-^2} |\f|^{\Delta_+ - \Delta_- \over\Delta_-}\left\{1+\mathcal{O}\left(\f, C |\f|^{\Delta_+ - \Delta_-\over\Delta_-}\right)\right\}
\label{b6}\ee
$$
+\mathcal{O}\left(q |\f|^{{2\Delta_+\over\Delta_-}+1}\right)\bigg],
$$
for the minus-branch. The plus-branch solutions are obtained by the replacement $\Delta_-\to\Delta_+$ and $C=0$.
When the UV dimension of $\f$ is $d$, then it corresponds to a marginally relevant coupling and the vev-related terms change.
As shown in \cite{bourdier,exotic},
\be
|\f|^{d\over \Delta_{-}}\to e^{-d\int{d\f\over \beta_{\rm pert}(\f)}}\sp \beta_{\rm pert}=-2(d-1){W_{p}'\over W_p} \, ,
\ee
 where $W_p$ is the power series part of the superpotential in (\ref{aUV2}).

Using \eqref{sub8}, \eqref{sub9} and \eqref{sub10}, the IR expansions are given by
\be
\beta_{\f} = - (d-1)
\bigg(
b - {D\over W_\infty} {\g \over {b\over2}+\g - {d\over(d-1)b}} e^{-\left(b+\g-{2d\over(d-1)b}\right)\f}
\bigg) + \cdots \, ,
\label{b7}\ee
\be
\beta_\theta = -(d-1) \sqrt{2b D\over W_\infty Y_\infty} e^{-\left({b\over2}+\g-{d\over(d-1)b}\right)\f}+\cdots \, .
\label{b8}\ee
Both $\beta_\f$ and $\beta_\theta$ vanish in the  UV, a fact  obvious from \eqref{b5}, \eqref{b6}.
Using \eqref{asub13} and $b>0$,  we observe that, in  the  IR, $\beta_\f$ goes to a constant value (related to the hyperscaling violating geometry, \cite{GK2,S}) and $\beta_\theta$ vanishes.
In section 6, we shall provide the numerical plot of the $\beta$-function in figure \ref{fig16}.

\section{The on-shell effective  action}
In this section, we calculate the on-shell action and free energy for holographic RG flow solutions with non-trivial axion profile in the bulk.

\subsection{General expressions for the on-shell action and free energy}
The on-shell action is given by the bulk action \eqref{A2} evaluated on a RG flow solution with metric ansatz \eqref{FE7}. As a first step, note that for the ansatz \eqref{FE7} the bulk curvature scalar $R$ can be re-expressed as
\be
R=-2d\ddot{A}-d(d+1)\dot{A}^2={1\over2}\dot{\f}^2+{1\over2}Y\dot{a}^2+{d+1\over d-1}V \, ,
\label{Free1}\ee
where in the second equality we used \eqref{a1} and \eqref{a2}. Substituting this into \eqref{A2} one obtains
\be
S_{\rm on-shell}={2\over d-1}M_p^{d-1} V_d \int^{\rm IR}_{\rm UV} du\, e^{dA} V+S_{GHY} \, ,
\label{Free2}\ee
where $V_d \equiv \int d^d x$ is the (formally infinite) volume of $d$-dimensional Minkowski space.
The integration over $u$ is performed from the UV fixed point at $u \rightarrow - \infty$ to the IR of the flow, which here is reached for to $u \rightarrow + \infty$. It will be convenient to rewrite \eqref{Free2} further. Using the equations of motion we can write
\be
V=-(d-1)\ddot{A}-d(d-1)\dot{A}^2.
\label{Free3}\ee
Further, as our bulk geometry exhibits a boundary, the Gibbons-Hawking-York term also contributes. For our ansatz \eqref{FE7} it can be shown to give
\be
S_{GHY}=-2dM_p^{d-1}V_d \left[e^{dA}\dot{A}\right]_{\textrm{UV}}.
\label{Free5}\ee
Inserting this, the expression for the on-shell action becomes
\begin{align}
\nonumber S_{\rm on-shell} &=-2(d-1)M_p^{d-1} V_d \left[e^{dA}\dot{A}\right]_{\textrm{UV}} -2M_p^{d-1} V_d \left[e^{dA}\dot{A}\right]_{\textrm{IR}} \\
&=M_p^{d-1} V_d \left[e^{dA} W \right]_{\textrm{UV}} + \frac{1}{(d-1)} M_p^{d-1} V_d \left[e^{dA} W \right]_{\textrm{IR}}.
\label{Free6}
\end{align}
The difference between the  coefficients in the UV and IR contributions arises because the GHY term is included only on the UV boundary.

Here we are exclusively interested in solutions which have a behaviour in the IR (i.e.~for $\f \rightarrow + \infty$) as described in section \ref{sub}. The corresponding expression for $W$ and $A$ as functions of $\f$ are given in \eqref{sub1} and \eqref{asub25}, respectively. Using these expressions, the IR contribution to \eqref{Free6} can be shown to give
\begin{align}
\left[e^{dA}W\right]_{\textrm{IR}} = \left[e^{dA} W \right]_{\f \rightarrow + \infty} \rightarrow W_{\infty} e^{dA_{\textrm{IR}}} \,  e^{- \frac{2d - (d-1) b^2}{2 (d-1) b} \f} \, .
\end{align}
Note that if the parameter $b$ satisfies the Gubser bound \eqref{a41} the exponent in the above is negative and the IR contribution vanishes. Hence, for flows in potentials satisfying the Gubser bound we are left with
\begin{align}
S_{\rm on-shell}=M_p^{d-1} V_d \left[e^{dA} W\right]_{\textrm{UV}}
\, .
\label{Free7}
\end{align}
This expression is formally the same that one finds without the axion (see e.g. \cite{kn,hrg}). The dependence on the axion enters indirectly through its modification of $A$ and $W$.

\vspace{0.3cm}
\noindent \textbf{The renormalized on-shell action and free energy.} The on-shell action as written in \eqref{Free7} is divergent and requires renormalization. As a first step we regularize by introducing a (dimensionless) cutoff as
\be
\Lambda\equiv \left.e^{A(u)}\over\ell \, |\f_-|^{1/\Delta_-}\right|_{u=\ell \log\epsilon} \, .
\label{Free8}\ee
Using this the regularized on-shell action is given by
\be
S_{\rm on-shell}
=(M_p \ell) ^{d-1} V_d \, |\f_-|^{d\over\Delta_-} \left[\Lambda^d \, \ell W(\f_\e)\right] ,
\label{Free12}\ee
with $\f_\e \equiv \f(\ell \log \e)$. Next, from \eqref{aUV2} it follows that $W(\f_\e)$ can be written as an expansion in powers of $\f_\e$. Using the near-UV expression for $e^{A(\f)}$ in \eqref{aUV11}  we can then write $W$ in terms of an expansion in increasingly negative powers of $\Lambda$. The combination $\Lambda^d \, \ell W(\f_\e)$ will hence contain a finite number of terms which diverge for $\Lambda \rightarrow \infty$, with the precise number of divergent terms depending on the exact values of $d$ and $\Delta_-$.

The divergences can then be removed by adding the following counterterm to the on-shell action (see e.g. \cite{Papa2}):
\be
S_{ct} = -M_p^{d-1} \Big[ \int d^dx \sqrt{|\g|}\, W_{ct}(\f) \Big]_{\substack{u=\ell \log \e \\ \f=\f(\ell \log \e)}} = - (M_p \ell) ^{d-1} V_d \, |\f_-|^{\frac{d}{\Delta_-}} \Big[ \Lambda^{d} \, \ell W_{ct}(\f_\e) \Big] \, ,
\label{Free13}\ee
where $\gamma_{\mu\nu}(u)=e^{2A(u)}\eta_{\mu\nu}$ is the induced metric on a constant-$u$-slice with $|\gamma| = e^{dA}$. The function $W_{ct}$ is defined as the solution of equation \eqref{a9} with $T=0$, i.e.
\be
{d\over4(d-1)}W_{ct}^2-{1\over2}(W_{ct}')^2+V=0.
\label{Free17}\ee
In the vicinity of the UV at $\f=0$ the function $W_{ct}$ then exhibits the following expansion in powers of $\f$:
\be
W_{ct}={2(d-1)\over\ell}+{\Delta_-\over2\ell}\f^2+{\cal O}(\f^3)+{C_{ct}\over\ell}|\f|^{d\over\Delta_-}+\cdots,
\label{Free18}\ee
i.e.~this takes the same form as the near-boundary expansion of $W$ given in \eqref{aUV2}, but with $q=0$. Here $C_{ct}$ is the integration constant introduced when solving for $W_{ct}$. This parameter can be freely chosen, with a particular choice of $C_{ct}$ corresponding to a choice of renormalization scheme.

Putting everything together, the renormalized on-shell action is given by
\begin{align}
S_{\textrm{on-shell}}^{\textrm{ren}} = \lim_{\epsilon \rightarrow 0} \Big( S_{\rm on-shell}
 + S_{ct} \Big) = \lim_{\epsilon \rightarrow 0} M_p^{d-1} V_d \left[e^{dA}\left(W-W_{ct}\right)\right]_{UV} \, .
\label{Free15}\end{align}
Inserting the near-UV expansions for $W$ and $W_{ct}$ in \eqref{aUV2} and \eqref{Free18} and taking the cutoff to infinity we finally arrive at
\begin{align}
S_{\textrm{on-shell}}^{\textrm{ren}} = \left(M_p \ell\right)^{d-1}  V_d \, |\f_-|^{d\over\Delta_-} \, \big( C(q)-C_{ct} \big).
\label{Free15b}\end{align}
Again, this is formally similar to the result one obtains  in the absence of the axion \cite{Papa1,kn,hrg}, with one important difference: we have written the integration constant $C$ coming from $W$ as a function of the other integration constant $q$, which parametrizes the axion running. The reason  is that only for certain combinations of $C$ and $q$ a flow emanating from the UV can be completed into a solution with regular IR behaviour. Hence on regular solutions the allowed values of $C$ and $q$ are correlated which allows us to consider one as the function of the other.

Given the renormalized on-shell action, the (Euclidean) free energy is defined as
\be
F_k \equiv - S_{\rm on-shell}^{\textrm{ren}}=- \left(M_p \ell\right)^{d-1}  V_d \, |\f_-|^{d\over\Delta_-} \, \big( C(q_k)-C_{ct} \big) \, ,
\label{Free16}\ee
$$
=- \left(M_p \ell\right)^{d-1}  V_d \, |\f_-|^{d\over\Delta_-} \, \left[ C\left({\theta_{UV}+2\pi k\over N_c}\right)-C_{ct} \right] \, .
$$
where we used (\ref{z}).
It should be noted that the free energy depends on the integer $k$ that labels the distinct saddle points of the theory.

Note that this exhibits several features familiar from QCD. The parameter $\f_-$ corresponds to the mass scale of the theory and is the analogue of $\Lambda_{\textrm{QCD}}$ \footnote{In the case at hand $\f_-$ is a dimensionful coupling, but one can also modify the setup so that the operator deforming the UV theory is marginally relevant like the QCD coupling. This can be achieved  by setting the mass term to zero the UV expansion of the potential, in which case the running is driven by the cubic or higher terms \cite{exotic}. Alternatively one can realize the UV as a runaway AdS solution, as in the Improved Holographic QCD models \cite{iQCD}. In either case, the scale $\Lambda_{\textrm{QCD}}$ is dynamically generated.}.  Further, like in QCD, there is another dimensionless coupling which here is given by $\theta_{UV}$. As $C$ is a dimensionless parameter, it only depends on the dimensionless coupling $\theta_{UV}$ (through $q$).
Then, we can recognize in \eqref{Free16} the structure of the free energy familiar from QCD, i.e.
\be
F_k \sim \Lambda_{\textrm{QCD}}^4 V\left({\theta_{UV}+2\pi k\over N_c}\right)
\label{F21}\ee
for $d=4$.
This is a general feature of holographic QCD-like theories \cite{glueball}.

By taking functional derivatives of $F$ with respect to the sources $\f_-$ and $\theta_{UV}$ we can obtain the vevs of the operators $O$ and $O_a$ dual to $\f$ and $a$ respectively. For $\langle O \rangle$ we obtain
\be
\langle O \rangle_k = - \frac{\delta F_k}{\delta \f_-} = (M_p \ell)^{(d-1)} |\f_-|^{{\Delta_+\over\Delta_-}}{d\over\Delta_-} (C(q_k)-C_{ct}).
\label{Free20}\ee
This is consistent with the expectation \eqref{UV26}, up to the scheme-dependence introduced through $C_{ct}$.

Similarly, for the  instanton density operator we find
\be
\langle O_a \rangle_k = - \frac{\delta F_k}{\delta \theta_{UV}}
= (M_p \ell)^{(d-1)} |\f_-|^{d\over\Delta_-} {1\over N_c} {\partial C(a_{UV})\over \partial a_{UV}}\Big|_{a_{UV}={\theta_{UV}+2\pi k\over N_c}} \,
\label{Free26}\ee
$$
= {1\over N_c}{\rm sign} (Q) (M_p \ell)^{(d-1)} \sqrt{q_k}\, |\f_-|^{d\over \Delta_-} + \mathcal{O}(q),
$$
The last equality follows from the small $q$ expansions \eqref{q18} and \eqref{q19} when eliminating $q$ in terms of $a_{UV}$. The final expression in equation  (\ref{Free2})  justifies the normalization in equation \eqref{UV26}.

The physical free energy is the minimization over $k$ of the free energies $F_k$ for fixed $\theta_{UV}$.
\be
F\left(\f_-,\theta_{UV}\right)={\rm Min_ {k\in\mathbb{Z}}}~F_k\left(\f_-,\theta_{UV}\right)
\label{F22}\ee
Finally, the topological susceptibility becomes
\be
\chi\equiv {1\over V_d}{\pa^2 F\over \pa \theta_{UV}^2}=- |\f_-|^{d\over\Delta_-}{(M_p \ell)^{d-1}\over N_c^2}{\partial ^2 C(a_{UV})\over \partial a_{UV}^2}\Big|_{a_{UV}={\theta_{UV}+2\pi k\over N_c}}
\label{F23}\ee

\subsection{The free energy and effective potential at small $q$}

It will be instructive to consider the case of a flow where the UV parameter $q$ is small so that it can be used as an expansion parameter. In this case we can evaluate the free energy and the $\theta$-angle in a perturbative expansion in $q$.

In particular, inserting the expansion of $C(q)$ in powers of $q$ introduced in \eqref{q18}, the free energy can be written as
\begin{align}
\nonumber F &=F^{(q0)}+q F^{(q1)}+\mathcal{O}(q^2) \\
&= - \left(M_p \ell\right)^{d-1}  V_d \, |\f_-|^{d\over\Delta_-} \, \Big[ \big(C^{(q0)} - C_{ct} \big) + C^{(q1)} q + \mathcal{O} ( q^2) \Big] \, .
\label{Free21}\end{align}
Then, inverting \eqref{q19} to write $q$ in terms of an expansion in $a_{UV}$ the free energy becomes
\be
F =F^{(q0)} - \frac{V_d \, |\f_-|^{d\over\Delta_-}}{4 \, C^{(q1)}} \, \left(M_p \ell\right)^{d-1} \, a_{UV}^2 + \mathcal{O} \big( a_{UV}^3 \big) \, .
\label{Free22}\ee
Finally, using the relation \eqref{UV9} between $a_{UV}$ and the theta-parameter $\theta_{UV}$ one obtains
\be
F =F^{(q0)} -  {\rm Min}_k \, \frac{V_d \, |\f_-|^{d\over\Delta_-}}{4 \, C^{(q1)}} \, \frac{\left(M_p \ell\right)^{d-1}}{N_c^2} \, \left(\theta_{UV}+2\pi k\right)^2 + \mathcal{O} \big( N_c^{-1} \big) \, ,
\label{Free23}\ee
where we have introduced a minimization over $k$ to take into account the many-to-one relation between $a_{UV}$ and $\theta$, see the discussion below equation (\ref{UV9}). The coefficient $C^{(q1)}$ can be calculated from quantities at order $\mathcal{O}(q^0)$ using \eqref{q17}.

The free energy (\ref{Free23})  is of the form expected in large $N_c$ YM, \cite{Witten-th,witten,iQCD}. We can identify the second term in \eqref{Free22} and \eqref{Free23} as the leading term in the effective potential $V_{\rm eff}(a_{UV})$ for the axion, defined as
\be
{F-F^{(q0)}\over V_d}=: |\f_-|^{d\over\Delta_-} V_{\rm eff}\left(a_{UV}\right) \, .
\label{Free27}\ee
Also, note that this leading contribution to $V_{\rm eff}$ is not suppressed by $N_c$. Since $\left(M_p \ell\right)^{d-1}\sim N_c^2$ the second term in \eqref{Free23} is of order $\mathcal{O}(N_c^{0}) \sim \mathcal{O}(1)$ while the higher order corrections are suppressed as $\mathcal{O}(N_c^{-1})$.

Given \eqref{Free22}, we can now derive an explicit expression for $\langle O_a \rangle$ by applying \eqref{Free26}. One finds
\begin{align}
\nonumber \langle O_a \rangle &=
- \left(M_p \ell\right)^{d-1} {1\over N_c} \frac{|\f_-|^{d\over\Delta_-}}{2 C^{(q1)}} \, a_{UV} + \mathcal{O} \big( a_{UV}^2 \big) \\
&= - {\rm Min}_k \left(M_p \ell\right)^{d-1} \frac{|\f_-|^{d\over\Delta_-}}{2 C^{(q1)}} {1\over N_c}\,  {\theta_{UV}+2\pi k \over N_c} +\mathcal{O}\left(\left(\theta_{UV}+2\pi k \over N_c\right)^2\right) \, .
\label{Free30}\end{align}
We can then make the following observations. For $\theta=0$, the instanton density does not develop a vev, and CP is unbroken. On the other hand, for $\theta=\pi$, the vev of $O_a$ is non-zero and CP is spontaneously broken.

From the definition (\ref{F23}) of the topological susceptibility, we obtain
\be
\chi
= -
 \frac{|\f_-|^{d\over\Delta_-}}{4 \, C^{(q1)}} \,
{\left(M_p \ell\right)^{d-1}\over N_c^2}
+\mathcal{O}(N_c^{-1}) \, ,
\label{Free24}\ee
at leading order in $q$. This expression depends only on background quantities (i.e. on the zeroth order solution with no running axion), as one can see by substituting for $C^{(q1)}$ the expression (\ref{q17}).

\section{Numerical results\label{Sec:Numerics}}

 \begin{figure}[t]
 \begin{center}
  \includegraphics[width=.45\textwidth]{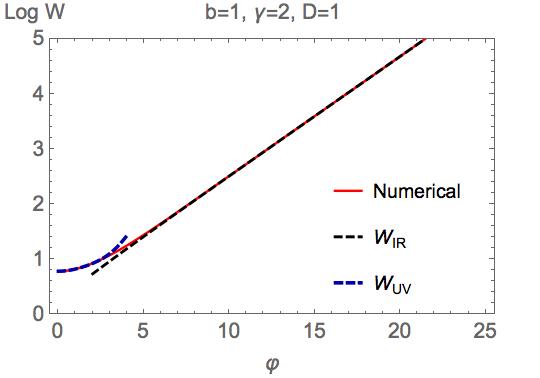}
   \includegraphics[width=.45\textwidth]{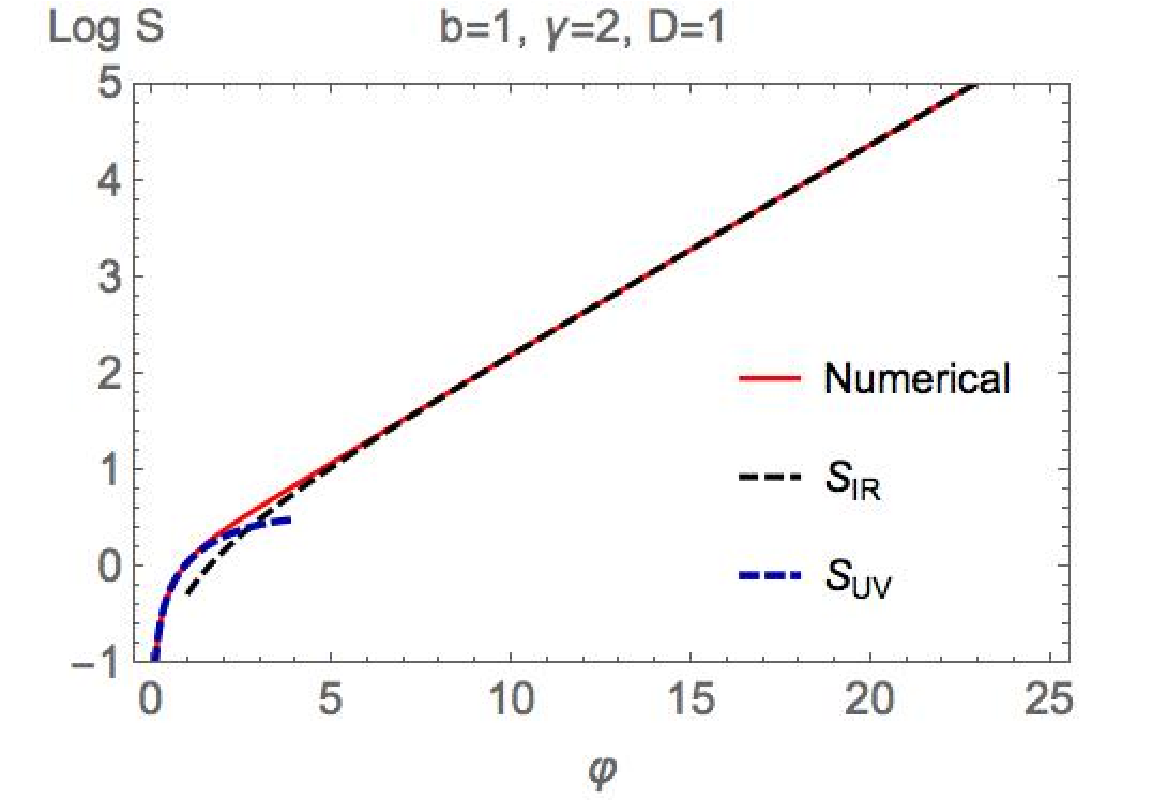}
 \end{center}
 \begin{center}
 \includegraphics[width=.55\textwidth]{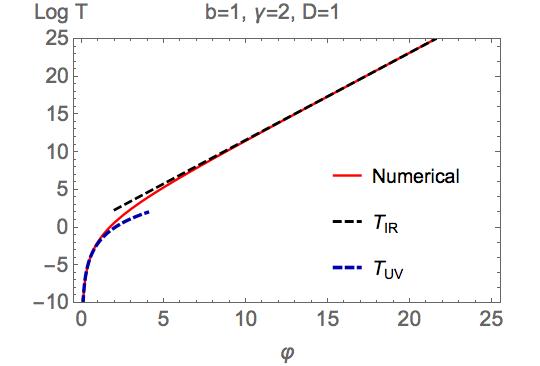}
  \end{center}
   \caption{Plots of $W$, $S$, $T$ vs.~$\f$ for a model with bulk functions \protect\eqref{Num1} and model parameters \protect\eqref{Num9}. \textbf{Top row, left:} Plot of $W(\f)$. The numerical result is well fitted by the UV and IR expansions with $C=-2\times10^{-2}$. Here $C$ is the integration constant appearing in the UV expansion of $W$, \protect\eqref{aUV2}. \textbf{Top row, right:} Plot of $S(\f)$. The numerical result is well fitted by the UV and IR expansion with $C=-2\times10^{-2}$.
\textbf{Bottom row:} Plot of $T(\f)$. The numerical result is well fitted by the UV and IR expansions with $q=10^{-2}$.}
  \label{fig2}
 \end{figure}

To find complete flow solutions that interpolate between the UV and IR, we turn to a numerical analysis. To this end,  we choose the  the bulk potential $V(\f)$ and the axion kinetic term $Y(\f)$ that enter in the action (\ref{A2}) to be
\be
V=
-{1\over\ell^2}
\left[d(d-1)+\left({1\over2}(d-\Delta_-)\Delta_--b^2V_\infty\right)\f^2+4V_\infty \sinh^2\left(b\f\over2\right)\right],
\, \quad
Y=e^{\gamma\f},
\label{Num1}\ee
These are convenient function which interpolate between the UV and IR behavior discussed in Section \ref{Sec:UV_and_IR}. The potential  $V(\f)$ has its AdS maximum at $\f=0$, which corresponds to the UV fixed point. Moreover, the potential exhibits the runaway behavior for $\f\to\infty$
\be
V\to
-{1\over \ell^2}V_\infty e^{b\f}+\mathcal{O}(\f^2).
\label{Num2}\ee
This corresponds to the IR region of the flow.
The other parameters that appear in (\ref{Num1}) are as follows: $\ell$ is the UV AdS length, $d$ is the dimension of the boundary, $\Delta_-=d-\Delta$,  with $\Delta$ the UV scaling dimension of the perturbing operator dual to $\f$. The parameters $b$ and  $V_{\infty}$ control the exponent and normalization of the asymptotic exponential behavior in the IR regime.  Finally, the axion kinetic function  was rescaled so that $Y_0=1$ in (\ref{Num1}). All parameters except $\ell$ are dimensionless.

   \begin{figure}[t]
 \begin{center}
  \includegraphics[width=.33\textwidth]{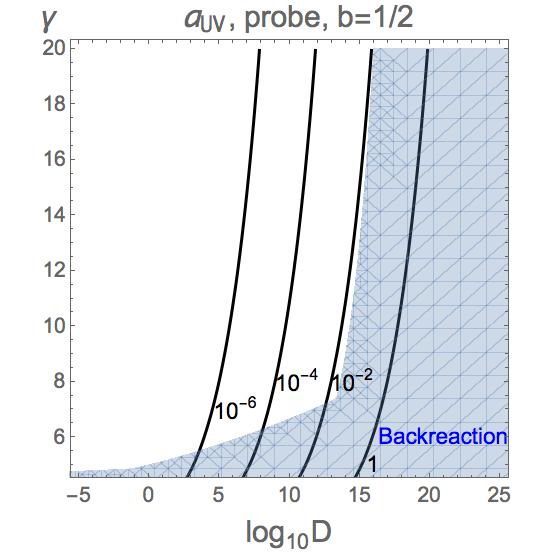}\hfil\hfil\hfil
  \includegraphics[width=.33\textwidth]{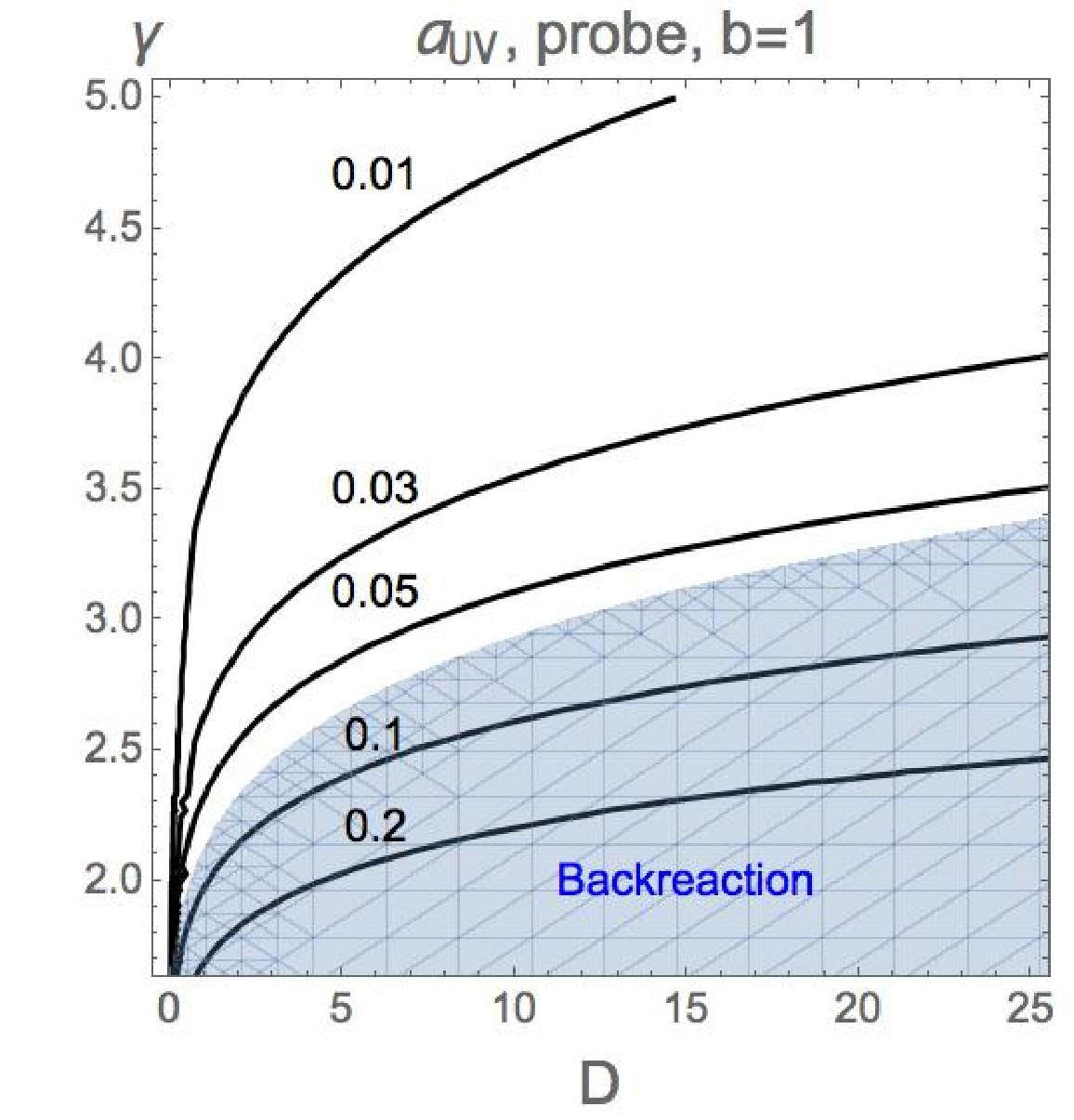}\hfil\hfil
  \includegraphics[width=.33\textwidth]{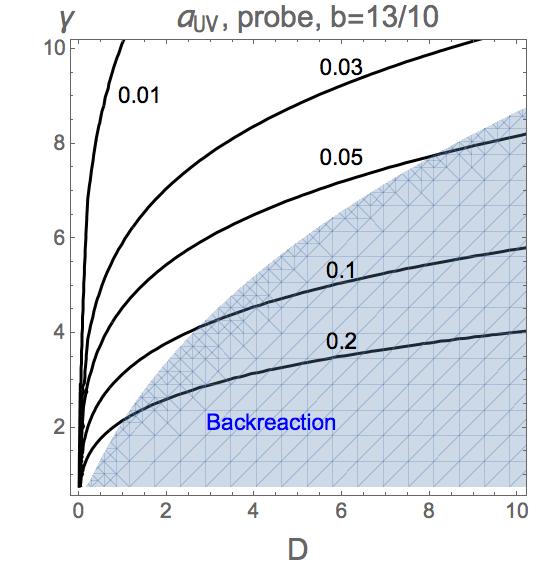}\\ \hfill
 \end{center}
 \caption{Contour plots of $a_{UV}$ (in the probe approximation) as a function of the integration parameter $D$ related to the axion vev and the exponent $\gamma$ of the function $Y(\f)$ in (\protect\ref{Num1}). The shaded region indicates the regime where axion backreaction becomes important. The different panels corresponds to three different values of the exponent $b$ that controls the asymptotic form of the bulk potential in (\protect\ref{Num1}).}
  \label{fig5}
 \end{figure}

The corresponding asymptotic forms of $W, S$ and $T$, as $\f\to\infty$, are given in \eqref{sub8}, \eqref{sub9} and \eqref{sub10}, respectively.

To be specific, as an initial example we consider a model with the following values for the model parameters:
\be
\Delta_-=1.2,\quad d=4,\quad b=1, \quad \gamma=2, \quad V_\infty=1, \quad D=1.
\label{Num9}\ee
Here we ensured that the choice for $\g$ is consistent with the bound from equation (\ref{sub11}) which reads
\be
\gamma\geq
{5\over3}.
\label{Num10}\ee
Our choice for $b$ also satisfies the Gubser bound $b<2\sqrt{2/3}\simeq1.63$.

Numerically, it is important to impose the boundary conditions in the IR (i.e.~large $\f$) regime, because the solution with  a ``good'' IR singularity is unique and not continuously connected to the generic solutions. This makes it practically impossible to reach the good solution by giving initial conditions in the UV.\footnote{Another way to see this more explicitly is that the good solution corresponds to a specific value (namely, zero) of the  integration constant appearing in the IR expansion of $W$ at the subleading order, as detailed in Appendix \ref{aB}, specifically equations  \eqref{asub16}, \eqref{asub17} and \eqref{asub18}. }
In practice, we solve the two equations (\ref{a8}-\ref{a9})
using the IR boundary conditions \eqref{sub8}, \eqref{sub9} and \eqref{sub10} at a point which is deep in the exponential regime of $V(\f)$. Specifically, here we chose the value $\f=25$.\footnote{Later in this section, the boundary condition is implemented at an even larger value of $\f$, depending on the parameter choice.}

In figure \ref{fig2}, $W$, $S$ and $T$ are plotted for the parameter choice in \eqref{Num9}. The solid lines corresponds to the numerical solutions while the dashed lines correspond to the analytic expressions in the UV and IR regions.

 \begin{figure}[t]
 \begin{center}
  \includegraphics[width=.49\textwidth]{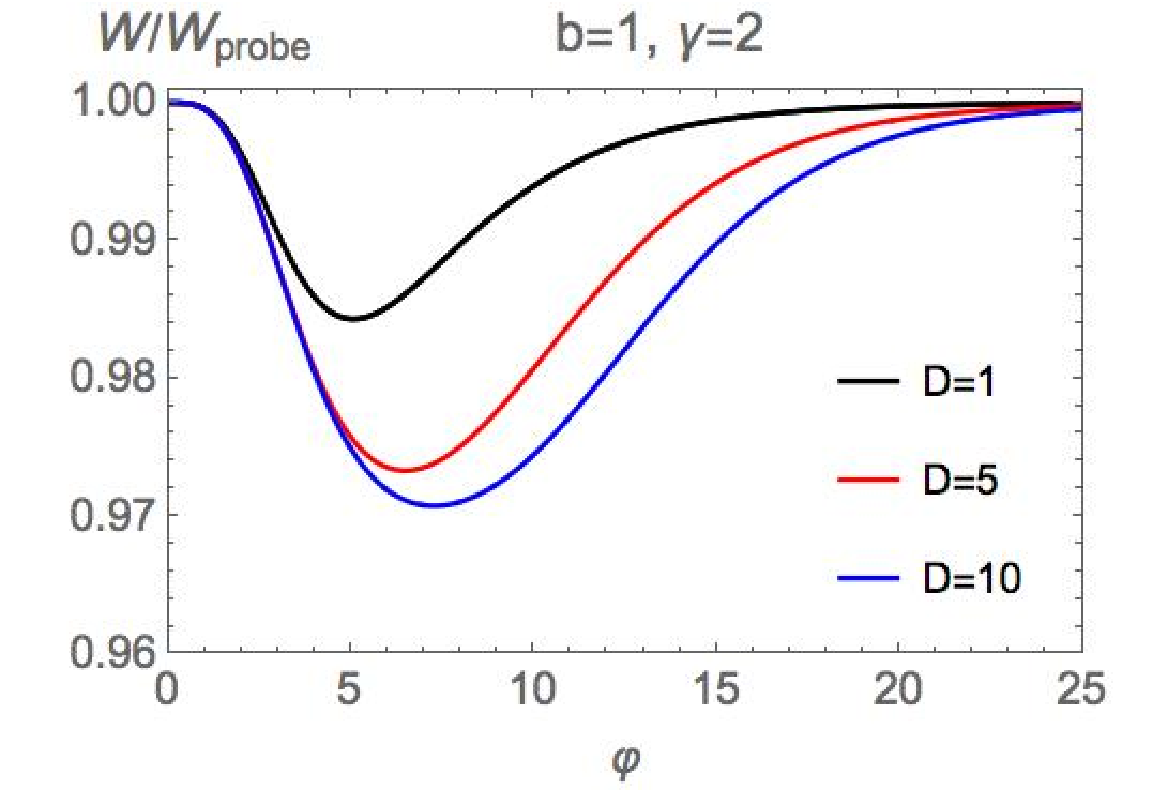}\hfil\hfil
  \includegraphics[width=.49\textwidth]{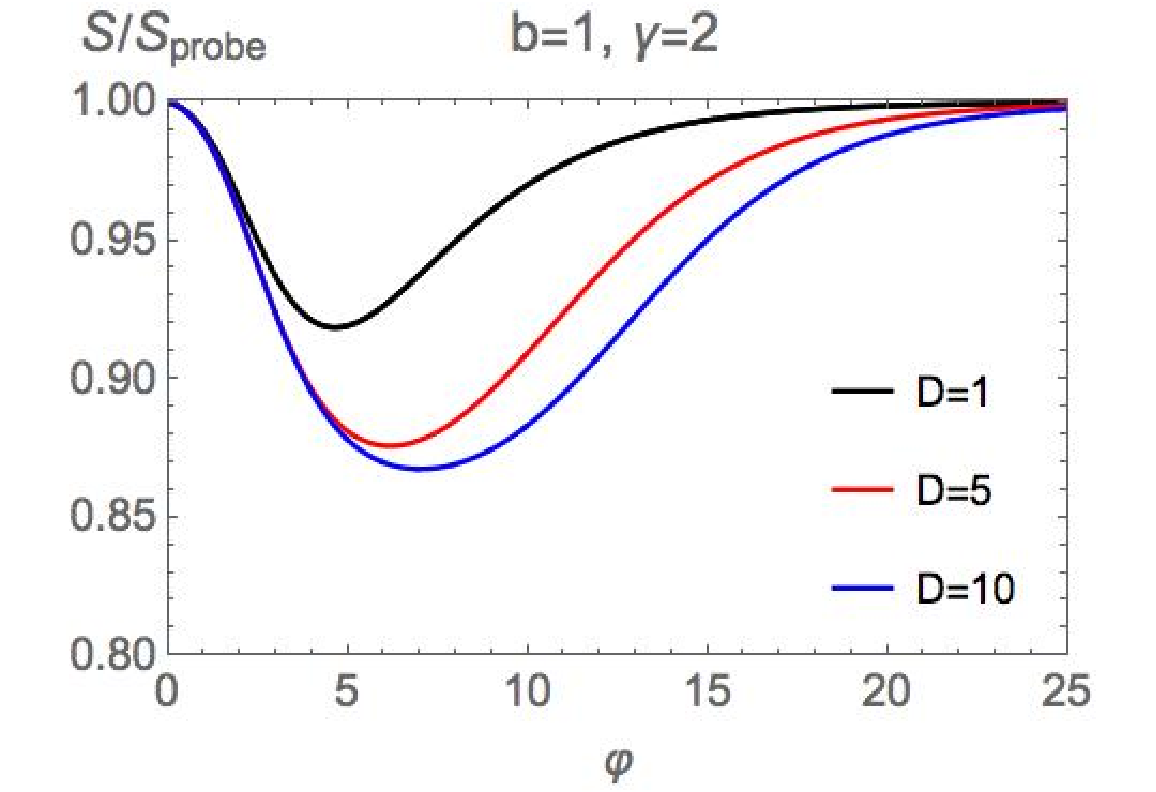}\\ \hfill \\
  \includegraphics[width=.49\textwidth]{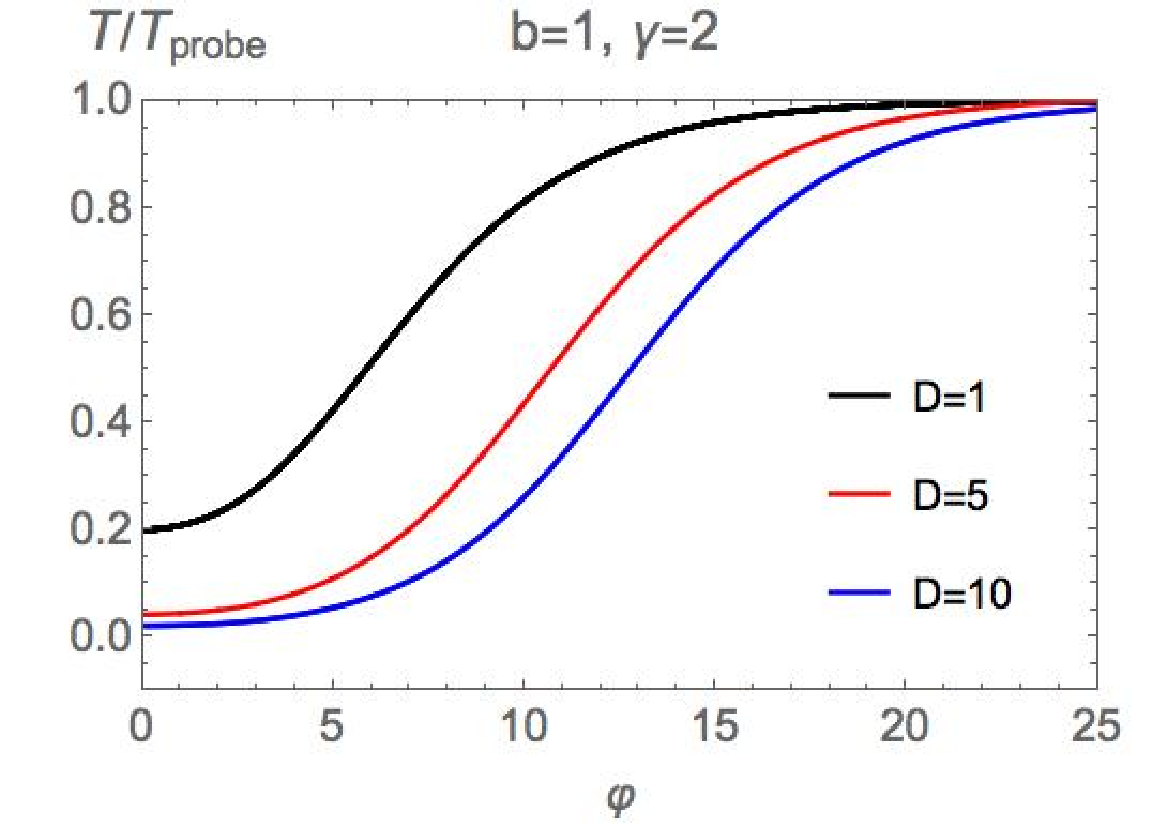}\hfill
  \includegraphics[width=.49\textwidth]{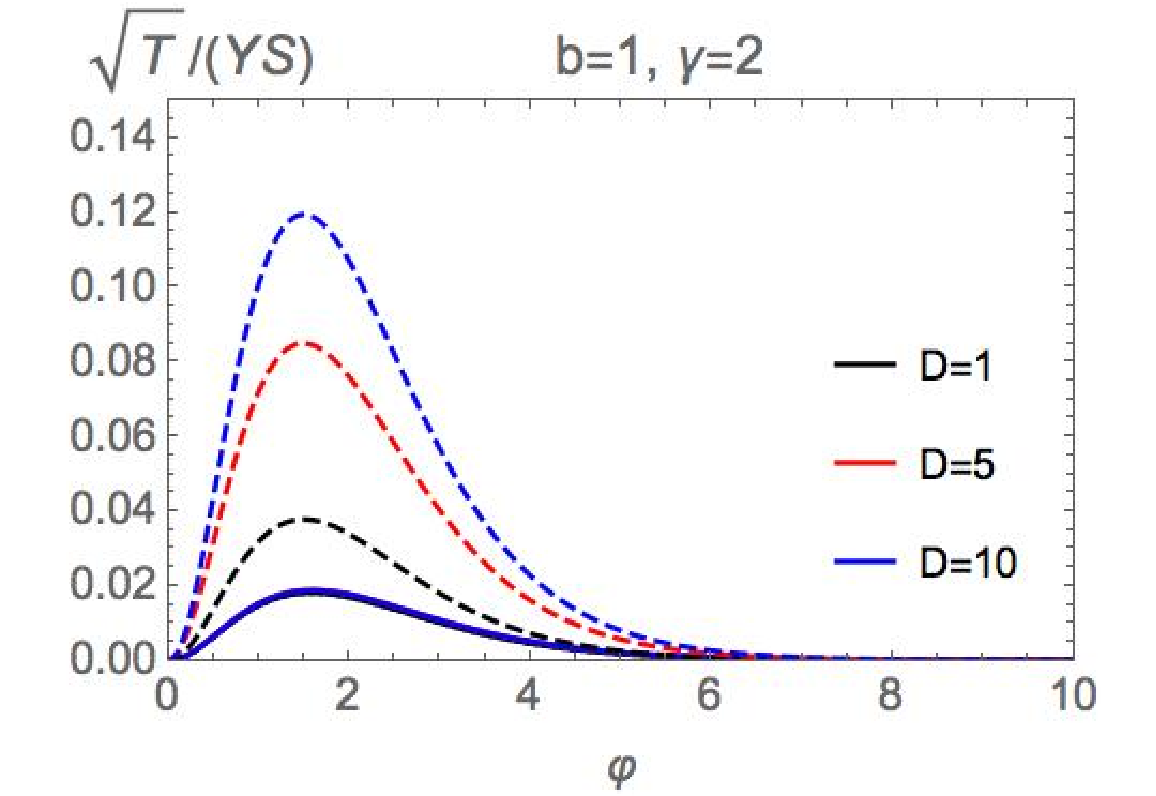}\\ \hfill \\
 \end{center}
 \caption{
Comparison of solutions in the probe approximation with fully backreacted solutions, for $b=1, \g=2$ with $D=1, 5$, and $10$.
\textbf{Upper left:} The ratio of full solution and probe solution for $W$. \textbf{Upper right:} The ratio of the full and probe solution for $S$. \textbf{Lower left:} The ratio of the full and probe solution for $T$. \textbf{Lower right:} The full and probe solutions for $\sqrt{T}/(YS)$, whose integration becomes $a_{UV}$ from \protect\eqref{UV8}. Solid and dashed lines correspond to the full and probe solutions, respectively. Here all the solid lines almost overlap one another.
 }
  \label{fig6}
 \end{figure}

We observe that the UV and IR behavior is well-fitted by the analytical formulae \eqref{aUV2}, \eqref{aUV3}, \eqref{aUV4}, \eqref{sub1}, \eqref{sub8}, \eqref{sub9} and \eqref{sub10}, as can be seen in figure \ref{fig2}.
Only in the narrow intermediate region around $\f\sim3$ the full solution cannot be described accurately by either the UV or IR formulae.
As we saw in section \ref{Sec:UV_and_IR}, there are two branches of the solution in the UV.
Because only the minus branch solutions contains the integration constant $C$, our numerical solution is connected to the minus branch unless the potential is fine-tuned.
One can observe that all the functions $W, S$ and $T$ are monotonically increasing.
The monotonicity of $W$  is related to the $c$-theorem of the holographic RG flow as shown in (\ref{a32a}). On the other hand $S$ can have both signs depending on whether the flow is going towards larger or smaller values of $\f$.
It can also change sign at ``bounces'', \cite{exotic}, i.e.~loci where the flow changes direction in $\f$.
Finally, $T$ is non-negative and always decreasing along the flow. When it vanishes,  it signals a singularity or a horizon. For the flows we consider in this paper, there is always a mild singularity, that is resolvable.
In holographic models like the black D$_4$ brane, discussed in section \ref{general}, this singularity is resolved by the KK states (i.e.~by going to one dimension higher).

 \begin{figure}[t]
 \begin{center}
  \includegraphics[width=.49\textwidth]{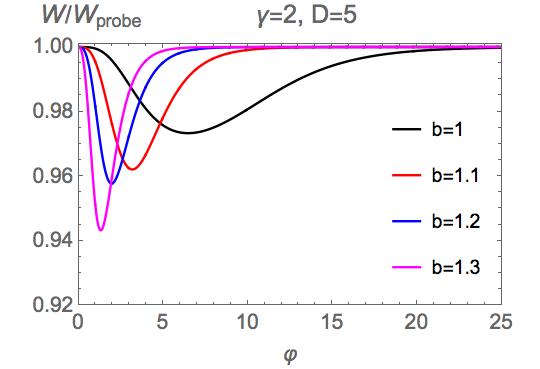}\hfil\hfil
  \includegraphics[width=.49\textwidth]{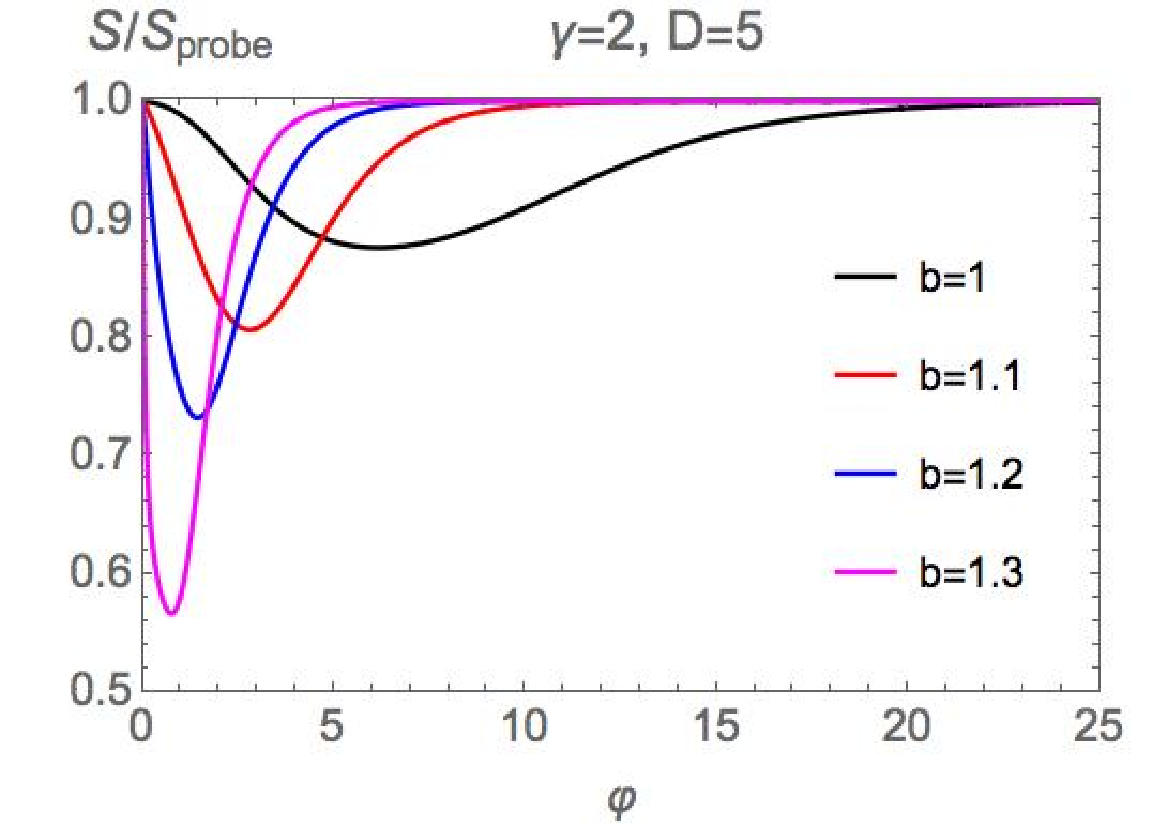}\\ \hfill \\
  \includegraphics[width=.49\textwidth]{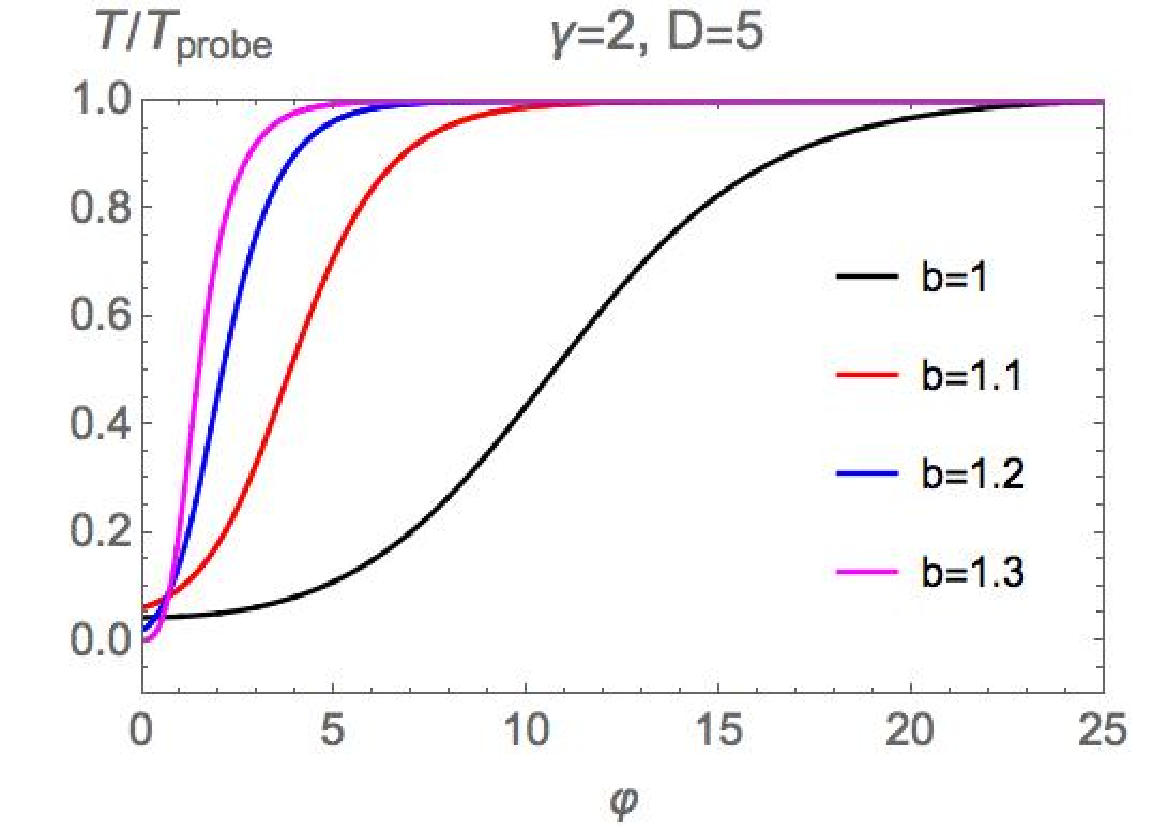}\hfill
  \includegraphics[width=.49\textwidth]{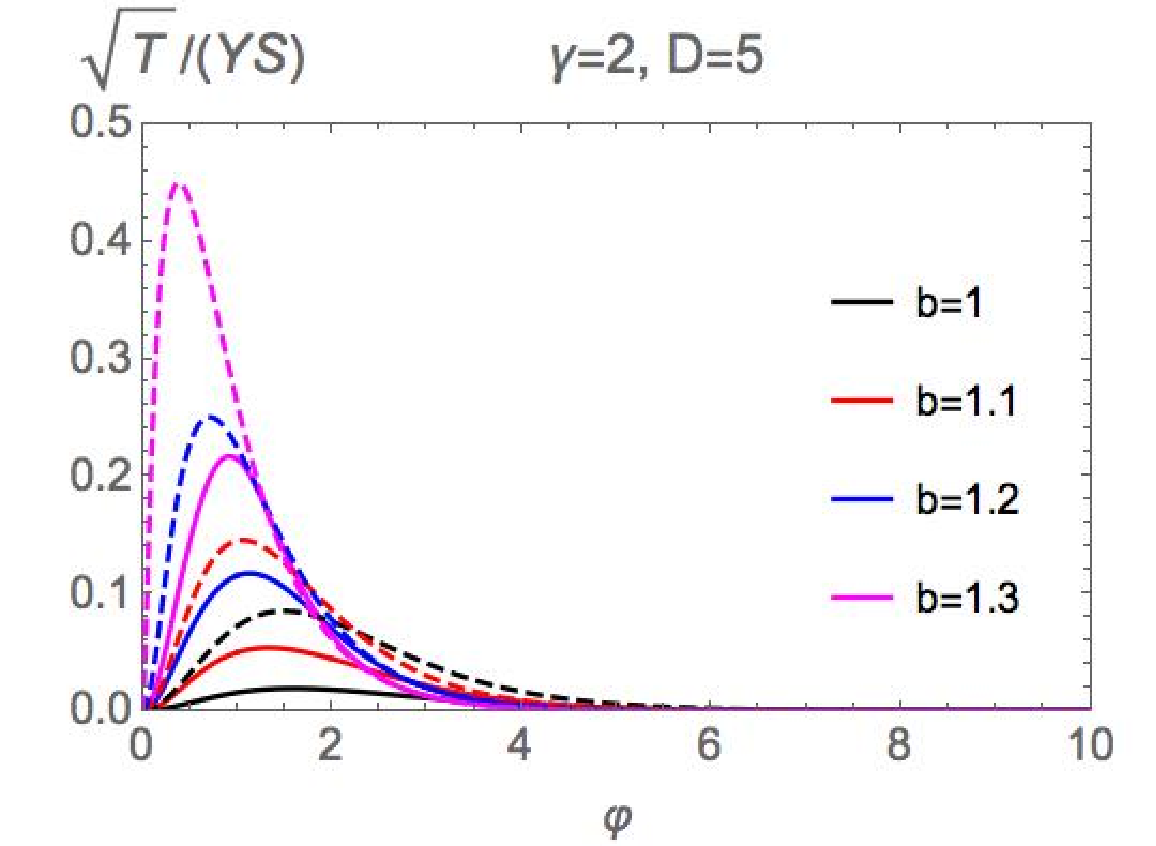}
 \end{center}
 \caption{Comparison of solutions in the probe approximation with fully backreacted solutions, for $\gamma=2, D=5$ and $b=1, 1.1, 1.2$ and $1.3$. For a description of the content of the 4 panels see the description below Fig.~\protect\ref{fig6}.}
  \label{fig7}
 \end{figure}

\subsection{The probe approximation}

 If $T(\f)$ is small, then the backreaction of the axion field on other fields is small. We can therefore use the probe approximation to calculate the axion source value  $a_{UV}$. This is the same as the small-$q$ approximation.
In this approximation, we first solve \eqref{a8} and \eqref{a9} with $T=0$. Then, $T$ is calculated by solving \eqref{a8-2} with the IR boundary condition \eqref{sub10}. Finally, we obtain the value of $a_{UV}$ by using \eqref{UV8}.

The resulting values of $a_{UV}$ as a function of $D$ and the exponent $\gamma$ are plotted in Fig.~\ref{fig5}.
The probe solution for $T$ is written as
\be
T_{\rm probe}(\f)=T_{\rm probe}(\f_{\rm ini}) \exp\left(\int^\f_{\f_{\rm ini}} d\f{d\over d-1}{W_{\rm probe}\over S_{\rm probe}}\right),
\ee
where $W_{\rm probe}, S_{\rm probe}$ and $T_{\rm probe}$ are the probe solution, and $\f_{\rm ini}$ is the initial condition which we take from \eqref{sub10} with a sufficiently large value of $\f_{\rm ini}$. Then, combining with \eqref{UV8}, we observe that, in the probe calculation, $a_{UV}$ is proportional to $\sqrt{D}$ and is a decreasing function of $\g$.
In figure \ref{fig5} and other figures in this section, we use \eqref{UV8} with ${\rm sign}(Q)=-1$ to calculate $a_{UV}$. By reversing the sign we can get the corresponding value of $a_{UV}$ for $Q>0$.

 \begin{figure}[t]
 \begin{center}
  \includegraphics[width=.49\textwidth]{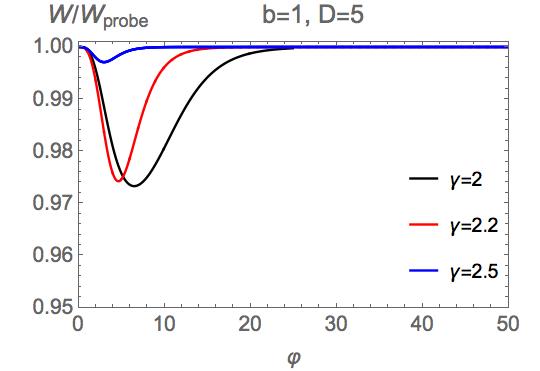}\hfil\hfil
  \includegraphics[width=.49\textwidth]{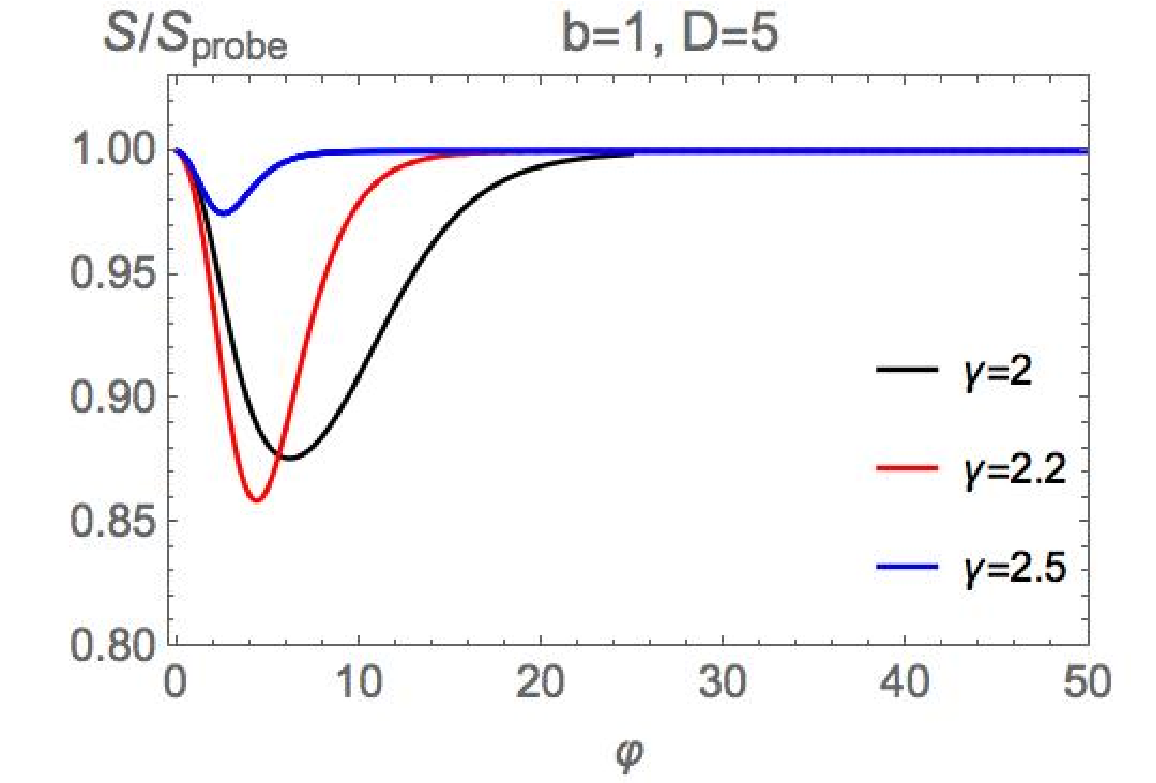}\\ \hfill \\
  \includegraphics[width=.49\textwidth]{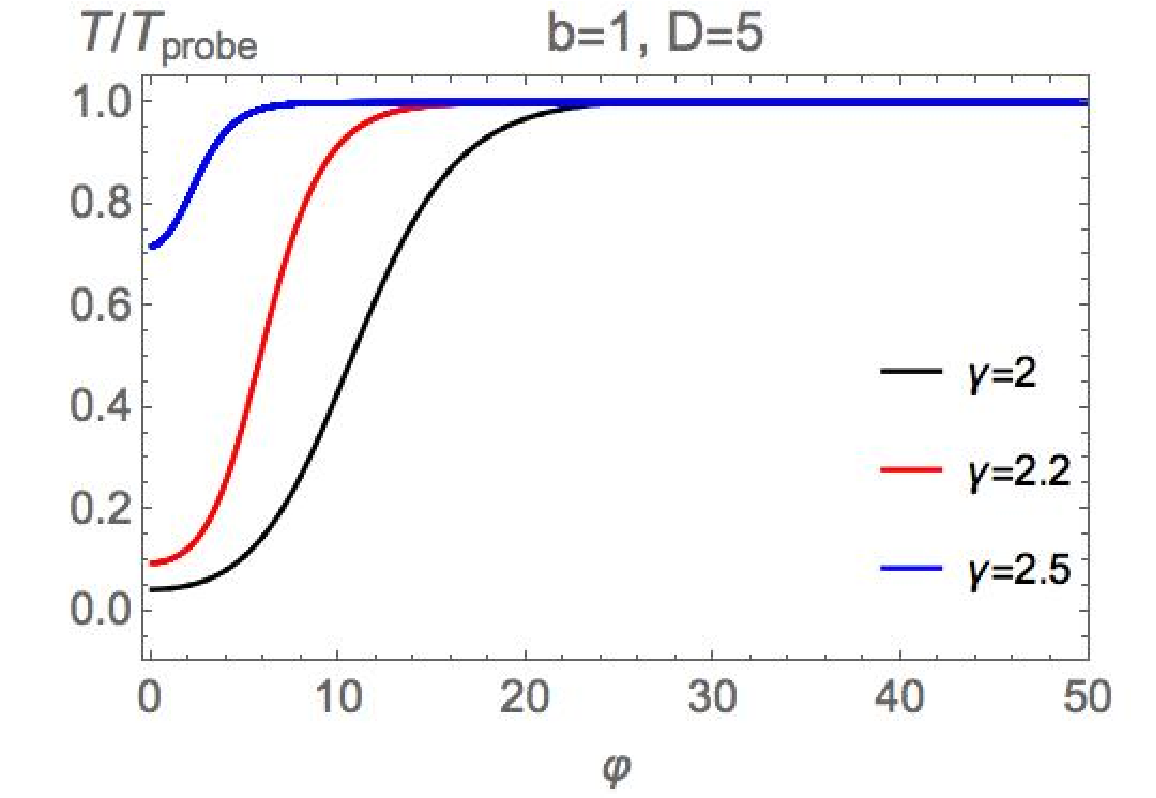}\hfill
  \includegraphics[width=.49\textwidth]{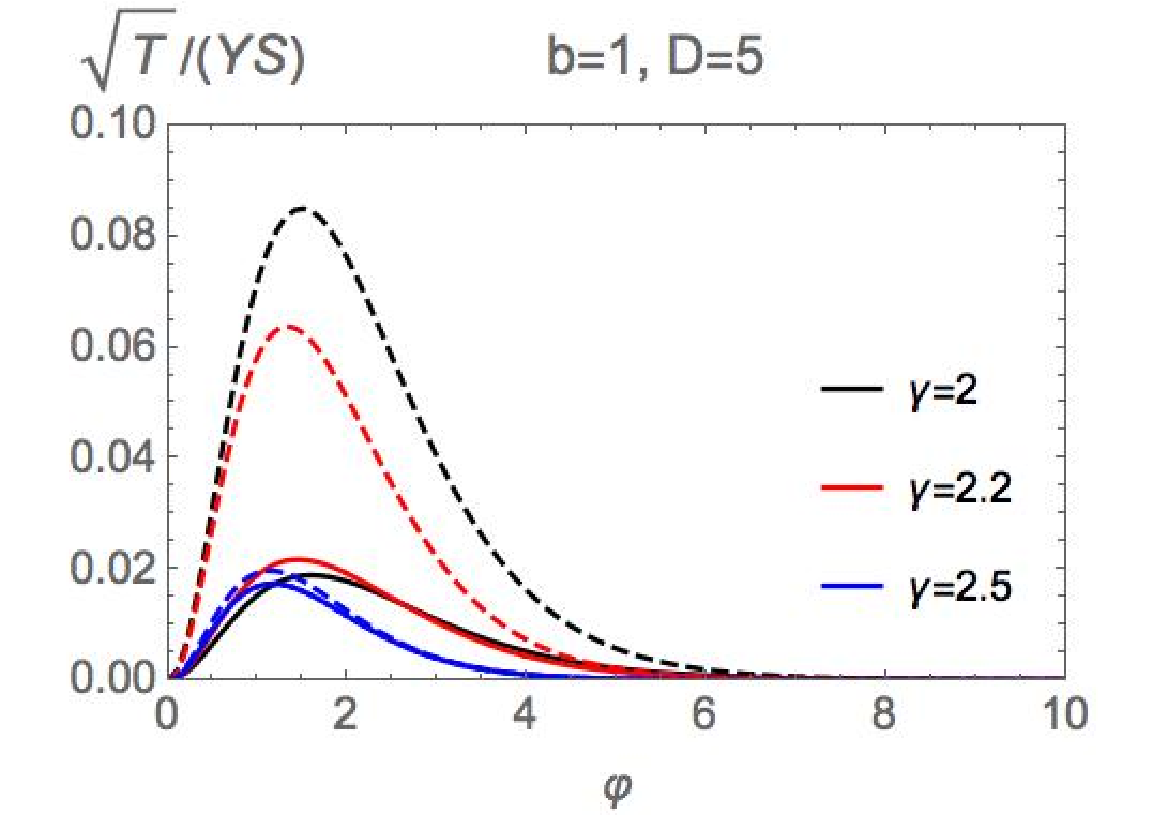}
 \end{center}
 \caption{
 Comparison of solutions in the probe approximation with fully backreacted solutions, for $b=1, D=5$ and $\g=2, 2.2$ and $2.5$. For a description of the content of the 4 panels see the description below Fig.~\protect\ref{fig6}.
 }
  \label{fig8}
 \end{figure}

This approximation is reliable as long as $T$ is small enough in \eqref{a8} and \eqref{a9}. The shaded region in figure \ref{fig5} indicates where backreaction becomes important  and the probe approximation cannot be trusted. In practice we identify this region by the inequality
\be
{\rm Min}\left({dW^2/[4(d-1)]\over T/(2Y)}, {S^2\over T/Y}\right)\leq 10,
\label{Num16}\ee
which corresponds the fact that the term containing $T$ becomes sizable (namely larger than $1/10$, taken as a reference value) compared with the other two terms in \eqref{a9}.

We observe that, the larger the value of the exponent $b$, the larger the maximum value of $a_{UV}$ in the region where the probe approximation is applicable.

\subsection{Backreacted solution}

 \begin{figure}[t]
 \begin{center}
  \includegraphics[width=.45\textwidth]{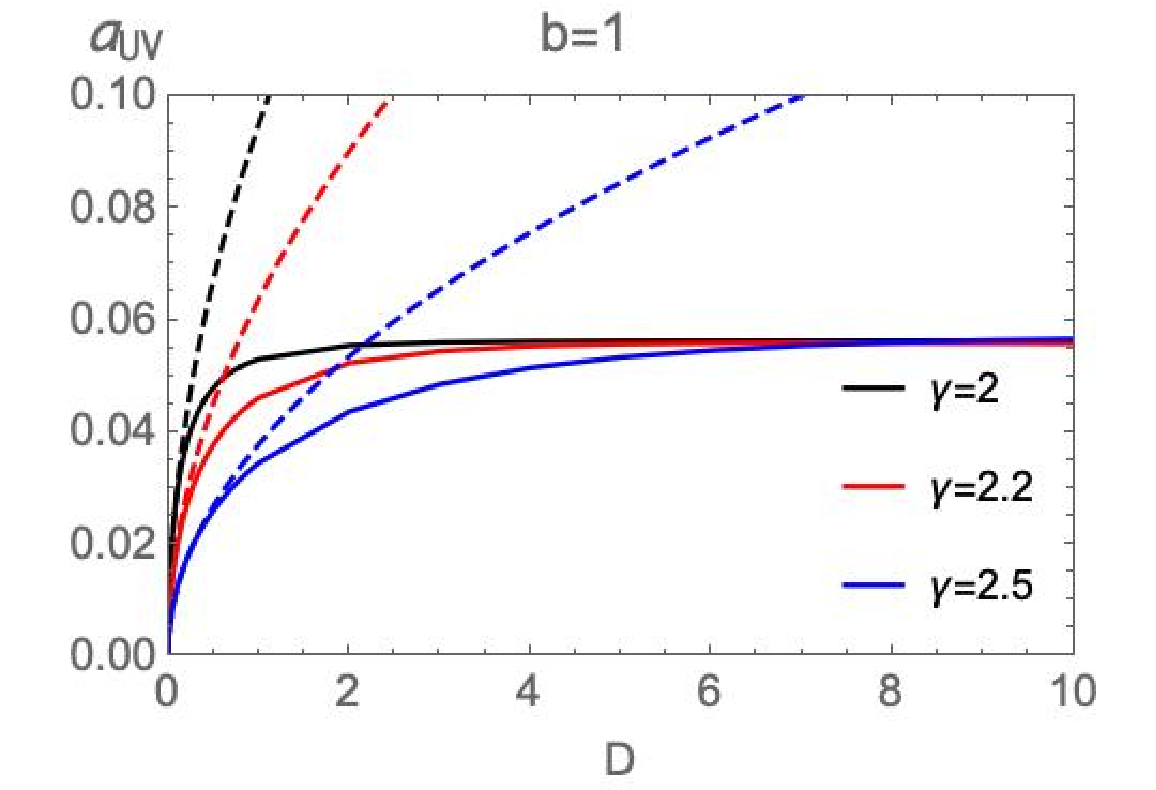}\hfil\hfil
  \includegraphics[width=.45\textwidth]{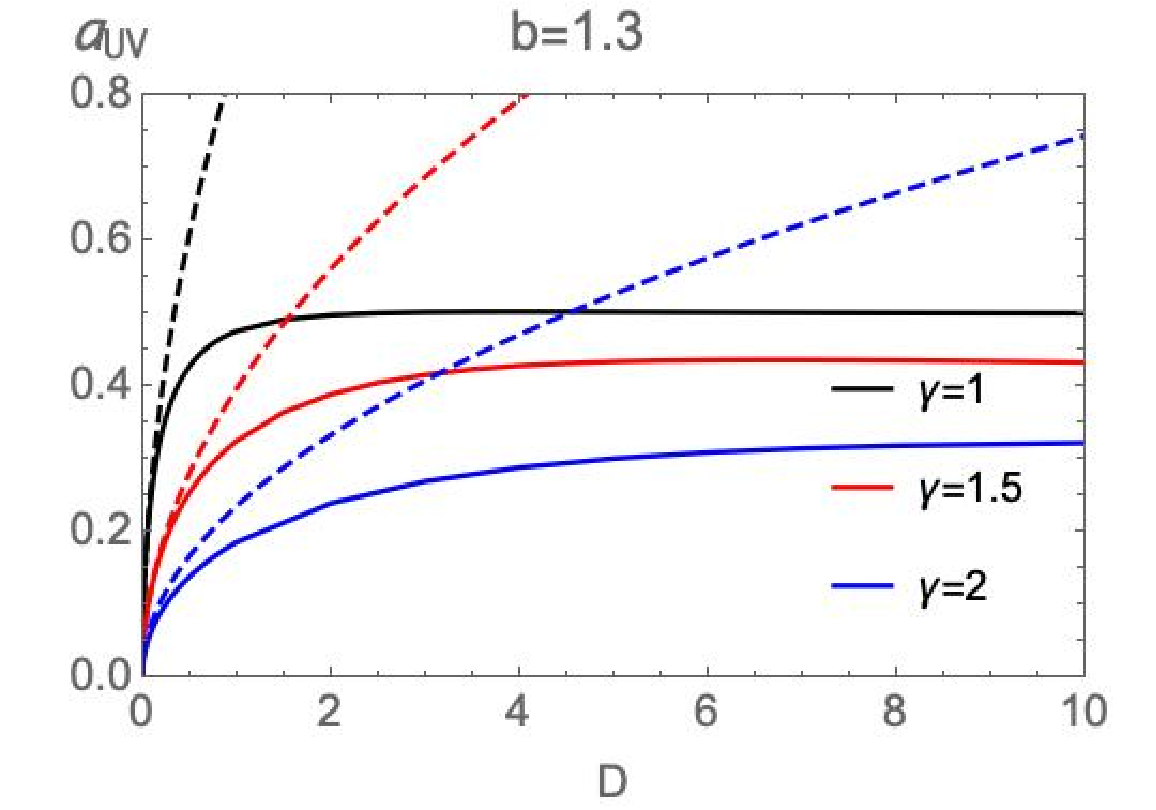}\\\hfil
 \end{center}
 \caption{
{Plot of $a_{UV}$ as a function of the constant $D$ that is related to the vev of the dual instanton density, for $b=1$, $\g=2, 2.2, 2.5$ (\textbf{left}) and $b=1.3$, $\g=1,1.5, 2$ ( \textbf{right}). The dashed and solid lines correspond to the probe and backreacted solutions, respectively. It is apparent from the backreacted results that as $D\to\infty$, $a_{UV}$ saturates, and the range of possible $a_{UV}$ values is compact. }
 }
  \label{fig10}
 \end{figure}

  \begin{figure}[t]
 \begin{center}
  \includegraphics[width=.45\textwidth]{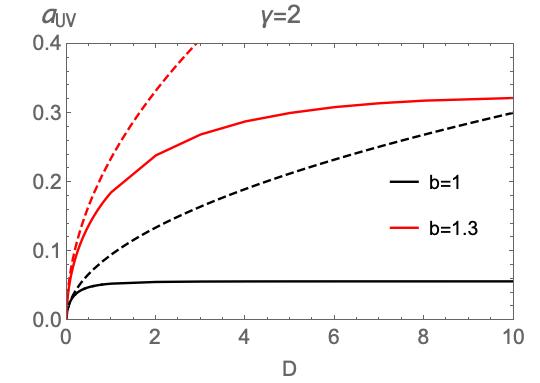}\hfil\hfil
   \includegraphics[width=.45\textwidth]{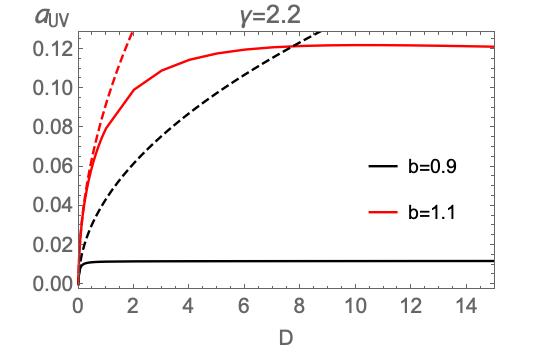}\hfil\hfil
 \end{center}
 \caption{
 {Plot of $a_{UV}$ as a function of $D$ for $b=1, 1.3$, $\g=2$ (\textbf{left}) and $b=0.9, 1.1$, $\g=2.2$ (\textbf{right}). The dashed and solid lines correspond to the probe and backreacted solutions, respectively.}
 }
  \label{fig11}
 \end{figure}

The backreaction of $T$ is important in the shaded region in Fig.~\ref{fig5}. The fully backreacted solution can be obtained solving numerically equations (\ref{a8}-\ref{a9}). Comparisons between probe solutions and corresponding backreacted solutions are shown in figures \ref{fig6}, \ref{fig7}, and \ref{fig8}.
In each figure, the ratio of the fully backreacted solution for $W, S$ and $T$ and the probe solution for $W_{\rm probe}, S_{\rm probe}$ and $T_{\rm probe}$ are plotted. The integrand in \eqref{UV8}, relevant for the determination of the axion profile, is plotted in the lower right panel of figures ~\ref{fig6}, \ref{fig7}, and \ref{fig8}.
The three figures correspond to different parameter choices for $b, \gamma$ and $D$. Backreaction on $W$ and $S$ in both the IR and UV regions is small, while backreaction can be significant in the intermediate region, as expected from the discussion in section~\ref{Sec:UV_and_IR}.

The effect of backreaction always decreases the values of $W, S$ and $T$.
Effectively this ``slows'' the RG flow. This phenomenon  can be understood analytically as follows.
Solving \eqref{a8} for $S$, we obtain
\be
S={1\over2} \left( W'\pm\sqrt{W'^2- {4T\over Y}}\right),
\label{Num17}\ee
Recall that $S$ has the same sign as $W'$, so the ($+$)-sign in equation (\ref{Num17}) corresponds to the situation $(W'\geq 0, S\geq 0)$, i.e.~$\f$ increases along the flow, and the ($-$)-sign to  $(W'\leq 0, S\leq 0)$, i.e.~$\f$ decreases along the flow.
Below we discuss in detail the case $S\geq 0$, which is also what we have assumed in the numerical solutions, and hence restrict to the ($+$)-sign in (\ref{Num17}).   One can argue along the same lines in the case $S<0$.

For $W'\geq 0$,  from equation \eqref{Num17} with the ($+$)-sign  we observe that
\be
{1\over2}W' \leq S \leq W' \, ,
\label{Num19}\ee
because $T,Y\geq 0$, and we assumed $W'\geq 0$. Next, the combination of \eqref{a8} and \eqref{a9} leads to
\be
{d\over 4(d-1)}W^2-{W'S\over2}+V=0,
\label{Num22}\ee
from which we obtain
\be
W'\geq\sqrt{{d\over2(d-1)}W^2+2V},
\label{Num24}\ee
where \eqref{Num19} was used.
On the other hand, the probe solution satisfies
\be
W'_{\rm probe} = \sqrt{{d\over2(d-1)}W_{\rm probe}^2+2V}.
\label{Num23}\ee
By comparing \eqref{Num24} and \eqref{Num23}, one can conclude that, if $W$ and $W_{\rm probe}$ are equal at any point $\f_*$, then  $W(\f)\leq W_{\rm probe}(\f)$ for any $\f< \f_*$. Since $W$ and $W_{\rm probe}$ become equal at $+\infty$, it follows that $W(\f)\leq W_{\rm probe}(\f)$ in the whole range $\f \in (0, +\infty)$. In practice, in the numerical integration to obtain the results shown in figures \ref{fig6}, \ref{fig7}, and \ref{fig8}, the IR boundary conditions are implemented at a finite but large value $\f_*$.

Having concluded that $W\leq W_{\rm probe}$, we can easily  reach the same conclusion for $S$, whose value is bounded as
\be
S\leq \sqrt{{d\over2(d-1)}W^2+2V} \leq
\sqrt{{d\over2(d-1)}W_{\rm probe}^2+2V}=W'_{\rm probe}=S_{\rm probe}.
\label{Num25}\ee
Finally, from \eqref{a9} as well as $T\geq0$, we deduce that
\be
{d\over2(d-1)}{W^2\over S^2}-1 \geq {S_{\rm probe}^2\over S^2}
\left({d\over2(d-1)}{W_{\rm probe}^2\over S_{\rm probe}^2}-1\right)
\geq {d\over2(d-1)}{W_{\rm probe}^2\over S_{\rm probe}^2}-1.
\label{Num26}\ee
Combined with \eqref{a8-2}, we obtain the relation $T\leq T_{\rm probe}$.

Therefore, the effect of the backreaction always decreases the values of $W, S$ and $T$. The same conclusion can be reached (for the absolute values) along similar lines when $W'$ and $S$ are negative. In this case we have to choose the $-$ sign in equation (\ref{Num17}), and the bounds (\ref{Num19}) are replaced by $W'\leq S  \leq W'/2$. The two branches with positive and negative $S$ can only meet at a ``bounce'', i.e. a point where $\dot{\f}=0$ and the flow direction of $\f(u)$ is inverted \cite{exotic}. In our specific numerical examples we have not encountered such a situation.

In Fig.~\ref{fig6}, we plot the ratio of fully backreacted solution and solution in the probe approximation for $W$ for fixed $b=1, \g=2$ and various values of $D$. One observation is that, as expected, the effect of backreaction increases for larger values of $D$.

In Fig.~\ref{fig7} we plot the same ratio, but now $\gamma=2$ and $D=5$ are kept fixed and plots for various values of $b$ are shown. Similarly, in Fig.~\ref{fig8} we plot this ratio for fixed $b=1, D=5$ and different values of $\g$. As we expect from the results shown in Fig.~\ref{fig5}, the probe solution becomes a better approximation to the full solution as the value of $\g$ increases.

In figures~\ref{fig10} and \ref{fig11}, the UV value of the axion field, $a_{UV}$, is plotted as a function of $D$. The probe calculation exhibits  good agreement with the full solution for small $D$. For large values of $D$, the effect of backreaction from $T$ cannot be neglected, and $a_{UV}$ flattens out.

It is not easy to derive the origin and the asymptotic value of the plateau, but we can argue that the value of $a_{UV}$ cannot be arbitrary large even if we take $D\to\infty$.
Combining equation \eqref{a8} with \eqref{UV8}\eqref{sub21a}, we deduce
\be
|a_{UV}|=\int^\infty_0 d\f {\sqrt{T}\over Y |S|}=\int^\infty_0 {d\f\over\sqrt{Y}}  \sqrt{{W'\over S}-1}
\leq \int^\infty_0 {d\f\over \sqrt{Y}} \, ,
\label{Num20}\ee
where \eqref{Num19} has been used to arrive at the inequality.
Note that this upper bound is independent of the choice of $V$, as long as $Y$ and $W'$ are non-negative functions.

For our choice of $Y$ in this section, \eqref{Num1}, we obtain
\be
|a_{UV}|\leq{2\over \g}.
\label{Num21}\ee
Numerically, as one can observe from figures~\ref{fig10} and \ref{fig11}, the asymptotic value is more suppressed compared to \eqref{Num21}. This is because $S$ is close to $W'$ in both the UV and IR region, and the integrand in \eqref{Num20} is non-suppressed only in the intermediate region. As a result \eqref{Num21} only gives a weak bound.

\begin{figure}[t]
 \begin{center}
  \includegraphics[width=.43\textwidth]{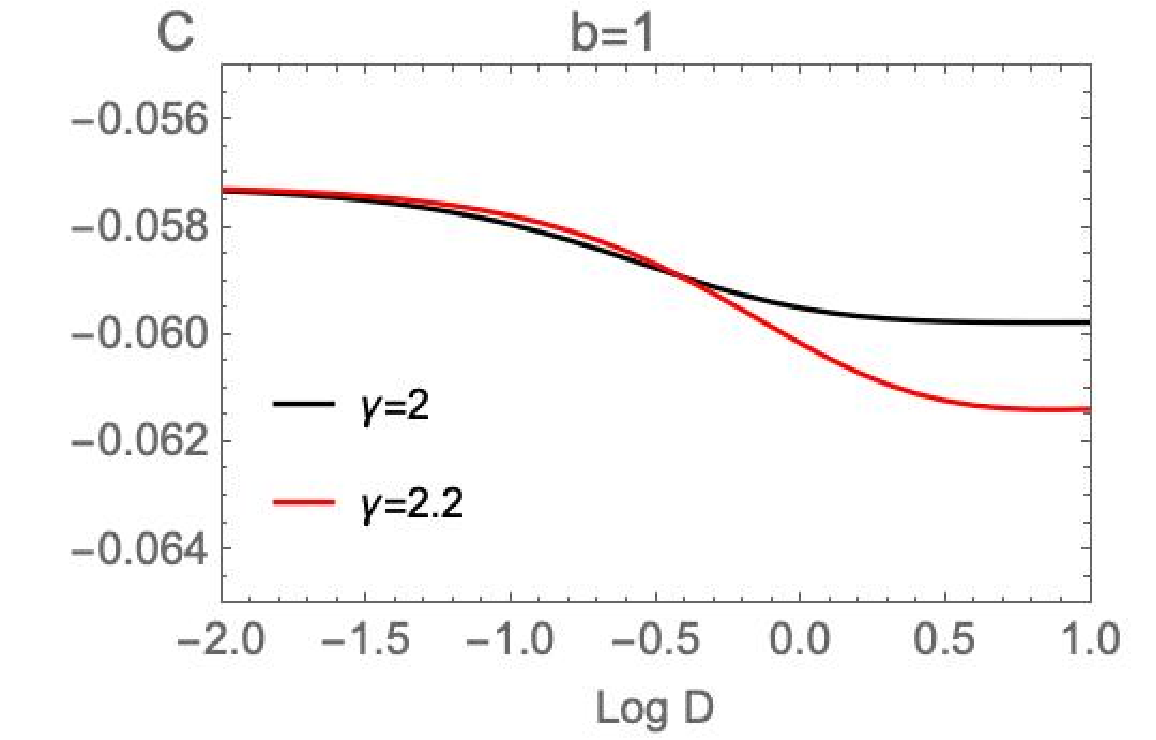}\hfil\hfil
   \includegraphics[width=.4\textwidth]{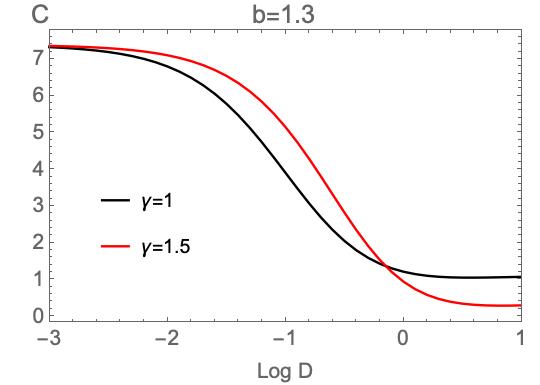}\hfil\hfil
 \end{center}
 \caption{
{
 Plot of $C$ as a function of $\log D$ for $b=1$ and $\g=2, 2.2$ (\textbf{left}) and  $b=1.3$ and $\g=1, 1.5$ (\textbf{right}).}
 }
  \label{fig13}
 \end{figure}

\begin{figure}[t]
 \begin{center}
  \includegraphics[width=.43\textwidth]{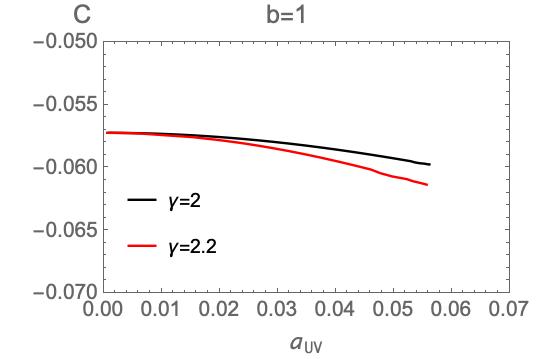}\hfil\hfil
   \includegraphics[width=.4\textwidth]{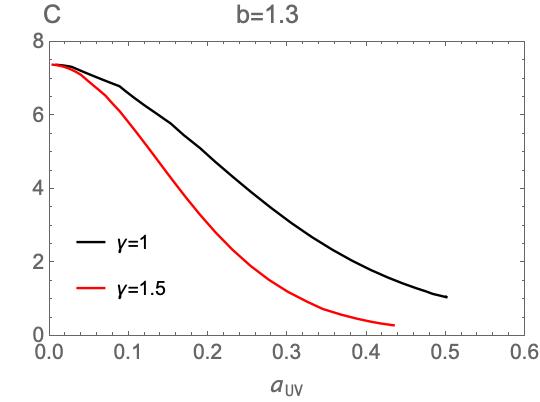}\hfil\hfil
   \\
   \includegraphics[width=.43\textwidth]{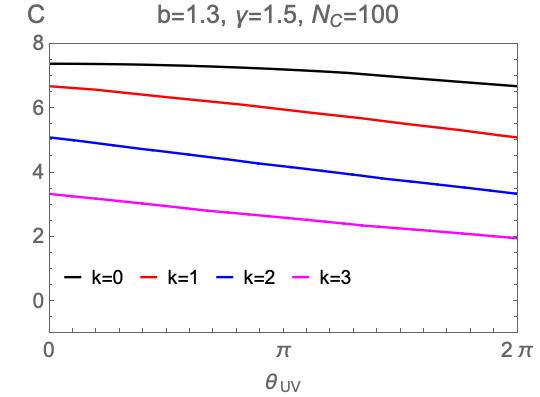}\hfil\hfil
 \end{center}
 \caption{
\textbf{Upper row, left:} Plot of $C$ vs.~$a_{UV}$ for $b=1$ and $\g=2, 2.2$. \textbf{Upper row, right:} Plot of $C$ vs.~$a_{UV}$ for $b=1.3$ and $\g=1, 1.5$.
\textbf{Bottom row:} Plot of $C$ vs.~$\theta_{UV}$ as given in \protect\eqref{z}, for $b=1.3, \g=1.5$ and $N_c=100$. The first four branches with $k=0,1,2,3$ are shown.
 }
  \label{fig14}
 \end{figure}

In figures \ref{fig13} and \ref{fig14} we show plots of $C$ as a function of $D$ and $a_{UV}$. Recall that $C$ controls the vev of the operator dual to $\f$. For small values of $D$ the parameter $C$ is not much affected by a change in $\g$ as the overall effect of the axion is small and changing $\g$ only affects axion-related effects.
For large values of $D$, $C$ asymptotes to different values depending  on the value of $\g$.

In the bottom panel of figure \ref{fig14}, $C$ is plotted as a function of $\theta_{UV}$ in \eqref{z}. As discussed around \eqref{z2}, there would be many distinct solutions corresponding to different values of the integer $k$. In the figure, the first four branches with $k=0,1,2,3$ are plotted. The function $C(\theta_{UV})$ satisfies $\left.C(2\pi)\right|_{k=i \text{ branch}}=\left.C(0)\right|_{k=i+1 \text{ branch}}$, as it should be.

\begin{figure}[t]
 \begin{center}
  \includegraphics[width=.43\textwidth]{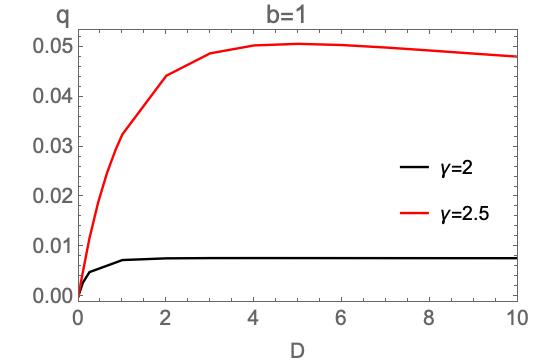}\hfil\hfil
  \includegraphics[width=.43\textwidth]{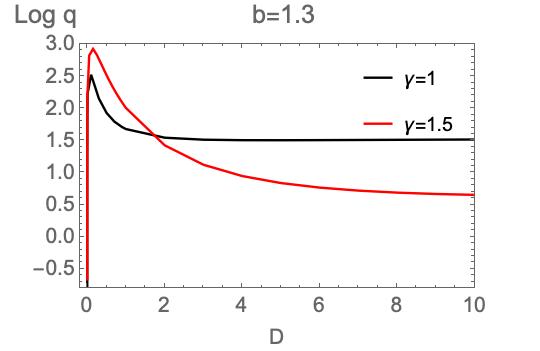}
 \end{center}
 \caption{
{
\textbf{Left:} Plot of $q$ vs.~$D$ for $b=1$ and $\g=2, 2.2$.
\textbf{Right:} Plot of Log $q$ vs.~$D$ for $b=1.3$ and $\g=1, 1.5$.
 }
  }
  \label{fig15}
 \end{figure}

The other UV integration constant, $q$, is plotted in figure \ref{fig15} as a function of $D$. For $D\to0$, this becomes zero, which is consistent with the absence of the axion. For larger values of $D$, the function $q(D)$ exhibits a plateau, as we saw in the plot of $a_{UV}$ in figures \ref{fig10} and \ref{fig11}.

  \begin{figure}[t]
 \begin{center}
  \includegraphics[width=.46\textwidth]{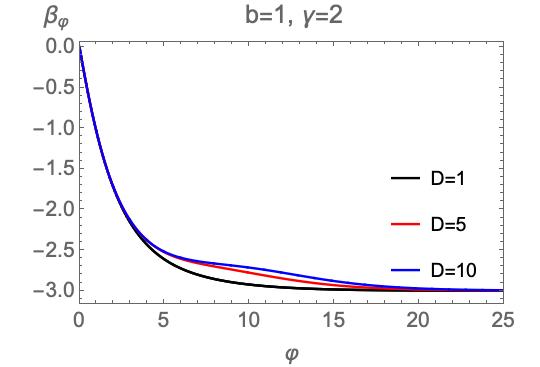}\hfil\hfil
   \includegraphics[width=.5\textwidth]{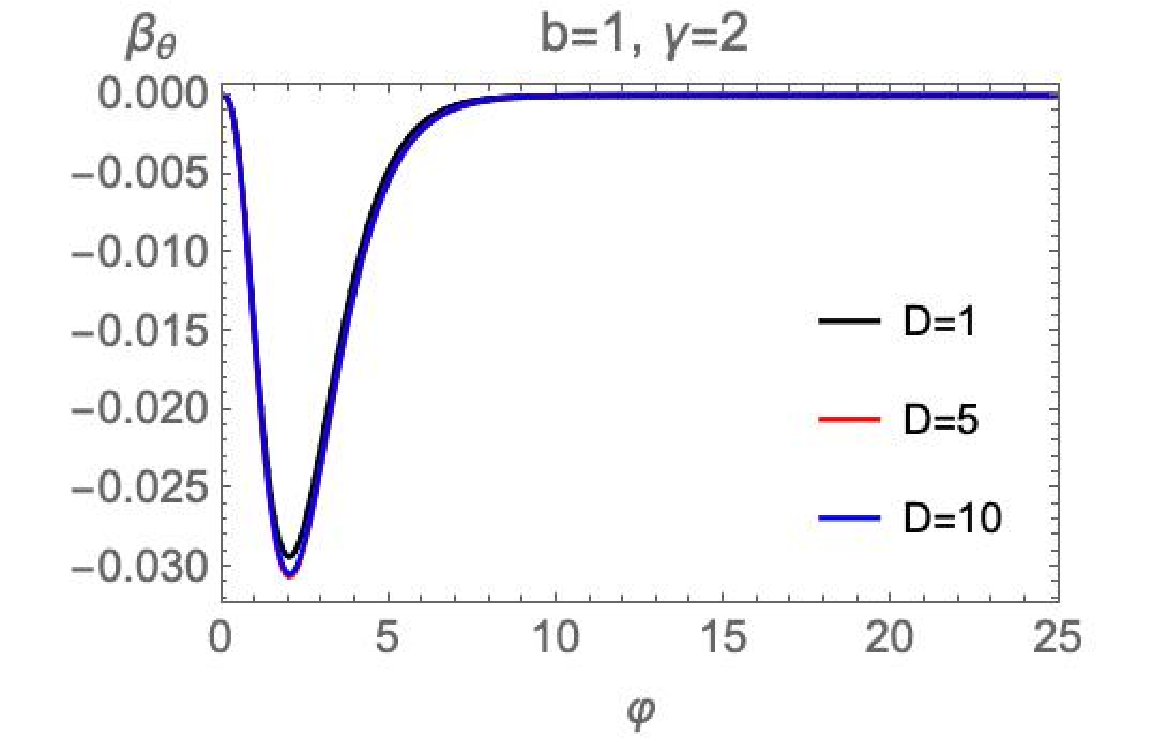}\hfil\hfil
   \\
  \includegraphics[width=.46\textwidth]{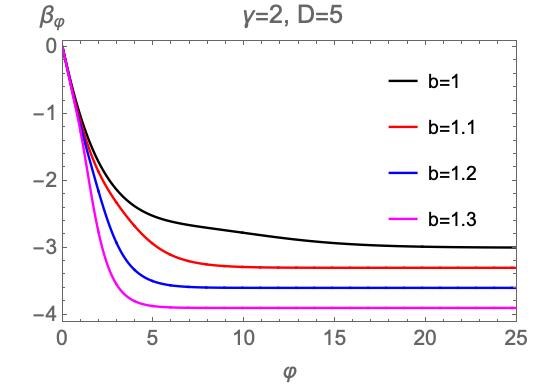}\hfil\hfil
   \includegraphics[width=.5\textwidth]{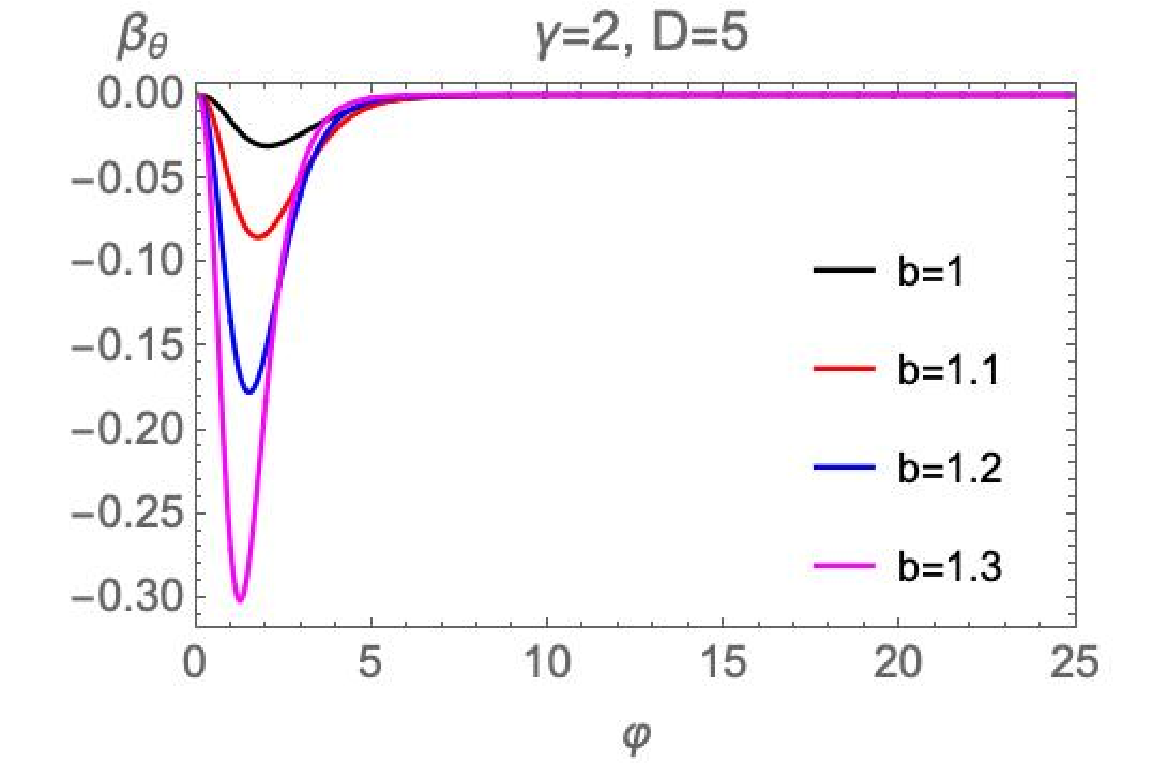}\hfil\hfil
   \\
  \includegraphics[width=.46\textwidth]{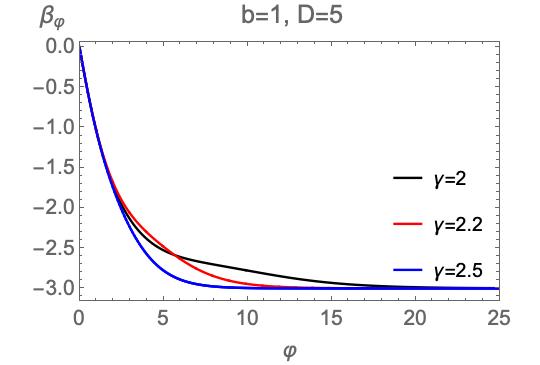}\hfil\hfil
   \includegraphics[width=.5\textwidth]{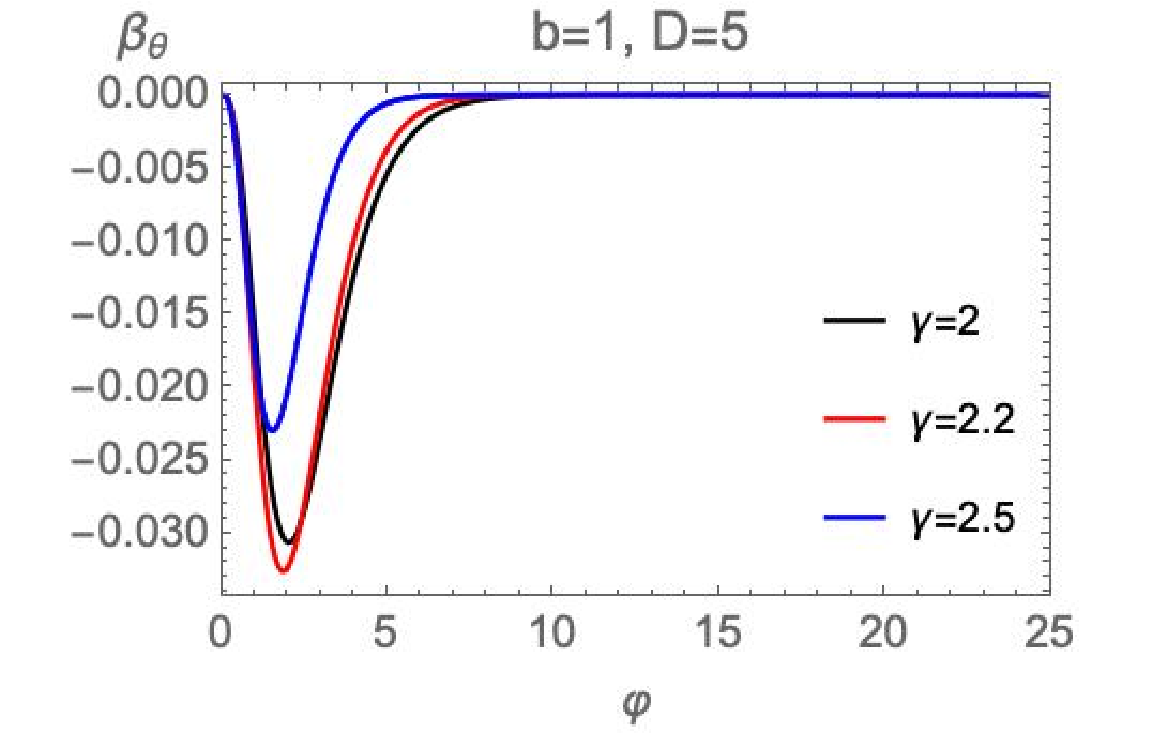}\hfil\hfil
 \end{center}
 \caption{
Plots of the holographic $\beta$ functions, $\beta_\f$ (\textbf{left}) and $\beta_\theta$ (\textbf{right}) as functions of $\f$. The three rows correspond to solutions with different values of $b, \g$ and $D$.
 }
  \label{fig16}
 \end{figure}

The holographic $\beta$-functions are plotted in the figure \ref{fig16} for the various values of $b, \g$ and $D$. The numerical results reproduce the expected asymptotic behavior recorded in \eqref{b5}, \eqref{b6}, \eqref{b7}, and \eqref{b8}. We can also observe that $\beta_\f$ is a monotonically decreasing function, and that $\beta_\theta$ quickly decays for larger values of $\f$.

\section*{Acknowledgements}
\addcontentsline{toc}{section}{Acknowledgements}

We would like to thank Matti Jarvinen for a discussion.

This work was supported in part by European Union's Seventh Framework Programme under grant agreements (FP7-REGPOT-2012-2013-1) no 316165 and the Advanced ERC grant SM-grav, No 669288.

\newpage
\appendix

\begin{appendix}
\renewcommand{\theequation}{\thesection.\arabic{equation}}
\addcontentsline{toc}{section}{Appendix}
\section*{APPENDIX}

\section{The near-boundary expansions\label{UV}}

In this appendix we shall determine the expansions of the functions $W,S,T$ near a maximum of the scalar potential $V(\f)$ that we shall put at $\f=0$ for convenience and without loss of generality. In the vicinity of an extremum the bulk functions $V(\f)$ and $Y(\f)$ can be expanded as (c.f.~\eqref{UV1}) in the main text:
\be
V=-{d(d-1)\over\ell^2}-{1\over2}{m^2\over\ell^2}\f^2+ V_3\f^3 +\mathcal{O}(\f^4) \, ,
\quad
Y= Y_0 + Y_1 \f +\mathcal{O}(\f^2) \, .
\label{aUV1}\ee
For completeness, we also reproduce the definition of the parameters $\Delta_\pm$, given in \eqref{UV29} in the main text:
\be
\Delta_{\pm}={d\over 2}\pm \sqrt{{d^2\over 4}-m^2\ell^2}
\label{aUV29}\ee
For a maximum,  $m^2>0$,  ${d\over 2}<\Delta_{+}<{d}$ and $0<\Delta_{-}<{d\over 2}$.

The expansions for $W,S,T$ can be found using similar techniques as in \cite{exotic}, \cite{curved}. There are two branches of solutions differing in the coefficient of $\f^2$ of $W$. In one (plus-branch) this coefficient is ${\Delta_+\over 2}$ while in the other (minus-branch) it is ${\Delta_-\over 2}$.   More specifically,

\begin{enumerate}
\item The terms in $W$ and $S$ independent of $q$ are determined by solving \eqref{a8} and \eqref{a9} with $T=0$, that is, ${d\over4(d-1)}W^2-{1\over2}W'^2+V=0$. It should be noted that $q$ will be related soon to the vev of the pseudoscalar operator dual to the axion.

\item The leading term of $T$ is calculated from \eqref{a8-2}.
\item The leading $\mathcal{O}(q)$ term of $W$ and $S$ are determined from \eqref{a8} and \eqref{a9}.
\item The subleading term of $T$ is then computed from \eqref{a8-2}.
\end{enumerate}
The analysis is performed independently for plus and minus type solutions.

For $W, S$ and $T$ we find for the minus-branch solutions :\footnote{The expansion is organized  as $\text{(analytic term)}+\text{(non-analytic term without $q$)}+\text{(non-analytic term including $q$)}$.}
\be
\ell W_{-}=
\left(2(d-1)+{\Delta_-\over2}\f^2 - {V_3\over d-\Delta_-}\f^3  +\mathcal{O}\left(\f^4\right)\right)
\label{aUV2}\ee
$$
+{C}|\f|^{d\over\Delta_-}
\left\{1+{d\over8(d-1)}\f^2
+\mathcal{O}\left(\f^3, C |\f|^{{\Delta_+-\Delta_-\over\Delta_-}}\right)\right\}
$$
$$
+{q\over 2d Y_0}|\f|^{2d\over\Delta_-}
\left\{1+\f \left( {6 d  V_3 \over (d-3\Delta_-)\Delta_-^2}  -  {d Y_1\over Y_0(d+\Delta_-)} \right)+\mathcal{O}\left(\f^2, C |\f|^{\Delta_+-\Delta_-\over\Delta_-}, q|\f|^{{2\Delta_+\over\Delta_-}+1}\right)
\right\},
$$
\be
\ell S_{-}=
\bigg({\Delta_-}\f - {3 V_3\over d-3\Delta_-}\f^2 +\mathcal{O}\left(\f^3\right)\bigg)
+{Cd\over\Delta_- }|\f|^{{\Delta_+\over\Delta_-}}
\left\{
1+{d+2\Delta_-\over8(d-1)}\f^2
+\mathcal{O}\left(\f^3, C |\f|^{{\Delta_+-\Delta_-\over\Delta_-}}\right)
\right\}
\label{aUV3}\ee
$$
+ q |\f|^{2d\over\Delta_-} {Y_1\over2Y_0^2(d+\Delta_-)} \bigg\{ 1 + \mathcal{O}\left(\f, C |\f|^{\Delta_+-\Delta_-\over\Delta_-}, q|\f|^{{2\Delta_+\over\Delta_-}+1}\right) \bigg\},
$$
\be
\ell^2 T_{-}={q\, |\f|^{2d\over\Delta_-}}
\bigg[\bigg(1 +  {6d V_3\over (d-3\Delta_-)\Delta_-^2}\f + \mathcal{O}(\f^2)\bigg)
\label{aUV4}\ee
$$
-{2Cd^2 \over (\Delta_+-\Delta_-)\Delta_-^2} |\f|^{\Delta_+ - \Delta_-\over\Delta_-}
\left\{1+\mathcal{O}\left(\f, C|\f|^{\Delta_+ - \Delta_-\over\Delta_-}\right)\right\}
+\mathcal{O}\left(q |\f|^{{2\Delta_+\over\Delta_-}+1}\right)\bigg],
$$

For the plus-branch we find,
\be
\ell W_{+}=
\left(2(d-1)+{\Delta_+\over2}\f^2+\mathcal{O}\left(\f^3\right)\right)
+{q\over 2d Y_0}|\f|^{2d \over\Delta_+}
\left\{1+\mathcal{O}\left(\f, q \,|\f|^{{2\Delta_-\over\Delta_+}+1}\right)\right\},
\label{aUV5}\ee
\be
\ell S_{+}=
\bigg(\Delta_+\f+\mathcal{O}\left(\f^2\right)\bigg)
+\mathcal{O}\left( q |\f|^{2d\over\Delta_+} \right),
\label{aUV6}\ee
\be
\ell^2 T_{+}=
q\, |\f|^{2d\over\Delta_+}
\left[
\bigg(1+\mathcal{O}(\f)\bigg)
+\mathcal{O}\left(q |\f|^{{2d\over\Delta_+}-1} \right)
\right],
\label{aUV7}\ee
where $C$ and $q$ are dimensionless integration constants. They are two of them in agreement with the fact that the system of equations for $W,S$, (\ref{a24}), (\ref{a30}) is second order, and $T$ is determined algebraically.  As we shall later see in \eqref{aUV28}, $q$ is related to $Q$, which was introduced in \eqref{a27} and it will be also related to the vev of the operator dual to the axion $a$. $C$ on the other hand will be related to the vev of the operator dual to the scalar $\f$.

Note also that the plus type solution only depends on one of the integration constants, namely  $q$. It will be interpreted as a flow driven by the vev of the operator dual to $\f$.

The higher order terms can be systematically calculated. The near-boundary expansion of the minus branch solution is schematically given by
\be
\ell W_{-}=
\sum_{n=0, n\neq1} \left[w_n \f^n
+\sum_{m=1} w_{n,m} \left( C^m  |\f|^{{\Delta_+-\Delta_-\over\Delta_-}m+2} \right) \f^{n}\right]
\label{aUV17}\ee
$$
+q \,|\f|^{2d\over\Delta_-} \left(\sum_{n,m=0}\tilde{w}_{n,m} \left(C |\f|^{\Delta_+-\Delta_-\over\Delta_-}\right)^m \f^n
+\sum_{n, m=0 \atop l=1} w_{n,m,l} \left(C |\f|^{\Delta_+-\Delta_-\over\Delta_-}\right)^m \left(q^l |\f|^{{2\Delta_+\over\Delta_-}l+1}\right) \f^n\right),
$$
\be
\ell S_{-}=
\sum_{n=1} s_n \f^n
+\sum_{n=0, \\ n\neq1 \atop m=1} s_{n,m} \left( C^m  |\f|^{{\Delta_+-\Delta_-\over\Delta_-}m+1} \right) \f^{n}
+\sum_{n,m=0\atop l=1} s_{n,m,l} \left(C |\f|^{\Delta_+-\Delta_-\over\Delta_-}\right)^m \left(q^l |\f|^{{2\Delta_+\over\Delta_-}l+2} \right) \f^n,
\label{aUV18}\ee
\be
\ell^2 T_{-}=q |\f|^{2d\over\Delta_-}
\left(
\sum_{n,m=0} t_{n,m} \left(C |\f|^{\Delta_+-\Delta_-\over\Delta_-}\right)^m \f^n
+\sum_{n,m=0,\atop l=1} t_{n,m,l} \left(C |\f|^{\Delta_+-\Delta_-\over\Delta_-}\right)^m \left(q^l |\f|^{{2\Delta_+\over\Delta_-}l+1}\right) \f^n
\right).
\label{aUV19}\ee
Similarly, the expansion of the plus branch solution is
\be
\ell W_{+}=
\sum_{n=0,2,3,\cdots} w_n \f^n
+q \,|\f|^{2d\over\Delta_-} \left(\sum_{n}\tilde{w}_{n}
+\sum_{n=0 \atop l=1} w_{n,l}  \left(q^l |\f|^{{2\Delta_-\over\Delta_+}l+1}\right) \f^n\right),
\label{aUV20}\ee
\be
\ell S_{+}=
\sum_{n=1} s_n \f^n
+\sum_{n=0\atop l=1} s_{n,l} \left(q^l |\f|^{{2\Delta_-\over\Delta_+}l+2} \right) \f^n,
\label{aUV21}\ee
\be
\ell^2 T_{+}=q |\f|^{2d\over\Delta_+}
\left(
\sum_{n=0} t_{n} \f^n
+\sum_{n=0,\atop l=1} t_{n,l} \left(q^l |\f|^{{2\Delta_-\over\Delta_+}l+1}\right) \f^n
\right).
\label{aUV22}\ee

The UV value of the axion will be determined through
\be
a_{UV}=-{\rm sign}(Q) \int^\infty_0 d\f~{\sqrt{T}\over YS}+a_{IR},
\label{aUV8}\ee
and  it will be related to the $\theta$-parameter of the UV CFT:
\be
\theta_{UV}+2\pi k=N_c a_{UV},
\quad k\in \mathbb{Z}, \quad 0\leq\theta_{UV}<2\pi.
\label{aUV9}\ee
For completeness, we also show the UV behavior of $A, \f$ and $a$. For the minus branch, from \eqref{a7} and \eqref{aUV3}, we obtain
\be
\f_{-}=
\f_- \ell^{\Delta_-}e^{\Delta_- {u/\ell}}\bigg(1+\mathcal{O}\left(\f_- \ell^{\Delta_-} e^{\Delta_- u/\ell}\right)\bigg)
\label{aUV10}\ee
$$
+{Cd\left(|\f_-|\,\ell^{\Delta_-}\right)^{{\Delta_+\over\Delta_-}}\over(\Delta_+-\Delta_-)\Delta_-}e^{\Delta_+{u/\ell}}\bigg[1+\mathcal{O}\left(\f_- \ell^{\Delta_-} e^{\Delta_- u/\ell}, C \left(|\f_-|\,\ell^{\Delta_-}\right)^{\Delta_+\over\Delta_-} e^{\Delta_+ u/\ell}\right)\bigg]
$$
$$
+\mathcal{O}\left(q \left(|\f_-|\,\ell^{\Delta_-}\right)^{2d\over\Delta_-} e^{2du/\ell}\right),
$$
where $\f_-$ is the integration constant whose mass dimension is $\Delta_-$.
From this we can read off the expectation value of the operator dual to $\f$:
\be
\langle O\rangle=C \, (M_p \ell)^{d-1}{d\over \Delta_-}|\f_-|^{\Delta_+\over\Delta_-}.
\label{aUV26}\ee

From \eqref{a6} and \eqref{a7}, we determine  that
\be
{d\over d\f} e^A=-{1\over2(d-1)}{W\over S} e^A\;,
\ee
 which leads to
\be
e^{A_{-}}=
e^{\bar{A}} \left(|\tilde{\f}_-|\over |\f| \right)^{1\over\Delta_-}
\bigg[
\bigg(1+\mathcal{O}(\f)\bigg)
+{Cd |\f|^{{\Delta_+ - \Delta_-\over\Delta_-}}\over(\Delta_+-\Delta_-)\Delta_-^2}
\left\{1+\mathcal{O}\left(\f, C |\f|^{{\Delta_+ - \Delta_-\over\Delta_-}} \right)\right\}
+\mathcal{O}\left(q |\f|^{{2d\over\Delta_-}-1}\right)
\bigg]
\label{aUV11}\ee
$$
=e^{\bar{A}-{u\over\ell}}
\bigg[
\bigg(1+ \mathcal{O}\left(\tilde{\f}_-^2 e^{2\Delta_- u/\ell}\right)\bigg)
+\mathcal{O}\left(C |\tilde{\f}_-|^{\Delta_+\over\Delta_-} e^{\Delta_+ u/\ell}\right)
+\mathcal{O}\left(q |\tilde{\f}_-|^{2d/\Delta_-} e^{2du/\ell}\right)
\bigg],
$$
where $\bar{A}$ is the integration constant and the dimensionless combination $\tilde{\f}_-:=\f_-\,\ell^{\Delta_-}$ is introduced. As mentioned below \eqref{a9}, the value of the $\bar{A}$ is not physical, and, in the following, we shall set $\bar{A}=0$. Finally, from \eqref{a5}, \eqref{aUV4} and \eqref{aUV10}, the near-boundary expansion of the axion field is
\be
a_{-}=a_{UV}
+ {\rm sign}(Q) {\sqrt{q}|\f|^{d\over\Delta_-}\over d Y_0}
\bigg[\bigg(1+\mathcal{O}(\f)\bigg)
\label{aUV23}\ee
$$
-{Cd^2 |\f|^{{\Delta_+\over\Delta_-}-1}\over (\Delta_+-\Delta_-)\Delta_-^2}
\left\{1+\mathcal{O}\left(\f, C|\f|^{{\Delta_+\over\Delta_-}-1}\right)\right\}
+\mathcal{O}\left(q |\f|^{{2\Delta_+\over\Delta_-}+1}\right)\bigg]
$$
$$
=a_{UV}
+{\rm sign}(Q){\sqrt{q} |\tilde{\f}_-|^{d\over\Delta_-} \over d Y_0} e^{du/\ell}
\bigg[
\bigg(1+\mathcal{O}\left(\tilde{\f}_-^2 e^{{2\Delta_-\over\ell}u}\right)\bigg)
+\mathcal{O}\left(C |\tilde{\f}_-|^{\Delta_+\over\Delta_-}e^{{\Delta_+\over\ell}u}\right)
+\mathcal{O}\left(q |\tilde{\f}_-|^{{2d\over\Delta_-}} e^{{2d\over\ell}u}\right)
\bigg].
$$
The expectation value of the operator dual\footnote{The dual operator is a topological density. In $d=4$ it is the standard instanton density.} to $a$ is
\be
\langle O_a\rangle=
{1\over N_c} {\rm sign}(Q) (M_p \ell)^{d-1} \, \sqrt{q}\,  |\f_-|^{d\over\Delta_-}.
\label{aUV27}\ee

Similarly, for the plus branch, we obtain from \eqref{a6}, \eqref{a7} and \eqref{aUV6}
\be
\f_{+}=
\f_+ \ell^{\Delta_+}e^{\Delta_+ {u/\ell}}\bigg(1+\mathcal{O}\left(\f_+ \ell^{\Delta_+} e^{\Delta_+ u/\ell}\right)\bigg)
+\mathcal{O}\left(q \left(|\f_+|\,\ell^{\Delta_+}\right)^{2d\over\Delta_+} e^{2du/\ell}\right),
\label{aUV12}\ee
\be
e^{A_{+}}=
\left(|\tilde{\f}_+|\over\f\right)^{1\over\Delta_+}
\left[
\bigg(1+\mathcal{O}(\f)\bigg)
+\mathcal{O}\left(q |\f|^{{2d\over\Delta_+}-1}\right)
\right]
\label{aUV13}\ee
$$
=e^{-{u\over\ell}}
\bigg[
\bigg(1+ \mathcal{O}\left(\tilde{\f}_+ e^{\Delta_+ u/\ell}\right)\bigg)
+\mathcal{O}\left(q |\tilde{\f}_+|^{2d/\Delta_+} e^{2du/\ell}\right)
\bigg],
$$
\be
a_{+}=a_{UV}
+{\rm sign}(Q){\sqrt{q}\over \ell Y_0}|\f|^{d\over\Delta_+}
\bigg[
\bigg(1+\mathcal{O}\left(\f\right)\bigg)
+\mathcal{O}\left(q |\tilde{\f}_+|^{2d/\Delta_+} e^{2du/\ell}\right)
\bigg]
\label{aUV24}\ee
$$
=a_{UV}
+{\rm sign}(Q) {\sqrt{q}\over\ell Y_0}|\tilde{\f}|^{d\over\Delta_+}e^{du/\ell}
\bigg[
\bigg(1+\mathcal{O}\left(\tilde{\f}_+ e^{{\Delta_+\over\ell}u}\right)\bigg)
+\mathcal{O}\left(q |\tilde{\f}_+|^{{2\Delta_-\over\Delta_+}+1} e^{{2d-\Delta_+\over\ell}u}\right)
\bigg].
$$
where $\f_+$ is the integration constant, and $\tilde{\f}_+:=\f_+ \ell^{\Delta_+}$.

Note that there exists a relation between the minus and plus branches of solutions. In particular, one can show that in the limit
\be
\f_-\to0,\quad
C, q\to\infty,\quad
C  |\tilde{\f}_-|^{\Delta_+\over\Delta_-}=
\text{const.},\quad
q |\tilde{\f}_-|^{2d\over\Delta_-}=\text{const.}\,,
\label{aUV14}\ee
the minus branch solutions asymptote to the corresponding solution on the plus branch, generalizing the same property valid in the presence of only scalar operators, \cite{Papa1,exotic}.
Indeed, in this limit, \eqref{aUV10}, \eqref{aUV11} and \eqref{aUV23} become
\be
\f_{-}\to
{Cd|\tilde{\f}_-|^{{\Delta_+\over\Delta_-}}\over(\Delta_+ - \Delta_-)\Delta_-}e^{\Delta_+{u/\ell}}
\left[1+\mathcal{O}\left(C |\tilde{\f}_-|^{\Delta_+\over\Delta_-} e^{\Delta_+ u/\ell}\right)\right]
+\mathcal{O}\left(q |\tilde{\f}_-|^{2d\over\Delta_-} e^{2du/\ell}\right),
\label{aUV15}\ee
\be
e^{A_{-}}\to
e^{-{u\over\ell}}
\left[
1
+\mathcal{O}\left(C |\tilde{\f}_-|^{\Delta_+\over\Delta_-} e^{\Delta_+ u/\ell}\right)
+\mathcal{O}\left(q |\tilde{\f}_-|^{2d/\Delta_-} e^{2du/\ell}\right)
\right],
\label{aUV16}\ee
\be
a_{-}\to
a_{UV}
+{\rm sign}(Q){\sqrt{q}\over\ell Y_0}|\tilde{\f}|^{d\over\Delta_-}e^{du/\ell}
\bigg[
1+\mathcal{O}\left(C |\tilde{\f}_-|^{\Delta_+\over\Delta_-}e^{{\Delta_+\over\ell}u}\right)
+\mathcal{O}\left(q |\tilde{\f}_-|^{{2d\over\Delta_-}} e^{{2d\over\ell}u}\right)
\bigg].
\label{aUV25}\ee
We observe that \eqref{aUV15}, \eqref{aUV16} and \eqref{aUV25} are of the same form as \eqref{aUV12}, \eqref{aUV13} and \eqref{aUV24}, respectively.

Finally, from \eqref{a5}, \eqref{aUV4} and \eqref{aUV11}, the integration constants $Q$ and $q$ are related as
\be
Q={\rm sign}(Q) {\sqrt{q}\over\ell}\left(\ell |\f_-|^{1/\Delta_-}\right)^d.
\label{aUV28}\ee

\vskip 1cm
\section{The subleading axion solution\label{aB}}
\vskip 1cm

In this appendix we shall determine the subleading axion solution presented in section \ref{sub} in detail. The starting point is the solution (\ref{a34}) for $W$. Substituting into (\ref{a36}) and (\ref{a37}) one finds
\be
S_\infty=\pm{b\sqrt{2(d-1)V_\infty}\over \sqrt{2d-(d-1)b^2}}= {b\over2} W_\infty \sp T_\infty=0.
\label{aa38}\ee
Therefore, this corresponds to a solution with a trivial (constant) axion.

This however is a starting point for obtaining a solution with a non-trivial axion,  that does not back-react on the geometry to leading order in the IR.
The regular axion solutions discussed and tested in \cite{iQCD} and
\cite{VQCD} are indeed of that type.

To setup the perturbation theory,  we write
\be \label{asub1}
\ell W \simeq W_\infty e^{{b\over2}\varphi} + \delta W,
\quad
\ell S \simeq {b\over2}W_\infty\, e^{{b\over2}\varphi} + \delta S,
\ee
where $W_\infty$ is given in (\ref{a34}), and $\delta W$ and $\delta S$ are small perturbations in the IR. $V$ and $Y$ are the same as in \eqref{a32}.

From \eqref{a8}, we obtain
\be \label{asub2}
\ell^2 T=-\ell^2 YS(S-W')=-{b\over2}W_\infty Y_\infty (\delta S - \delta W') e^{({b\over2}+\gamma)\f}.
\ee
and \eqref{a8-2} becomes
\be \label{asub3}
{({b\over2}+\gamma)(\delta S- \delta W')+(\delta S'- \delta W'')\over\delta S- \delta W'}
={2d\over (d-1)b},
\ee
from which we obtain
\be \label{asub4}
\ell\left(\delta S- \delta W'\right)= -D e^{\left({2d\over (d-1)b} -{b\over2}-\gamma\right)\f},
\quad
\ell^2 T={b\over2} D W_\infty Y_\infty e^{{2d\over (d-1)b} \f},
\ee
where $D$ is an integration constant.

Finally, \eqref{a9} becomes
\be \label{asub5}
W_\infty e^{{b\over2}\f} \left\{{d\over2(d-1)}\delta W- {b\over2} \delta S - {b\over4}{D\over\ell} e^{\left({2d\over (d-1)b} - {b\over2}-\gamma\right)\f}\right\}=0.
\ee
Substituting \eqref{asub4} into \eqref{asub5}, we obtain
\be \label{asub6}
{d\over2(d-1)}\delta W - {b\over2} \delta W' + {b\over4} {D\over\ell} e^{\left({2d\over (d-1)b}-{b\over2}-\g\right)\f}=0,
\ee
whose solution is
\be \label{asub7}
\ell \delta W=
E e^{{d\over(d-1)b}\f}
-{D\over b+2\g-{2d\over (d-1)b}} e^{\left({2d\over (d-1)b}-{b\over2}-\g\right)\f},
\ee
where $E$ is another integration constant. Collecting everything, we have
\begin{align}\label{asub16}
&\ell W=
W_\infty \,e^{{b\over2}\f}
-{D\over2}{1\over {b\over2}+\g-{d\over (d-1)b}} e^{\left({2d\over (d-1)b}-{b\over2}-\g\right)\f}
+E \,e^{{d\over(d-1)b}\f}+\cdots,
\\ \label{asub17}
&\ell S=
{b\over2}W_\infty\, e^{{b\over2}\varphi}
-{D\over2}{{b\over2}+\g\over {b\over2}+\g-{d\over (d-1)b}} e^{\left({2d\over (d-1)b}-{b\over2}-\g\right)\f}
+{Ed\over(d-1)b} \,e^{{d\over(d-1)b}\f}+\cdots,
\\ \label{asub18}
&\ell^2 T=
{b\over2}D W_\infty Y_\infty \,e^{{2d\over(d-1)b}\f}+\cdots,
\end{align}
with $D$ and $E$ the integration constants introduced above. Here $D$ should be $D\geq0$ from \eqref{a27} and $b, W_\infty, Y_\infty, e^{dA}\geq0$.
However, the term in $W$ containing $E$, can be shown to lead to a violation of the Gubser bound~\cite{Gubser}, giving rise to an unacceptable singularity.
The reason is, that when $b$ satisfies the Gubser bound in (\ref{a41}) then
\be
{d\over(d-1)b}~>~\sqrt{d-1\over 2d}
\ee
and therefore this term violates the Gubser bound.
To avoid this, this term must be absent, which we ensure by setting $E=0$ in the following.

Therefore the final solution can be written as
\be
\label{asub8}
\ell W=
W_\infty \,e^{{b\over2}\f}
	-{D\over2}{ e^{-\left({b\over2}+\g-{2d\over (d-1)b}\right)\f} \over {b\over2}+\g-{d\over (d-1)b}}
+\cdots
\equiv
W_\infty e^{{b\over2}\f}+ W_{\infty,1} e^{-\left({b\over2}+\g-{2d\over (d-1)b}\right)\f}+\cdots,
\ee
\be
\ell S=
{b\over2}W_\infty\, e^{{b\over2}\varphi}
	-{D\over2}{{b\over2}+\g \over {b\over2}+\g-{d\over (d-1)b}}
	e^{-\left({b\over2}+\g-{2d\over (d-1)b}\right)\f}
	+\cdots\equiv
S_\infty e^{{b\over2}\f}
+ S_{\infty,1} e^{-\left({b\over2}+\g-{2d\over (d-1)b}\right)\f}
+\cdots	,
\label{asub9}
\ee
\be
\ell^2 T=
{b\over2}D W_\infty Y_\infty \,e^{{2d\over(d-1)b}\f}+\cdots.
\label{asub10}\ee

In order for the subleading corrections to be truly subleading, we must have
\be
\gamma\geq {2d\over(d-1)b}-b={2d-(d-1)b^2\over (d-1)b},
\label{asub11}\ee
assuming $b>0$.
In the case of equality, the solution coincides with the non-trivial solution found in (\ref{a39}).

Now we can compute the axion solution $a$
as
\be
a(u)={\rm sign}(Q) \int du {\sqrt{T}\over Y}
={\rm sign}(Q)\int d\f~{\sqrt{T}\over YS}
={\rm sign}(Q)\sqrt{2D\over bW_\infty Y_\infty}\int d\f ~e^{-\left({b\over 2}+\gamma-{d\over (d-1)b}\right)\f}+\cdots
\label{asub12}\ee
$$
=-{\rm sign}(Q)\sqrt{2D\over bW_\infty Y_\infty}{1\over {b\over 2}+\gamma-{d\over (d-1)b}}
e^{-\left({b\over 2}+\gamma-{d\over (d-1)b}\right)\f}+\cdots.
$$
Equation (\ref{sub11}) implies that the exponent above is negative
\be
-\left({b\over 2}+\gamma-{d\over (d-1)b}\right)\leq -{d\over (d-1)b}\leq 0
\label{asub13}\ee
and therefore the integral in (\ref{asub12}) is convergent as $\f\to\infty$.
Imposing that $a_{IR}=0$ as in \eqref{sub21a}, the source $a_{UV}$ is obtained as
\be
a_{UV}=-{\rm sign}(Q)\int^\infty_0 d\f~{\sqrt{T}\over YS}.
\label{asub21}\ee

We shall also calculate the behavior of $\f$ and $A$ in the IR.
From \eqref{a7} we have
\be
e^{{b\over2}\f}=
{2\over b S_\infty (u_{IR}-u)/\ell}
-{S_{\infty,1}\over S_\infty^2} {1\over {3\over2}b+\g-{2d\over (d-1)b}}\left(2\over bS_\infty(u_{IR}-u)/\ell\right)^{-2\left(1+{\g\over b}-{2d\over(d-1)b^2}\right)}+\cdots.
\label{asub22}\ee
Similarly, from \eqref{a6}, we obtain
\be
A=
A_{IR}
-{\f\over(d-1)b}
+{W_{\infty,1}\over W_\infty}{1\over(d-1)b^2} e^{-\left(b+\gamma-{2d\over(d-1)b}\right)\f}+\cdots,
\label{asub25}\ee
where $A_{IR}$ is the integration constant. As we shall see in \eqref{asub26}, this can be written in terms of other integration constants.

We note that the two constants,  $D$ defined by IR expansion and $q$ defined by UV expansion are related. From \eqref{a8-2} and \eqref{sub10}, we obtain the relation as
\be
{q\over D}
=\lim_{\f_{UV}\to0\atop \f_{IR}\to\infty}
{{b\over2}W_\infty Y_\infty e^{{2d\over(d-1)b}\f_{IR}} \over |\f_{UV}|^{2d/\Delta_-} \exp\left({d\over d-1}\int^{\f_{IR}}_{\f_{UV}}d\f\,{W\over S}\right)}.
\label{asub14}\ee
This does not necessary mean that $q$ is proportional to $D$,  as both $W$ and $S$ on the right-hand side also depend on this integration constant.
The integration constant $A_{IR}$ is also determined by $q$,
\be
e^{2d A_{IR}}=
{2\over b D W_\infty Y_\infty} \ell^2 Q^2=
{2\over b D W_\infty Y_\infty} q \left(\ell |\f_-|^{1/\Delta_-}\right)^{2d},
\label{asub26}\ee
where \eqref{a27}, \eqref{aUV28}, \eqref{asub18} and \eqref{asub25} have been used.

\vskip 1cm
\section{Perturbation theory for small $q$\label{aq}}

\vskip 1cm

In this appendix we provide several calculational details for the small $q$ expansions described in section \ref{q}.

For small $q$, we can derive an analytical formula for the correction to the solution with $q=0$.

Assuming small $q$, the functions $W, S$ and $T$ can be expanded as
\be
W=W^{(q0)}+ q W^{(q1)} + q^2 W^{(q2)}+\mathcal{O}(q^3),
\label{aq1}\ee
\be
S=S^{(q0)} + q S^{(q1)}+ q^2 S^{(q2)}+\mathcal{O}(q^3),
\label{aq2}\ee
\be
T=q T^{(q1)} + q^2 T^{(q2)}+\mathcal{O}(q^3),
\label{aq3}\ee
where $S^{(q0)}=W'^{(q0)}$. In the following, we shall show that $W^{(q1)}$, $S^{(q1)}$ and $T^{(q1)}$ can be expressed in terms of $W^{(q0)}$ and $S^{(q0)}$.

First, from \eqref{a8-2}, it follows that
\be
\ell^2 T^{(q1)}=\lim_{\f_{UV}\to0} |\f_{UV}|^{2d/\Delta_-}\exp\left({d\over d-1}\int^{\f}_{\f_{UV}}d\f'\,{W^{(q0)}\over S^{(q0)}}\right).
\label{aq4}\ee
Then, from \eqref{a8} and \eqref{a9}, we obtain\footnote{Here $S^{(q0)}\neq0$ is assumed.}
\be
S^{(q1)}=W'^{(q1)}-{T^{(q1)}\over YS^{(q0)}},
\label{aq7}\ee
and
\be
S^{(q0)}W'^{(q1)}={T^{(q1)}\over 2Y}+{d\over2(d-1)}W^{(q0)}W^{(q1)}.
\label{aq5}\ee
We want to solve \eqref{aq7} and \eqref{aq5} while keeping the regularity condition in the IR.
The general solution of \eqref{aq5} is
\be
\ell W^{(q1)}=
\lim_{\f_{UV}\to0} e^{{d\over2(d-1)}\int^\f_{\f_{UV}}d\f'{W^{(q0)}\over S^{(q0)}}}
\left(E^{(q1)}
+{1\over2} |\f_{UV}|^{2d/\Delta_-}\int^\f_{\f_{UV}}d\f' {e^{{d\over2(d-1)}\int^{\f'}_{\f_{UV}}d{\f''}{W^{(q0)}\over S^{(q0)}}}\over Y \ell S^{(q0)}}\right),
\label{aq6}\ee
where $E^{(q1)}$ is an integration constant, which should be fixed by IR regularity.
By using \eqref{asub13}, we observe that the $\f$ integration in the second term in the parenthesis is convergent for $\f\to\infty$. Then, IR regularity fixes $E^{(q1)}$ as
\be
E^{(q1)}=
-\lim_{\f_{UV}\to0}{1\over2} |\f_{UV}|^{2d/\Delta_-}\int^\infty_{\f_{UV}}d\f' {e^{{d\over2(d-1)}\int^{\f'}_{\f_{UV}}d{\f''}{W^{(q0)}\over S^{(q0)}}}\over Y \ell S^{(q0)}}.
\label{aq8}\ee
From \eqref{aq7} and \eqref{aq6}, the leading corrections are
\be
\ell W^{(q1)}=
-\lim_{\f_{UV}\to0}{1\over2} |\f_{UV}|^{2d/\Delta_-}e^{{d\over2(d-1)}\int^\f_{\f_{UV}}d\f'{W^{(q0)}\over S^{(q0)}}}
\int^{\infty}_{\f}d\f' {e^{{d\over2(d-1)}\int^{\f'}_{\f_{UV}}d{\f''}{W^{(q0)}\over S^{(q0)}}}\over Y \ell S^{(q0)}},
\label{aq9}\ee
\be
S^{(q1)}=
{d\over2(d-1)}{W^{(q0)}\over S^{(q0)}}W^{(q1)}-{T^{(q1)}\over2YS^{(q0)}},
\label{aq10}\ee
and $T^{(q1)}$ is given in \eqref{aq4}. Thus every correction at order $\mathcal{O}(q^1)$ can be written in terms of the $\mathcal{O}(q^0)$ solutions $W^{(q0)}$ and $S^{(q0)}$.

Now we examine the UV and IR limits of \eqref{aq9}, \eqref{aq10} and \eqref{aq4}.
The IR asymptotics of the corrections are
\be
W^{(q1)}=
-{1\over{b\over2}+\g-{d\over(d-1)b}}{T^{(q1)}\over2YS^{(q0)}}+\cdots,
\label{aq11}\ee
\be
S^{(q1)}=
\left({b\over2}+\g\right)W^{(q1)}+\cdots,
\label{aq12}\ee
\be
\ell^2 T^{(q1)}=
\lim_{\f_{UV}\to0\atop \f_{IR}\to\infty}
|\f_{UV}|^{2d/\Delta_-}{e^{{d\over d-1}\int^{\f_{IR}}_{\f_{UV}}d\f'\,{W^{(q0)}\over S^{(q0)}}}\over e^{{2d\over(d-1)b}\f_{IR}}}e^{{2d\over(d-1)b}\f}
+\cdots,
\label{aq13}\ee
Therefore, we obtain
\be
D=
q \lim_{\f_{UV}\to0\atop \f_{IR}\to\infty}
{|\f_{UV}|^{2d/\Delta_-}e^{{d\over d-1}\int^{\f_{IR}}_{\f_{UV}}d\f'\,{W^{(q0)}}}\over{b\over2}W_\infty Y_\infty e^{{2d\over(d-1)b}\f_{IR}}}
+\mathcal{O}(q^2),
\label{aq20}\ee
from \eqref{asub18}.

The UV asymptotics are
\be
\ell W^{(q1)}=
-{1\over2}
\lim_{\f_{UV}\to0}\left(|\f_{UV}|^{d/\Delta_-}\int^\infty_0 d\f' {e^{{d\over2(d-1)}\int^{\f'}_{\f_{UV}}d{\f''}{W^{(q0)}\over S^{(q0)}}}\over Y \ell S^{(q0)}}\right) |\f|^{d/\Delta_-}
+{1\over2d Y_0} |\f|^{2d/\Delta_-}
+\cdots,
\label{aq14}\ee
\be
\ell S^{(q1)}=
-{d\over2\Delta_-}\lim_{\f_{UV}\to0}\left(|\f_{UV}|^{d/\Delta_-}\int^\infty_0 d\f' {e^{{d\over2(d-1)}\int^{\f'}_{\f_{UV}}d{\f''}{W^{(q0)}\over S^{(q0)}}}\over Y \ell S^{(q0)}}\right) |\f|^{{d\over\Delta_-}-1}
+\cdots,
\label{aq15}\ee
\be
\ell^2 T^{(q1)}= |\f|^{2d\over\Delta_-}
+\cdots.
\label{aq16}\ee
By comparing \eqref{aUV2} and \eqref{aq14}, we observe that the correction to $C$ is given by
\be
C^{(q1)}=
-\lim_{\f_{UV}\to0}{1\over2} |\f_{UV}|^{d/\Delta_-}\int^\infty_0 d\f' {e^{{d\over2(d-1)}\int^{\f'}_{\f_{UV}}d{\f''}{W^{(q0)}\over S^{(q0)}}}\over Y \ell S^{(q0)}},
\label{aq17}\ee
where $C^{(q1)}$ is defined as the $q$-expansion of $C$:
\be
C=C^{(q0)}+ q\, C^{(q1)} + q^2 C^{(q2)}+\mathcal{O}(q^3).
\label{aq18}\ee

From \eqref{sub21}, the small $q$ perturbation of the source of the axion $a_{UV}$ is
\be
a_{UV}=
-{\rm sign}(Q)\sqrt{q} \lim_{\f_{UV}\to0} |\f_{UV}|^{d/\Delta_-} \int^\infty_0 d\f {e^{{d\over2(d-1)}\int^{\f}_{\f_{UV}}d\f' {W^{(q0)}\over S^{(q0)}}}\over Y \ell S^{(q0)}}+\mathcal{O}(q)
=2{\rm sign}(Q) \sqrt{q} \, C^{(q1)}+\mathcal{O}(q).
\label{aq19}\ee

\end{appendix}



\begin{thebibliography}{100}

\bibitem{QCD-axion}
  R.~D.~Peccei and H.~R.~Quinn,
  {\em ``CP Conservation in the Presence of Instantons,''}
  \href{https://doi.org/10.1103/PhysRevLett.38.1440}{Phys.\ Rev.\ Lett.\  {\bf 38} (1977) 1440}.

  \bibitem{Kim}
  J.~E.~Kim and G.~Carosi,
  {\em ``Axions and the Strong CP Problem,''}
  Rev.\ Mod.\ Phys.\  {\bf 82} (2010) 557
  doi:10.1103/RevModPhys.82.557
\hri{0807.3125}{[hep-ph]}.

  \bibitem{KN}
  M.~Kawasaki and K.~Nakayama,
  {\em ``Axions: Theory and Cosmological Role,''}
  Ann.\ Rev.\ Nucl.\ Part.\ Sci.\  {\bf 63} (2013) 69
  doi:10.1146/annurev-nucl-102212-170536
 \hri{1301.1123}{[hep-ph]}.


    \bibitem{Baer}
  H.~Baer, K.~Y.~Choi, J.~E.~Kim and L.~Roszkowski,
  {\em ``Dark matter production in the early Universe: beyond the thermal WIMP paradigm,''}
  Phys.\ Rept.\  {\bf 555} (2015) 1
  doi:10.1016/j.physrep.2014.10.002
\hri{1407.0017}{ [hep-ph]}.

  \bibitem{Marsh}
  D.~J.~E.~Marsh,
  {\em ``Axion Cosmology,''}
  Phys.\ Rept.\  {\bf 643} (2016) 1
  doi:10.1016/j.physrep.2016.06.005
 \hri{1510.07633}{[astro-ph.CO]}.

  \bibitem{relaxion}
  P.~W.~Graham, D.~E.~Kaplan and S.~Rajendran,
  {\em ``Cosmological Relaxation of the Electroweak Scale,''}
  Phys.\ Rev.\ Lett.\  {\bf 115} (2015) no.22,  221801
  doi:10.1103/PhysRevLett.115.221801
  \hri{1504.07551}{[hep-ph]}.

  \bibitem{GILLV}
  P.~W.~Graham, I.~G.~Irastorza, S.~K.~Lamoreaux, A.~Lindner and K.~A.~van Bibber,
  {\em ``Experimental Searches for the Axion and Axion-Like Particles,''}
  Ann.\ Rev.\ Nucl.\ Part.\ Sci.\  {\bf 65} (2015) 485
  doi:10.1146/annurev-nucl-102014-022120
\hri{1602.00039}{[hep-ex]}.

    \bibitem{IR}
  I.~G.~Irastorza and J.~Redondo,
  {\em ``New experimental approaches in the search for axion-like particles,''}
  Prog.\ Part.\ Nucl.\ Phys.\  {\bf 102} (2018) 89
  doi:10.1016/j.ppnp.2018.05.003
\hri{1801.08127}{[hep-ph]}.

\bibitem{ws}
  P.~Svrcek and E.~Witten,
  {\em ``Axions In String Theory,''}
  JHEP {\bf 0606} (2006) 051
  doi:10.1088/1126-6708/2006/06/051
  \hre{hep-th}{0605206}.

\bibitem{OV}
  H.~Ooguri and C.~Vafa,
  {\em `Summing up D instantons,''}
  Phys.\ Rev.\ Lett.\  {\bf 77} (1996) 3296
  doi:10.1103/PhysRevLett.77.3296
  \hre{hep-th}{9608079}.

  \bibitem{W}
  E.~Witten,
  {\em ``World sheet corrections via D instantons,''}
  JHEP {\bf 0002} (2000) 030
  doi:10.1088/1126-6708/2000/02/030
  \hre{hep-th}{9907041}.

  \bibitem{BCKW}
  R.~Blumenhagen, M.~Cvetic, S.~Kachru and T.~Weigand,
  {\em ``D-Brane Instantons in Type II Orientifolds,''}
  Ann.\ Rev.\ Nucl.\ Part.\ Sci.\  {\bf 59} (2009) 269
  doi:10.1146/annurev.nucl.010909.083113
  \hri{0902.3251}{[hep-th]}.

  \bibitem{BK}
  M.~Bianchi and E.~Kiritsis,
  {\em ``Non-perturbative and Flux superpotentials for Type I strings on the Z$_3$ orbifold,''}
  Nucl.\ Phys.\ B {\bf 782} (2007) 26
  doi:10.1016/j.nuclphysb.2007.05.006
  \hre{hep-th}{0702015}].

  \bibitem{panago}
  E.~Vicari and H.~Panagopoulos,
  {\em ``$\Theta$-dependence of SU(N) gauge theories in the presence of a topological term,''}
  Phys.\ Rept.\  {\bf 470} (2009) 93
  doi:10.1016/j.physrep.2008.10.001
\hri{0803.1593}{ [hep-th]}.

    \bibitem{SW}
  N.~Seiberg and E.~Witten,
  {\em ``Electric - magnetic duality, monopole condensation, and confinement in N=2 supersymmetric Yang-Mills theory,''}
  Nucl.\ Phys.\ B {\bf 426} (1994) 19
   Erratum: [Nucl.\ Phys.\ B {\bf 430} (1994) 485]
  doi:10.1016/0550-3213(94)90124-4, 10.1016/0550-3213(94)00449-8
  \hre{hep-th}{9407087}.

  \bibitem{witten}
  E.~Witten,
  {\em ``$\theta$ dependence in the large N limit of four-dimensional gauge theories,''}
  Phys.\ Rev.\ Lett.\  {\bf 81} (1998) 2862
  doi:10.1103/PhysRevLett.81.2862
  \hre{hep-th}{9807109}.

\bibitem{iQCD}
 U.~Gursoy and E.~Kiritsis,
  {\em ``Exploring improved holographic theories for QCD: Part I,''}
  JHEP {\bf 0802} (2008) 032
  doi:10.1088/1126-6708/2008/02/032
\hri{0707.1324}{[hep-th]};\\
  U.~Gursoy, E.~Kiritsis and F.~Nitti,
  {\em ``Exploring improved holographic theories for QCD: Part II,''}
  JHEP {\bf 0802} (2008) 019
  doi:10.1088/1126-6708/2008/02/019
  \hri{0707.1349}{[hep-th]};\\
  U.~Gursoy, E.~Kiritsis, L.~Mazzanti, G.~Michalogiorgakis and F.~Nitti,
  {\em ``Improved Holographic QCD,''}
  Lect.\ Notes Phys.\  {\bf 828} (2011) 79
  doi:10.1007/978-3-642-04864-7
\hri{1006.5461}{[hep-th]}.

  \bibitem{disect}
  E.~Kiritsis,
  {\em ``Dissecting the string theory dual of QCD,''}
  Fortsch.\ Phys.\  {\bf 57} (2009) 396
  doi:10.1002/prop.200900011
  \hri{0901.1772}{[hep-th]}.

  \bibitem{VQCD}
  D.~Arean, I.~Iatrakis, M.~Jarvinen and E.~Kiritsis,
  {\em ``CP-odd sector and $\theta$ dynamics in holographic QCD,''}
  Phys.\ Rev.\ D {\bf 96} (2017) no.2,  026001
  doi:10.1103/PhysRevD.96.026001
\hri{1609.08922}{[hep-ph]}.


\bibitem{smgrav}
  E.~Kiritsis,
  {\em ``Gravity and axions from a random UV QFT,''}
  EPJ Web Conf.\  {\bf 71} (2014) 00068;
  \hri{1408.3541}{[hep-ph]}.

  \bibitem{axion}
  P.~Anastasopoulos, P.~Betzios, M.~Bianchi, D.~Consoli and E.~Kiritsis,
  {\em ``Emergent/Composite axions,''}
\hri{1811.05940}{[hep-ph]}.

  \bibitem{dudas}
  K.~R.~Dienes, E.~Dudas and T.~Gherghetta,
  {\em ``Invisible axions and large radius compactifications,''}
  Phys.\ Rev.\ D {\bf 62} (2000) 105023
  doi:10.1103/PhysRevD.62.105023
  \hre{hep-ph}{9912455}.

\bibitem{self-tuning}
C.~Charmousis, E.~Kiritsis and F.~Nitti,
  {\em ``Holographic self-tuning of the cosmological constant,''}
  JHEP {\bf 1709} (2017) 031
  doi:10.1007/JHEP09(2017)031
 \hri{1704.05075}{[hep-th]}.

  \bibitem{hknw}
  Y. Hamada, E. Kiritsis, F. Nitti, L. Witkowski, work in progress.

\bibitem{iQCD-data}
U.~Gursoy, E.~Kiritsis, L.~Mazzanti and F.~Nitti,
  {\em ``Improved Holographic Yang-Mills at Finite Temperature: Comparison with Data,''}
  Nucl.\ Phys.\ B {\bf 820}, 148 (2009)
  \hri{0903.2859}{[hep-th]}.

\bibitem{Tada}
  T.~Azeyanagi, W.~Li and T.~Takayanagi,
  {\em ``On String Theory Duals of Lifshitz-like Fixed Points,''}
  JHEP {\bf 0906} (2009) 084
  doi:10.1088/1126-6708/2009/06/084
\hri{0905.0688}{[hep-th]}.

\bibitem{Mateos}
  D.~Mateos and D.~Trancanelli,
  {\em ``The anisotropic N=4 super Yang-Mills plasma and its instabilities,''}
  Phys.\ Rev.\ Lett.\  {\bf 107} (2011) 101601
  doi:10.1103/PhysRevLett.107.101601
\hri{1105.3472}{[hep-th]};\\
{\em ``Thermodynamics and Instabilities of a Strongly Coupled Anisotropic Plasma,''}
  JHEP {\bf 1107} (2011) 054
  doi:10.1007/JHEP07(2011)054
\hri{1106.1637}{[hep-th]}.

\bibitem{linear-axion}
  Y.~Bardoux, M.~M.~Caldarelli and C.~Charmousis,
  {\em ``Conformally coupled scalar black holes admit a flat horizon due to axionic charge,''}
  JHEP {\bf 1209} (2012) 008
  doi:10.1007/JHEP09(2012)008
  \hri{1205.4025}{[hep-th]}.

\bibitem{monodromy}
  E.~Silverstein and A.~Westphal,
  {\em ``Monodromy in the CMB: Gravity Waves and String Inflation,''}
  Phys.\ Rev.\ D {\bf 78} (2008) 106003
  doi:10.1103/PhysRevD.78.106003
  \hri{0803.3085}{[hep-th]}.

\bibitem{monodromy2}
  L.~McAllister, E.~Silverstein and A.~Westphal,
  {\em ``Gravity Waves and Linear Inflation from Axion Monodromy,''}
  Phys.\ Rev.\ D {\bf 82} (2010) 046003
  doi:10.1103/PhysRevD.82.046003
  \hri{0808.0706}{[hep-th]}.

\bibitem{1011.4521}
  X.~Dong, B.~Horn, E.~Silverstein and A.~Westphal,
  {\em ``Simple exercises to flatten your potential,''}
  Phys.\ Rev.\ D {\bf 84} (2011) 026011
  doi:10.1103/PhysRevD.84.026011
  \hri{1011.4521}{[hep-th]}.

\bibitem{1411.2032}
  A.~Hebecker, P.~Mangat, F.~Rompineve and L.~T.~Witkowski,
  {\em ``Tuning and Backreaction in F-term Axion Monodromy Inflation,''}
  Nucl.\ Phys.\ B {\bf 894} (2015) 456
  doi:10.1016/j.nuclphysb.2015.03.015
  \hri{1411.2032}{[hep-th]}.

\bibitem{1602.06517}
  F.~Baume and E.~Palti,
  {\em ``Backreacted Axion Field Ranges in String Theory,''}
  JHEP {\bf 1608} (2016) 043
  doi:10.1007/JHEP08(2016)043
  \hri{1602.06517}{[hep-th]}.

\bibitem{1611.00394}
  I.~Valenzuela,
  {\em ``Backreaction Issues in Axion Monodromy and Minkowski 4-forms,''}
  JHEP {\bf 1706} (2017) 098
  doi:10.1007/JHEP06(2017)098
  \hri{1611.00394}{[hep-th]}.

\bibitem{Gubser}
S.~S.~Gubser,
  {\em ``Curvature singularities: The Good, the bad, and the naked,''}
  Adv.\ Theor.\ Math.\ Phys.\  {\bf 4} (2000) 679
  \hre{hep-th}{0002160}.
  
  \bibitem{witten1}
  E.~Witten,
  {\em ``Large N Chiral Dynamics,''}
  \href{http://doi.org/10.1016/0003-4916(80)90325-5}{Annals Phys.\  {\bf 128} (1980) 363}.

\bibitem{VW}
  C.~Vafa and E.~Witten,
  {\em ``Parity Conservation in QCD,''}
  \href{http://doi.org/10.1103/PhysRevLett.53.535}{Phys.\ Rev.\ Lett.\  {\bf 53} (1984) 535}.



  \bibitem{Papa2}
  I.~Papadimitriou,
  {\em ``Holographic Renormalization of general dilaton-axion gravity,''}
  JHEP {\bf 1108} (2011) 119
  doi:10.1007/JHEP08(2011)119
\hri{1106.4826}{[hep-th]}.

\bibitem{CHS}
  T.~Crisford, G.~T.~Horowitz and J.~E.~Santos,
  {\em ``Testing the Weak Gravity - Cosmic Censorship Connection,''}
  Phys.\ Rev.\ D {\bf 97} (2018) no.6,  066005
  doi:10.1103/PhysRevD.97.066005
\hri{1709.07880}{[hep-th]}.

\bibitem{ArkaniHamed:2006dz}
  N.~Arkani-Hamed, L.~Motl, A.~Nicolis and C.~Vafa,
  {\em ``The String landscape, black holes and gravity as the weakest force,''}
  JHEP {\bf 0706}, 060 (2007)
  doi:10.1088/1126-6708/2007/06/060
  \hre{hep-th}{0601001}.

  \bibitem{Rudelius:2014wla}
  T.~Rudelius,
  {\em ``On the Possibility of Large Axion Moduli Spaces,''}
  JCAP {\bf 1504} (2015) no.04,  049
  doi:10.1088/1475-7516/2015/04/049
  \hri{1409.5793}{[hep-th]};
{\em ``Constraints on Axion Inflation from the Weak Gravity Conjecture,''}
  JCAP {\bf 1509} (2015) no.09,  020
  doi:10.1088/1475-7516/2015/09/020, 10.1088/1475-7516/2015/9/020
  \hri{1503.00795}{[hep-th]}.

  \bibitem{Montero:2015ofa}
  M.~Montero, A.~M.~Uranga and I.~Valenzuela,
  {\em ``Transplanckian axions!?,''}
  JHEP {\bf 1508} (2015) 032
  doi:10.1007/JHEP08(2015)032
  \hri{1503.03886}{[hep-th]}.

  \bibitem{Brown:2015iha}
  J.~Brown, W.~Cottrell, G.~Shiu and P.~Soler,
  {\em ``Fencing in the Swampland: Quantum Gravity Constraints on Large Field Inflation,''}
  JHEP {\bf 1510} (2015) 023
  doi:10.1007/JHEP10(2015)023
   \hri{1503.04783}{[hep-th]}.

  \bibitem{Long:2016jvd}
  C.~Long, L.~McAllister and J.~Stout,
  {\em ``Systematics of Axion Inflation in Calabi-Yau Hypersurfaces,''}
  JHEP {\bf 1702} (2017) 014
  doi:10.1007/JHEP02(2017)014
  \hri{1603.01259}{[hep-th]}.

  \bibitem{Dolan:2017vmn}
  M.~J.~Dolan, P.~Draper, J.~Kozaczuk and H.~Patel,
  {\em ``Transplanckian Censorship and Global Cosmic Strings,''}
  JHEP {\bf 1704} (2017) 133
  doi:10.1007/JHEP04(2017)133
  \hri{1701.05572}{[hep-th]}.

  \bibitem{Hebecker:2017lxm}
  A.~Hebecker, P.~Henkenjohann and L.~T.~Witkowski,
  {\em ``Flat Monodromies and a Moduli Space Size Conjecture,''}
  JHEP {\bf 1712} (2017) 033
  doi:10.1007/JHEP12(2017)033
  \hri{1708.06761}{[hep-th]}.

\bibitem{Vafa:2005ui}
  C.~Vafa,
  {\em ``The String landscape and the swampland,''}
  \hre{hep-th}{0509212}.

\bibitem{Klaewer:2016kiy}
  D.~Klaewer and E.~Palti,
  {\em ``Super-Planckian Spatial Field Variations and Quantum Gravity,''}
  JHEP {\bf 1701} (2017) 088
  doi:10.1007/JHEP01(2017)088
  \hri{1610.00010}{[hep-th]}.
  

  \bibitem{witten2}
  E.~Witten,
  {\em ``Anti-de Sitter space, thermal phase transition, and confinement in gauge theories,''}
  Adv.\ Theor.\ Math.\ Phys.\  {\bf 2} (1998) 505
  doi:10.4310/ATMP.1998.v2.n3.a3
  \hre{hep-th}{9803131}.

  \bibitem{gg}
  M.~B.~Green and M.~Gutperle,
  {\em ``Effects of D instantons,''}
  Nucl.\ Phys.\ B {\bf 498} (1997) 195
  doi:10.1016/S0550-3213(97)00269-1
  \hre{hep-th}{9701093}.

  \bibitem{kp}
  E.~Kiritsis and B.~Pioline,
  {\em ``On R$^4$ threshold corrections in IIb string theory and (p, q) string instantons,''}
  Nucl.\ Phys.\ B {\bf 508} (1997) 509
  doi:10.1016/S0550-3213(97)00645-7
  \hre{hep-th}{9707018}.

\bibitem{GK2}
  B.~Gouteraux and E.~Kiritsis,
  {\em ``Generalized Holographic Quantum Criticality at Finite Density,''}
  JHEP {\bf 1112} (2011) 036
  doi:10.1007/JHEP12(2011)036
\hri{1107.2116}{[hep-th]}.


  \bibitem{CS}
  U.~Gursoy, I.~Iatrakis, E.~Kiritsis, F.~Nitti and A.~O'Bannon,
  {\em ``The Chern-Simons Diffusion Rate in Improved Holographic QCD,''}
  JHEP {\bf 1302} (2013) 119
  doi:10.1007/JHEP02(2013)119
  \hri{1212.3894}{[hep-th]}.

\bibitem{DeWolfe:1999cp}
  O.~DeWolfe, D.~Z.~Freedman, S.~S.~Gubser and A.~Karch,
  {\em ``Modeling the fifth-dimension with scalars and gravity,''}
  Phys.\ Rev.\ D {\bf 62}, 046008 (2000);
\hre{hep-th}{9909134}.

\bibitem{exotic}
  E.~Kiritsis, F.~Nitti and L.~S.~Pimenta,
  {\em ``Exotic RG Flows from Holography,''}
  Fortsch.\ Phys.\ {\bf 65} (2017) no.2, 1600120;
  doi:10.1002/prop.201600120
\hri{1611.05493}{[hep-th]}.

\bibitem{curved}
  J.~K.~Ghosh, E.~Kiritsis, F.~Nitti and L.~T.~Witkowski,
  {\em ``Holographic RG flows on curved manifolds and quantum phase transitions,''}
  JHEP {\bf 1805} (2018) 034
  doi:10.1007/JHEP05(2018)034
\hri{1711.08462}{[hep-th]}.

\bibitem{thermo}
U.~Gursoy, E.~Kiritsis, L.~Mazzanti and F.~Nitti,
  {\em ``Holography and Thermodynamics of 5D Dilaton-gravity,''}
  JHEP {\bf 0905}, 033 (2009)
\hri{0812.0792}{[hep-th]}.

  \bibitem{GK}
  C.~Charmousis, B.~Gouteraux, B.~S.~Kim, E.~Kiritsis and R.~Meyer,
  {\em ``Effective Holographic Theories for low-temperature condensed matter systems,''}
  JHEP {\bf 1011} (2010) 151
  doi:10.1007/JHEP11(2010)151
\hri{1005.4690}{[hep-th]};\\
  B.~Gouteraux and E.~Kiritsis,
  {\em ``Quantum critical lines in holographic phases with (un)broken symmetry,''}
  JHEP {\bf 1304} (2013) 053
  doi:10.1007/JHEP04(2013)053
  \hri{1212.2625}{[hep-th]}.

  \bibitem{DG}
  R.~A.~Davison, S.~A.~Gentle and B.~Gouteraux,
  {\em ``Slow relaxation and diffusion in holographic quantum critical phases,''}
  \hri{1808.05659}{[hep-th]};
  {\em ``Impact of irrelevant deformations on thermodynamics and transport in holographic quantum critical states,''}
  \hri{1812.11060}{ [hep-th]}.

\bibitem{Witten-th}
  E.~Witten,
  {\em ``Large N Chiral Dynamics,''}
  \href{http://doi.org/doi:10.1016/0003-4916(80)90325-5}{Annals Phys.\  {\bf 128} (1980) 363}.



  \bibitem{bourdier}
  J.~Bourdier and E.~Kiritsis,
  {\em ``Holographic RG flows and nearly-marginal operators,''}
  Class.\ Quant.\ Grav.\  {\bf 31} (2014) 035011
  doi:10.1088/0264-9381/31/3/035011
\hri{1310.0858}{[hep-th]}.

\bibitem{S}
  L.~Huijse, S.~Sachdev and B.~Swingle,
  {\em ``Hidden Fermi surfaces in compressible states of gauge-gravity duality,''}
  Phys.\ Rev.\ B {\bf 85} (2012) 035121
  doi:10.1103/PhysRevB.85.035121
 \hri{1112.0573}{[cond-mat.str-el]}.



  \bibitem{kn}
  E.~Kiritsis and V.~Niarchos,
  {\em ``Interacting String Multi-verses and Holographic Instabilities of Massive Gravity,''}
  Nucl.\ Phys.\ B {\bf 812} (2009) 488;
  \hri{0808.3410}{[hep-th]}.


\bibitem{hrg}
 E.~Kiritsis, W.~Li and F.~Nitti,
  {\em ``Holographic RG flow and the Quantum Effective Action,''}
  Fortsch.\ Phys.\  {\bf 62} (2014) 389
  doi:10.1002/prop.201400007
  \hri{1401.0888}{[hep-th]}.


\bibitem{glueball}
E.~Kiritsis, W.~Li and F.~Nitti,
  {\em ``On the gluonic operator effective potential in holographic Yang-Mills theory,''}
  JHEP {\bf 1504}, 125 (2015)
  doi:10.1007/JHEP04(2015)125
  \hri{1410.1091}{[hep-th]}.

\bibitem{Papa1}
  I.~Papadimitriou,
  {\em ``Multi-Trace Deformations in AdS/CFT: Exploring the Vacuum Structure of the Deformed CFT,''}
  JHEP {\bf 0705} (2007) 075
  doi:10.1088/1126-6708/2007/05/075
  \hre{hep-th}{0703152}.



























\end{thebibliography}
\end{document}